\documentclass[%
reprint,
superscriptaddress,
 amsmath,amssymb,
aps,
floatfix,
pra]{revtex4-2}

\usepackage{graphicx}
\usepackage{dcolumn}
\usepackage{color}
\usepackage{bm}
\usepackage{hyperref}
\usepackage{upgreek}

\usepackage[normalem]{ulem}

\usepackage{tcolorbox}
\usepackage{braket}
\newcommand{\mean}[1]{\ensuremath{\left\langle#1\right\rangle}}
\newcommand{\nbth}{n_b^{\mathrm{th}}}

\usepackage{setspace}

\renewcommand{\selectlanguage}[1]{}

\begin{document}

\title{Molecular optomechanics with atomic antennas}

\author{Miko\l aj K. Schmidt}
\affiliation{
School of Mathematical and Physical Sciences, Macquarie University, North Ryde NSW 2109, Australia
}
\author{Alexander A. High}
\affiliation{Pritzker School of Molecular Engineering$,$ University of Chicago$,$ Chicago$,$ Illinois 60637$,$ USA}
\affiliation{Center for Molecular Engineering and Materials Science Division$,$ Argonne National Laboratory$,$\\Lemont$,$ Illinois 60439$,$ USA}
\author{Michael J. Steel}

\affiliation{
School of Mathematical and Physical Sciences, Macquarie University, North Ryde NSW 2109, Australia
}

\date{\today}

\begin{abstract}
A typical surface-enhanced Raman scattering (SERS) system relies on deeply subwavelength field localization in nanoscale plasmonic cavities to enhance both the excitation and emission of Raman-active molecules. Here, we demonstrate that a germanium-vacancy (GeV) defect in diamond can efficiently mediate the excitation process, by acting as a bright \textit{atomic antenna}. At low temperatures, the GeV’s low dissipation allows it to be efficiently populated by the incident field, resulting in a thousand-fold increase in the efficiency of Raman scattering. We show that atomic antenna-enhanced Raman scattering can be distinguished from conventional SERS by tracing the dependence of Stokes intensity on input power.
\end{abstract}

\maketitle


\section{Introduction}

Raman spectroscopy is a highly versatile technique for detecting and studying molecules and proteins, and despite over four decades of research, it continues to provoke new research questions and yield surprising experimental results. In particular, in the last decade a new class of plasmonic nano- and picocavities~\cite{jakob_giant_2023,benz_single-molecule_2016,urbieta_atomic-scale_2018,wu_bright_2021} offering extreme field localisation has enabled a new level of capability. This includes  imaging the  internal structure of molecules with sub-nanometer resolution~\cite{zhang_chemical_2013}, observing the stretching of individual atomic bonds~\cite{lee_visualizing_2019}, and ultimately controlling their chemistry~\cite{roslawska_submolecular-scale_2024}, while relying on only  visible light. These studies have also extended beyond plasmonic systems, with several groups pursuing the realisation of hybrid cavities, which inherit strong localisation of light from plasmonics, and small dissipation from dielectrics~\cite{shlesinger_integrated_2023,shlesinger_integrated_2021,dezfouli_molecular_2019}. 

\begin{figure}[ht!]
	\centering
    \includegraphics[width=\linewidth]{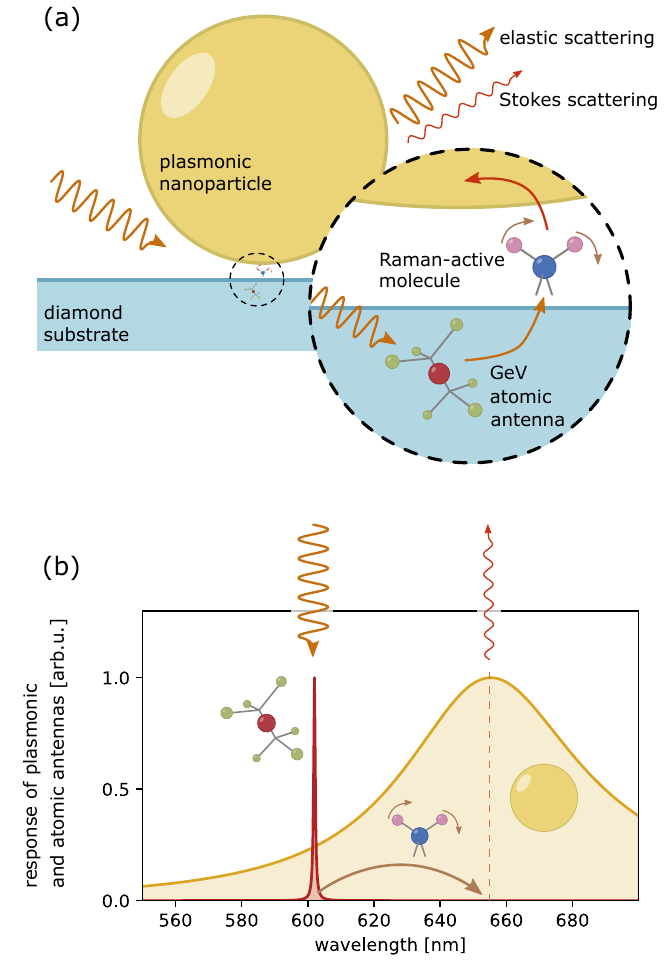}
	\caption{(a) Schematic and (b) idealized spectrum of atomic antenna-enhanced Raman scattering. The incident laser at frequency $\omega_l$ couples to an atomic antenna, which localizes the electric field on the nearby Raman-active molecule. The Raman emission (here shown as a Stokes line) from the molecule is mediated by the plasmonic antenna.}\label{fig:1}
\end{figure}

These developments in alternative platforms have been accompanied by novel theoretical formalisms. In particular, ``molecular optomechanics''~\cite{roelli_molecular_2016,schmidt_quantum_2016,schmidt_linking_2017} aims to describe cavity-enhanced Raman scattering by embracing a quantum-mechanical picture of both the optical mode (or modes) and the molecular vibrations. To date, molecular optomechanics has provided insights into various fundamental phenomena such as the optical spring effect and linewidth narrowing {and broadening}~\cite{jakob_giant_2023}, dynamical instabilities~\cite{lombardi_pulsed_2018}, and correlations between the Stokes and anti-Stokes emission~\cite{schmidt_quantum_2016,schmidt_frequency-resolved_2021}. It has also provided a convenient formalism for describing upconversion of IR photons via Raman transitions~\cite{roelli_molecular_2020,chen_continuous-wave_2021,xomalis_detecting_2021}, and offered insights into the effects of molecular anharmonicities~\cite{schmidt_molecular_2024,kalarde_photon_2024}, and coherences between vibrations of neighboring molecules~\cite{zhang_optomechanical_2020}.

The framework of molecular optomechanics also transparently describes the dual effect of plasmonic cavities in Tip-Enhanced (TERS) and Surface-Enhanced (SERS) Raman Spectrosocpy systems~\cite{itoh_toward_2023,hoppener_tip-enhanced_2024}. Specifically, cavities serve to (i) localize the field around the molecule, enhancing its driving to a virtual state, and (ii) boost the Stokes and anti-Stokes emission rates by modifying the density of electromagnetic states. In conventional plasmonic SERS, both effects are associated with either one, or several modes of the plasmonic cavity~\cite{zhang_addressing_2021-1,kamandar_dezfouli_quantum_2017}, which have linewidths $\kappa$ similar to slightly smaller than the mechanical frequency of the molecular vibrations ($\kappa \lesssim \omega_b$). The efficiency of this process can be described by a simple figure of merit $Q/V_\text{eff}^2$, defined through the effective mode volume $V_\text{eff}$ and quality factor $Q$ of the optical mode.

Here we propose an alternative platform to these designs in the form of a system incorporating an atomic defect --- specifically, a germanium vacancy (GeV) in diamond --- and a plasmonic nanoparticle (see schematic in Fig.~\ref{fig:1}(a)). At low temperatures, GeV defects offer two key characteristics ideal for a Raman-enabling antenna: a very low decoherence rate of order $\gamma/2\pi\approx 100$~MHz (about 4 times larger than its lifetime transformed decay rate $\gamma_0$), and an unprecedented degree of localisation of the electromagnetic fields. The former means that by acting as an \textit{atomic antenna}~\cite{li_atomic_2024}, GeV can be far more efficiently excited than any mode of a plasmonic cavity, while the latter ensures that it can drive Raman processes in nearby molecules. At first sight, one might hope to replace both resonant functions of the plasmonic nanoparticle described above with the atomic antenna. However, the atomic linewidth is too narrow to accommodate both the pump and Raman-shifted lines, and it must be partnered with another component, e.g. a gold nanoparticle.

In this work we assess the viability of this setup, which we dub atomic antenna-enhanced Raman scattering. We ask whether atomic antennas can provide effective excitation of the Raman processes, and how is this reflected by the parameters of molecular optomechanics? Can the system offer stronger perturbation of vibrational populations, perhaps leading to dynamical instabilities? How will the efficiency of the atomic antenna-enhanced Raman scattering compare to that of conventional plasmonic SERS, and will the intrinsic nonlinearity of the atomic antenna bring about new effects that would differentiate its effects from those of a conventional plasmonic setup?

The paper is constructed as follow: in Section~\ref{sec:formulation} we modify the formalism of molecular optomechanics to explore the differences between the atomic antenna and conventional SERS systems, and compare the two for a set of characteristic parameters in Section~\ref{sec:comparison.to.sers}. Finally, in Section~\ref{sec:nonlinearities} we discuss the effects of the incoherently nonlinear response of the atomic antenna and the resulting characteristics of Raman response.

\section{Results and discussion}

\subsection{Atomic antenna-enhanced Raman scattering}\label{sec:formulation}

{A quantum-mechanical description of the off-resonant Raman scattering from a system schematically depicted in Fig.~\ref{fig:1}(a) can be formulated by explicitly decomposing the optical response of the environment into modes of the plasmonic cavity, and the intrinsically nonlinear atomic antenna. Below we present an abbreviated description of the atomic antenna-enhanced Raman scattering, and direct the Reader to Appendix~\ref{app:derivation} for a more complete formulation of this problem.}

\subsubsection{Coherent dynamics}\label{subsec:coherent}

We consider a Raman dipole $\hat{\mathbf{p}}_R$, induced in the molecule positioned at $\mathbf{r}_m$ by the electric field $\hat{\mathbf{E}}_\sigma$ generated by the excited atomic antenna: $\hat{\mathbf{p}}_R = \hat{x} R:\hat{\mathbf{E}}_\sigma$. Here $R$ is the Raman tensor, and $\hat{x}$ is the \textit{generalized coordinate} operator which describes the coordinates of the atoms in the molecule. In the lowest-order approximation~\cite{schmidt_molecular_2024-1}, we quantize it as a one-dimensional harmonic oscillator, with zero-point fluctuation $Q^0=\sqrt{\hbar/(2\omega_b)}$, frequency $\omega_b$, and bosonic operators $\hat{b}/\hat{b}^\dagger$ describing the annihilation/creation of the optical phonon. The Raman dipole interacts with the electric field of a particular mode of the plasmonic cavity $\hat{\mathbf{E}}_a$, as described by the interaction Hamiltonian~\cite{roelli_molecular_2016,schmidt_quantum_2016}
\begin{align}
    \hat{H}_I &= -\frac{1}{2}\hat{\mathbf{p}}_R \cdot \hat{\mathbf{E}}_a(\mathbf{r}_m) \nonumber\\
    & = -\frac{1}{2} \Big(\hat{\mathbf{E}}_\sigma(\mathbf{r}_m)\Big)_i R_{ij} \hat{x} \Big(\hat{\mathbf{E}}_a(\mathbf{r}_m)\Big)_j \nonumber \\
    & = -\frac{1}{2} R_{ij} Q^0 (\hat{b}+\hat{b}^\dag) \Big(\hat{\mathbf{E}}_\sigma(\mathbf{r}_m)\Big)_i \Big(\hat{\mathbf{E}}_a(\mathbf{r}_m)\Big)_j.
\end{align}
To express the electric field from the atomic antenna positioned at $\mathbf{r}_\sigma$, we model it as a two-level system with frequency $\omega_\sigma$, and pseudo-spin operators $\hat{\sigma} = |g\rangle\langle e|$. Re-interpreting the field amplitudes as being generated by the atomic transition dipole moment with matrix element $\mathbf{d}_\sigma = \langle e|\hat{\mathbf{d}}|g\rangle$, and using the electromagnetic Green's function of the system 
$\mathbf{G}(\mathbf{r}_\text{src},\mathbf{r}_\text{dest}, \omega)$, we find
\begin{equation}\label{eq:Greens}
    \hat{\mathbf{E}}_{\sigma}(\mathbf{r}_m) = \omega_\sigma^2 \mu_0 \mathbf{G}(\mathbf{r}_\sigma,\mathbf{r}_m,\omega_\sigma) \cdot \mathbf{d}_\sigma \hat{\sigma} + \text{h.c.}.
\end{equation}
The electric field operator of the plasmonic cavity mode is expressed in terms of the associated classical mode function $\mathbf{e}_{a}$ and annihilation/creation operators $\hat{a}/\hat{a}^\dagger$ for the plasmon polariton, such that  
\begin{equation}
    \hat{\mathbf{E}}_{a}(\mathbf{r}_m) = \mathbf{e}_{a}(\mathbf{r}_m) \hat{a} + \text{h.c.},
\end{equation}
where $\mathbf{e}_{a}$ carries the fast harmonic time dependence, information on the spatial distribution, and the normalisation conditions. 

The interaction Hamiltonian (we omit the position and frequency dependencies)
\begin{align}\label{eq:interaction.hamiltonian}
    \hat{H}_I= -&\frac{1}{2} R_{ij} Q^0 (\hat{b}+\hat{b}^\dag) \left[\omega_\sigma^2 \mu_0 (\mathbf{G} \cdot \mathbf{d}_\sigma)_i \hat{\sigma} + \text{h.c.}\right] \nonumber \\
    &\times\left[(\mathbf{E}_a)_j \hat{a} + \text{h.c.}\right],
\end{align}
is accompanied by the free-energy Hamiltonian $\hat{H}_0$ terms describing the energy of the vibrations, atomic antenna and the optical cavity
\begin{equation}\label{eq:H0}
    \hat{H}_0= \hbar\omega_b \hat{b}^\dag\hat{b} + \hbar\omega_\sigma \hat{\sigma}^\dag \hat{\sigma} + \hbar\omega_a \hat{a}^\dag\hat{a},
\end{equation}
and a term $\hat{H}_\text{coh}$ describing the coherent driving of the atomic antenna by an incident laser
\begin{equation}
    \hat{H}_\text{coh}= \hbar\Omega\left[e^{-i\omega_lt} \hat{\sigma} + e^{i\omega_lt} \hat{\sigma}^\dag\right],
\end{equation}
where
\begin{equation}\label{eq:Omega}
    \hbar\Omega = |\mathbf{d}_\sigma|t_p \sqrt{\frac{4I}{c \varepsilon_0}},    
\end{equation}
is the coherent driving amplitude from the laser with intensity $I$. Note that we explicitly account for the change in the intensity of the $p$-polarized incident laser due to the transmission from air to the diamond substrate, by including the Fresnel transmission coefficient $t_p$ in the definition of $\Omega$. 

{
In Appendix~\ref{app:derivation} we discuss additional effects that will influence the Raman response of the system, including the coherent driving of the plasmonic mode by incident laser, conventional Raman scattering mediated by that plasmonic mode, and finally its resonant Jaynes-Cummings coupling to the atomic antenna.}


\subsubsection{Dissipative dynamics}\label{subsec:dissipation}
To account for the dissipation in the system, we embrace the formalism of open quantum systems, in which the state of the quantum system, comprising the two-level system modeling the atomic antenna, plasmonic cavity mode, and vibrations, is described with a density matrix $\hat{\rho}$, and its dynamics is governed by a master equation. The details of this description are discussed in previous contributions on molecular, and cavity optomechanics~\cite{schmidt_linking_2017,aspelmeyer_cavity_2014,roelli_nanocavities_2024,marquardt_quantum_2008,clerk_introduction_2010}. Here we only clarify that the dissipation of the optical cavity mode is incorporated into the master equation via the Gorini–Kossakowski–Sudarshan–Lindblad (GKSL) operators $(\kappa/2)\mathcal{D}[\hat{a}]$, where  $\mathcal{D}[\hat{O}] \rho = 2\hat{O}\rho \hat{O}^\dag - \hat{O}^\dag\hat{O}\rho- \rho\hat{O}^\dag\hat{O}$. Similarly, the interaction of the vibrational mode with the thermal environment --- both decay and incoherent pumping --- is described by terms $(\Gamma/2)(\nbth + 1)\mathcal{D}[\hat{b}]$ and $(\Gamma/2)\nbth\mathcal{D}[\hat{b}^\dagger]$, respectively, where $\nbth$ is the thermal population of the environment at the frequency $\omega_b$. The decay of the atomic antenna is  included as $(\gamma/2)\mathcal{D}[\hat{\sigma}]$. 

\subsubsection{Linearization and effective optomechanical coupling}\label{subsec:linearisation}
The complete Hamiltonian $\hat{H}_0+\hat{H}_I+\hat{H}_\text{coh}$ can be \textit{linearized} by realising that in the limit of weak coherent pumping (the so-called \textit{Heitler} regime) and weak optomechanical interaction, the atomic antenna ($\hat{\sigma}$) will be driven by the incident laser into a predominantly coherent state with amplitude
\begin{equation}\label{eq:coherence}
    \alpha_\sigma = \frac{i\Omega}{-(\omega_\sigma-\omega_l)+i\gamma/2}\approx \mean{\hat{\sigma}}.
\end{equation}
The original papers on molecular optomechanics considered only one plasmonic mode, and attributed the Raman-generated sidebands in the optical mode with the fluctuations in $\hat{a}$ around its mean. We revisit this approximation in the last section of the paper, where we explore the \textit{Mollow} limit of strong pumping of the atomic antenna. 

The linearized Hamiltonian, in the rotating wave approximation, and with some elementary assumptions about the isotropic nature of the Raman tensor, takes on a simpler form (see Appendix~\ref{app:derivation} for derivation)
\begin{align}\label{eq:hamiltonian.1}
    \hat{H}_{I,\text{lin}} &=  -\underbrace{\frac{1}{2} R Q^0 \omega_\sigma^2 \mu_0 \mathbf{e}_a \cdot (\mathbf{G} \cdot \mathbf{d}_\sigma)^*}_{\hbar g_{0,\sigma}} \alpha_\sigma \left(\hat{a} + \hat{a}^\dag\right) (\hat{b}+\hat{b}^\dag)\nonumber \\
    & =-\hbar g_{0,\sigma} \alpha_\sigma \left(\hat{a} + \hat{a}^\dag\right) (\hat{b}+\hat{b}^\dag).
\end{align}
The coefficient $g_{\sigma} = g_{0,\sigma} \alpha_\sigma$ is the \textit{effective optomechanical coupling}, which accounts for the population of the atomic antenna. In conventional cavity optomechanics, these populations can be very large, and compensate for the normally weak $g_0$~\cite{aspelmeyer_cavity_2014}. Conversely, plasmonic cavities are characterized by very large losses, and cannot be easily populated far above one plasmon ($|\alpha_a|=1$). For an atomic antenna, the coherence similarly cannot exceed $|\alpha_\sigma| = 2^{-3/2}$~\cite{carmichael2013statistical}. 

{ The exceptionally strong single-photon coupling identified in molecular optomechanics~\cite{benz_single-molecule_2016,roelli_molecular_2016} is due to the strong confinement of the electromagnetic field in deeply subwavelength volumes $V_\text{eff}$ (via the definition of $g_{0,a}\propto V_\text{eff}^{-1}$, see Appendix~\ref{app:conventional}). In the hybrid system discussed here, the single-photon optomechanical coupling $g_{0,\sigma}$ is proportional to the field profile $|\mathbf{e}_a|\propto V_\text{eff}^{-1/2}$.}

\subsubsection{Emission spectrum}\label{subsec:emission}
Let us consider the spectrum of emission from the system, assuming that the inelastically scattered Stokes and anti-Stokes photons are emitted from the molecule \textit{into the plasmonic cavity}. 
If the cavity emits light as a dipolar scatterer, its incoherent emission spectrum will be given by
\begin{equation}\label{eq:emission.a}
    S_a(\omega) = \omega^4 \int_{-\infty}^\infty \textrm{d}\tau~e^{i\omega \tau}\braket{\hat{a}^\dag(\tau)  \hat{a}(0)}.
\end{equation}
We appreciate that the frequency prefactor is typically not included in the definitions of the emission spectra, but we include it here to directly relate $S_a(\omega)$ with the spectral density of the emitted power (see discussion in Appendix~\ref{app:powers}). In Appendix~\ref{app:spectrum}, we show that this spectrum is given by
\begin{align}\label{eq:spectrum.final}
    S_a&(\omega) = 2\omega^4 |g_\sigma|^2  \nonumber \\ 
    & ~~ \times  \Bigg[
    \frac{n_b}{(\Delta_a - \omega_b)^2 +(\kappa/2)^2} \frac{\Gamma}{(\omega_b + \omega_l - \omega)^2+(\Gamma/2)^2}  \nonumber \\
    &~~~~~~+\frac{n_b+1}{(\Delta_a + \omega_b)^2 +(\kappa/2)^2} \frac{\Gamma}{(\omega_b - \omega_l +\omega)^2+(\Gamma/2)^2}\Bigg],
\end{align}
where $\Delta_a = \omega_a - \omega_l$, and $n_b=\langle \hat{b}^\dagger \hat{b}\rangle $ is the population of molecular vibrations, discussed below. At the Stokes ($\omega_S= \omega_l-\omega_b$) and anti-Stokes frequencies ($\omega_{aS}=\omega_l + \omega_b$) the second, and first terms dominate, respectively:
\begin{align}\label{eq:Stokes.b}
    S_a(\omega_S) & \approx \frac{2\omega_S^4 |g_\sigma|^2 (n_b+1)}{[(\Delta_a + \omega_b)^2 +(\kappa/2)^2]\Gamma}, \\
    S_a(\omega_{aS}) & \approx \frac{2\omega_{aS}^4 |g_\sigma|^2 n_b}{[(\Delta_a - \omega_b)^2 +(\kappa/2)^2]\Gamma}.
\end{align}

\subsubsection{Vibrational populations}\label{subsec:populations}
In the absence of optical driving, the population of vibrations is determined by the thermal environment, leading to $n_b=n_b^\text{th}$. However, the Stokes and anti-Stokes transitions continuously add, and remove phonons from the molecule, respectively. The rates of these processes ($\Gamma_\pm$) are proportional to the emission intensities $S_a(\omega_S)$ and $S_a(\omega_{aS})$ respectively, and can be calculated as~\cite{wilson-rae_theory_2007,aspelmeyer_cavity_2014} 
\begin{equation}
    \Gamma_\pm = \frac{g_\sigma^2 \kappa}{(\Delta_a \pm \omega_b)^2+(\kappa/2)^2}.
\end{equation}
We can then formulate the rate equation for the  phonon population  as (see derivation in Ref.~\citenum{schmidt_linking_2017})
\begin{equation}
    \frac{\textrm{d}}{\textrm{d}t} n_b = -(\Gamma + \Gamma_-) n_b + \Gamma_+ (n_b+1) + \Gamma n_b^\text{th},
\end{equation}
with the steady-state solution
\begin{equation}\label{eq:rate.solution}
    n_{b,ss} = \frac{\Gamma n_b^\text{th}}{\Gamma + \Gamma_- - \Gamma_+} + \frac{\Gamma_+}{\Gamma + \Gamma_- - \Gamma_+}.
\end{equation}
The effective dissipation rate of the phonon mode is also modified by these processes as $\Gamma \rightarrow \Gamma_\text{eff} = \Gamma + \Gamma_- - \Gamma_+$. Both effects (the changes in the steady-state phonon population  and the linewidth) should be included into (Eq.~\eqref{eq:spectrum.final}) for the spectrum by substituting $n_b \rightarrow n_{b,ss}$ and $\Gamma \rightarrow \Gamma_\text{eff}$.

Typical Raman scattering experiments are carried out at room temperature, where the thermal phonon populations $n_b^\text{th}$ are of the order of $10^{-1}$, and at laser intensities where the modification of the dissipation rate is negligible. In fact, {to our best knowledge the first} conclusive observation of the change of phonon population due to \textit{vibrational pumping} (described by the second term on the right-hand side of Eq.~\eqref{eq:rate.solution}), { indicated by a quadratic dependence of the anti-Stokes intensity on pump laser power, was reported by Kneipp \textit{et al.} in Ref.~[\citenum{PhysRevLett.76.2444}]. This study spurred a decade-long effort to understand the different mechanisms that lead to the population of molecular vibrations, and to confirm the central role of vibrational pumping \cite{maher2006conclusive,vitkova_cryogenically_2022,yi_surface-enhanced_2025,shen_optomechanical_2023}}. Observation of a build-up of phonon populations due to the optomechanical backaction, represented by the reduction of the effective phonon damping rate $\Gamma_\text{eff}$, required the use of pulsed lasers with large peak powers, to avoid destroying the plasmonic cavities~\cite{lombardi_pulsed_2018}.

Atomic defects, like the GeV discussed here, require low temperatures (optimally around $50$~K~\cite{li_atomic_2024}) to suppress the decoherence, and operate as high-$Q$ antennas. At those temperatures the thermal phonon population $n_b^\text{th}$ is further reduced. To estimate how strong that mechanisms can be, in the next section we estimate the parameters of a realistic scheme for atomic-antenna enhanced Raman scattering.


\subsection{Comparison to a conventional plasmonic SERS}\label{sec:comparison.to.sers}

Having developed the general framework for describing atomic antenna-enhanced Raman scattering, we now seek to compare its performance to that of a conventional SERS system, for which the formalism of molecular optomechanics was originally developed~\cite{roelli_molecular_2016,schmidt_quantum_2016}. For completeness, we list the expressions for relevant parameters in the latter setup in Appendix~\ref{app:conventional}.

In Table~\ref{tab:estimates} we present a comparison of parameters characterising an atomic antenna-enhanced Raman scattering scheme (in a setup schematically represented in Fig.~\ref{fig:1}), and a conventional plasmonic SERS setup with the same nominal geometry, but \textit{without} the atomic antenna, where the plasmonic cavity mode mediates both the excitation of the Raman dipole, and its emission (see the schematic of both systems in Fig.~\ref{fig:numerics}(a)). For simplicity, both systems use the same properties of the plasmonic cavity mode (frequency, linewidth and effective mode volume). We note that these parameters are not intended to describe a particular specific plasmonic system, but are characteristic of various nanoparticle-on-mirror and nanoparticle-in-grove platforms~\cite{roelli_nanocavities_2024}. To obtain simple expressions, we also invoke substantial approximations, most significantly (i) assuming that the Green's function of the system can be approximated by the free-space Green's function, (ii) taking the coherent driving amplitude $\Omega$ as defined in Eq.~\eqref{eq:Omega}, without accounting for additional focusing due to the nanoparticle, and (iii) treating the decay rate of the atomic antenna $\gamma$ as independent of the presence of the nanoparticle. We revisit all these approximations in Subsection~\ref{subsec:modeling}.

Parameters of the atomic antenna, including the transition dipole moment, are directly taken, or inferred, from Ref.~\cite{li_atomic_2024}~(see Appendix~\ref{app:parameters} for the detailed derivation). We assume a shallow location of the defect, 2~nm below the surface of the diamond. {Like other group-IV centres, germanium vacancies GeV exhibit inversion symmetry which decouples the electronic orbitals from the fluctuations of the electric field due to surface charges, protecting the shallow atomic antennas from dephasing. Having said that, we acknowledge that the quoted decoherence rate of $100$~MHz have been measured from bulk defects that are deposited deeper in the crystal.} We also emphasize that while this work focuses on coherent color centers in diamond, optically coherent defects are observed in several systems, including molecules, rare earth dopants, nanodiamonds, and atomically-thin layered materials such as hexagonal boron nitride~\cite{gottscholl2021room,anderson2022five,PhysRevX.12.031028,zhong2015nanophotonic,PhysRevLett.62.2535}. These systems are all suitable candidates for atomic-antenna enhanced Raman spectroscopy.

\begin{table*}
    \begin{tabular}{l|c|c}
                                           & Conventional plasmonic system                 & Atomic antenna \\\hline 
        Plasmon mode                         & \multicolumn{2}{c}{{(freq. $\omega_a$, linewidth $\kappa$)}$/2\pi = (498, 50)$~THz,} \\                      & \multicolumn{2}{c}{{mode volume} $V_\text{eff} = 10^3~\text{nm}^3 \approx \lambda^3/(2\times 10^5)$} \\
        & \multicolumn{2}{c}{{ext. cross section} $\sigma_\text{ext}= 10^{-13}~\text{m}^2$} \\ \hline
        Atomic antenna                          &        & {(freq. $\omega_\sigma$, rad. decay rate $\gamma_0$)}$/2\pi$ \\
        & & $= (498~\text{THz}, 25~\text{MHz})$\\
                    &              & {tot. decay rate} $\gamma = 4\gamma_0$ (at 50~K~\cite{li_atomic_2024})\\
                    & & {transition dip. mom.} $|\mathbf{d}_\sigma|\approx 6.7$~D\\ 
                    & & {GeV-mol. distance} $|\mathbf{r}_\sigma-\mathbf{r}_m|=3$~nm \\ \hline
        {Molecular vibration} & \multicolumn{2}{c}{{(freq. $\omega_b$, linewidth $\Gamma$)}$/2\pi = (40, 1)$~THz} \\ 
                    & \multicolumn{2}{c}{{Raman strength} $R Q^0 \approx 4\pi \varepsilon_0 2\times 10^{-30}~\text{m}^3$} \\  \hline
        Electric field from                    & $e_a \approx 1.4\times 10^8$~V/m                                                                                                                             & $e_\sigma \approx 7.5 \times 10^6$~V/m \\
        1 plasmon/exciton                 & (Eq.~(43) in SI)  & (Eq.~(38) in SI) \\ \hline
        Optomechanical             & $\hbar g_{0,a} \approx 1.4\times 10^{-5}$~eV  & $\hbar g_{0,\sigma} \approx 7\times 10^{-7}$~eV  (Eq.~\eqref{eq:hamiltonian.1}) \\
        coupling                   & $g_{0,a}/2\pi \approx 3.3\times 10^{9}$~Hz         & $g_{0,\sigma}/2\pi \approx 1.8\times 10^{8}$~Hz \\ 
                                    & (Eq.~(44) in SI) &  \\ 
        {(cooperativity)}        & {$C_{0,a} \approx 8.6\times 10^{-7}$} & {$C_{0,\sigma} \approx 1.3\times 10^{-3}$} \\ \hline
        Laser intensity required            & $I \approx 10^3~\upmu \text{W}\cdot \upmu\text{m}^{-2}$  
                                            & $I \approx 6.9\times 10^{-3}~\upmu \text{W}\cdot \upmu \text{m}^{-2}$  \\
        for 1 plasmon/exciton               &  &       
    \end{tabular}
    \caption{Parameters of the systems used to compare the performance of the conventional plasmonic system and atomic antenna. The properties of the plasmonic antenna are shared between the two models, although in the case of the latter system the role of the plasmonic cavity is limited to controlling the Raman emission. Details on the properties of the atomic antenna and the addressed Raman activity are given in Appendix~\ref{app:parameters}; equations used to analyse the response of a conventional plasmonic system are given in Appendix~\ref{app:conventional}. {Laser intensities given in the last row are calculated using Eq.~\eqref{eq:plasmon.threshold}, and $|\alpha_\sigma|=1$, $60$~deg. incidence and $t_p = 0.46$ in Eq.~\eqref{eq:Omega}.}}\label{tab:estimates}
\end{table*}

The laser intensity which would be required to excite a single coherent plasmon (i.e. to induce a coherent population $|\langle \hat{a}\rangle|=1$) is about $10^2~\upmu\text{W}\cdot \upmu\text{m}^{-2}$ (see discussion in Appendix~\ref{app:conventional}), similar to the maximum CW intensities reported in Ref.~[\citenum{lombardi_pulsed_2018}]. For comparison, the atomic antenna becomes saturated for laser intensities about five orders of magnitude smaller (we discuss the effects of saturation of the atomic dipolar transition in the antenna in Section~\ref{sec:nonlinearities}). These estimates are consistent with the reported values of power required to saturate other group-IV centers in diamond, e.g. SnV~\cite{arjona_martinez_photonic_2022}.

The single-photon optomechanical coupling in conventional SERS ($g_{0,a}$) is about $20$ times larger than for the atomic antenna ($g_{0,\sigma}$), and both are far smaller than the record values estimated for plasmonic picocavities~\cite{benz_single-molecule_2016}. {Additionally, to compare these couplings with the relevant dissipation and decoherence rates, we estimate the single-photon cooperativities $C_{0,a}=4g_0^2/(\kappa \Gamma)$ and $C_{0,\sigma}=4g_0^2/(\gamma \Gamma)$ for the plasmonic and atomic antennas, respectively, and find the latter about 3 orders of magnitude larger. Since both system can only operate below, or near the unitary coherent population of the plasmon or exciton ($|\alpha_\sigma|<1$, $|\alpha_a|\lesssim 1$), the effective cooperativities $C_{a}=|\alpha_a|^2C_{0,a}$ and $C_{\sigma}=|\alpha_\sigma|^2C_{0,\sigma}$ are bounded by similar numbers.}

This contrast of laser intensities required for driving the two systems, and the similarity of the single-photon optomechanical coupling, is the first key message of our work. It demonstrates that atomic antenna-enhanced Raman scattering is a viable alternative for conventional SERS, in particular in systems in which the use of high-intensity lasers is undesirable due to fragility of the sample, for instance to photo-induced bleaching or charge instability, or the focusing element. 

We can also estimate the \textit{efficiency of down-conversion} of the incident to emitted Stokes photons (a similar question of the conversion efficiency of IR to visible photon was considered by Roelli and co-authors in Ref.~\cite{roelli_molecular_2020}). 
We derive the complete expression for this efficiency in Appendix~\ref{app:powers}, and find that it can be expressed as
\begin{equation}\label{eq:efficiency.atomic}
    \eta_\sigma = \eta' \frac{\left|g_{0,\sigma}\alpha_\sigma \right|^2}{I},
\end{equation} 
where $\eta'$ is a prefactor given in Eq.~\eqref{eq:etaprime}.
Similarly, we quantify the conversion efficiency for the conventional plasmonic Raman system as
\begin{equation}\label{eq:efficiency.plasmon}
    \eta_a = \eta' \frac{\left|g_{0,a}\alpha_a \right|^2}{I}.
\end{equation}

It is clear from Eqs.~\eqref{eq:efficiency.atomic} and \eqref{eq:efficiency.plasmon}, that the difference in efficiencies between the two systems is dictated by the effective optomechanical couplings $g_{0,\sigma} \alpha_\sigma$ and $g_{0,a} \alpha_a$, normalized by the intensity of the incident laser. Exciting plasmons requires much larger laser intensities than exciting the atomic antennas (see Table~\ref{tab:estimates}), meaning that \textit{per intensity} $|\alpha_a|^2/I \ll |\alpha_\sigma|^2/I$; on the other hand, conventional plasmonic cavities yield larger single-photon optomechanical couplings $|g_{0,a}|\gg|g_{0,\sigma}|$. 

Therefore to make meaningful assessments about the 
resulting contrast in efficiencies, we need to perform  a more careful analysis, investigating the effects of the three approximations listed above. 

\subsubsection{Modification of the properties of atomic antenna by a nanoparticle}\label{subsec:modeling}

In the list of parameters given in Table~\ref{tab:estimates}, we have neglected the effects of the nanoparticle on the characteristics of the atomic antenna. Here we carry out a more quantitative discussion of these three effects, and their impact on the overall Raman scattering efficiency, supported by numerical solutions implemented in COMSOL Multiphysics software. A schematic representation of these systems is shown in Fig.~\ref{fig:numerics}(a), and the details of the numerical model are given in Appendix~\ref{app:comsol}. We stress that these calculations are only aimed at establishing the \textit{relative corrections} to the driving amplitude, decay rate, and field of the atomic antenna, rather than re-assessing all the parameters given in Table~\ref{tab:estimates}. Therefore, we keep the remaining values, most critically the frequency and mode volume of the cavity, as constant in the following discussion, meaning that the efficiency of the conventional plasmonic SERS setup remains unchanged.

The first two effects are commonly discussed in the context of plasmon-enhanced or quenched fluorescence~\cite{anger_enhancement_2006,kuhn_enhancement_2006}:
\begin{itemize}
    \item \textit{Enhancement of the coherent driving amplitude}\\ The plasmonic nanoparticle will modify the intensity of the incident laser at the site of the atomic antenna by a factor $K_\text{inc}^2$, enhancing its driving amplitude, and lowering the threshold for the excitation of the atomic antenna. To quantify this effect, we calculate the field distribution for illumination with  a $p$-polarized plane wave incident at an angle of $60^\circ$, for a range of substrate-nanoparticle distances. We plot the field intensity normalized by the field intensity calculated in the absence of the nanoparticle (which already accounts for the imperfect transmission of the incident light into the substrate) in Fig.~\ref{fig:numerics}(b) with a solid blue line. 
    
    \item \textit{Modification of the atomic antenna dissipation via the Purcell effect}\\ The plasmonic nanoparticle will also modify the intrinsic dissipation pathways for the atomic antenna, both by enhancing its coupling to the radiation modes, and by introducing nonradiative decay associated with absorption in the metal. The resulting decay rate of the atomic antenna will be increased, meaning that  a larger laser intensity is required for its excitation. We model this effect by calculating the scattering from  a dipolar electric source embedded in the diamond substrate, and oriented vertically, and quantify the enhancement of the decay rate by calculating the sum of the radiated and absorbed power ($P_\text{rad}+P_\text{abs}$), normalized by the emitted power in the absence of the nanoparticle $P_0$. Since this enhancement will only modify the rate of the radiative decay of the emitter $\gamma_0$, {(see the discussion in SI Section F.1)} the overall decay rate can be estimated as 
    \begin{align}\label{eq:total.rate}
        \gamma \rightarrow \gamma_\text{tot} &= \underbrace{\frac{P_\text{rad}+P_\text{abs}}{P_0} \gamma_0}_{\text{radiative decay}} \,\, + \underbrace{3\gamma_0}_{\text{non-radiative decay}} \nonumber \\
        & = \left(\frac{1}{4}\frac{P_\text{rad}+P_\text{abs}}{P_0} + \frac{3}{4}\right)\gamma,
    \end{align}
    where we have again assumed that at the low temperature of 50~K, and in the absence of the nanoparticle, $\gamma=4\gamma_0$~\cite{li_atomic_2024}. We plot the resulting shortening of the antenna's lifetime $(\gamma_\text{tot}/\gamma)^{-1}$ in Fig.~\ref{fig:numerics}(b) with a solid orange line. We note the limiting case of the decay rate enhancement from a buried electric dipole was considered in Ref.~\citenum{chen_metallodielectric_2012}, where the enhancement in a similar system, with a plasmonic nanoparticle positioned on the dielectric substrate, exceeded $10^3$. Interestingly, due to the reflection from the interface between the diamond substrate and air, which modifies the radiative properties of the atomic antenna compared to its bulk response, this ratio does not converge to 1 in the limit of large spacings $d$.
    \item \textit{Modification of the field at the position of the Raman-active molecule} \\ Finally, the presence of the nanoparticle will modify the field from the atomic antenna, experienced by the molecule. To quantify this effect, we calculate the  electric field from the electric dipolar source positioned 2~nm below the surface, at the distance of 1~nm above the surface of the substrate, in the gap below the nanoparticle, and normalize it by the same quantity in the absence of the nanoparticle, to find the field enhancement factor $K_\text{mol}$ as a function of the substrate-nanoparticle spacing $d$. We plot the result in Fig.~\ref{fig:numerics}(b) with a dashed green line.
\end{itemize}

The above effects enter the conversion efficiency $\eta_\sigma$ by (i) boosting the effective coherent population $|\alpha_\sigma|^2$ by a factor given by the field intensity enhancement $K_\text{inc}^2$, (ii) increasing the decay rate $\gamma \rightarrow \gamma_\text{tot}$ and thus lowering $|\alpha_\sigma|^2$ by $\gamma/\gamma_\text{tot}$, and (iii) modifying the optomechanical coupling coefficient $g_{0,\sigma}$ by $K_\text{mol}$. In Fig.~\ref{fig:numerics}(c) we plot the resulting \textit{corrected} conversion efficiency for the atomic antenna-enhanced Raman scattering, 
\begin{align}\label{eq:efficiency.atomic.mod}
    \eta_\sigma \rightarrow \eta' \frac{\left|g_{0,\sigma}\alpha_\sigma \right|^2}{I} K_\text{inc}^2 K_\text{mol}^2 \left(\frac{\gamma}{\gamma_\text{tot}}\right)^2,    
\end{align}
and the conventional plasmonic system with parameters given in Table~\ref{tab:estimates}. 
We find that while the nanoparticle can modify the efficiency of the atomic antenna-enhanced Raman scattering via the three effects we investigated, this overall change in efficiency shown in Fig,~\ref{fig:numerics}(c) is {within a factor of 4}.  

\begin{figure*}[ht!]
	\centering
	\includegraphics[width=.5\linewidth]{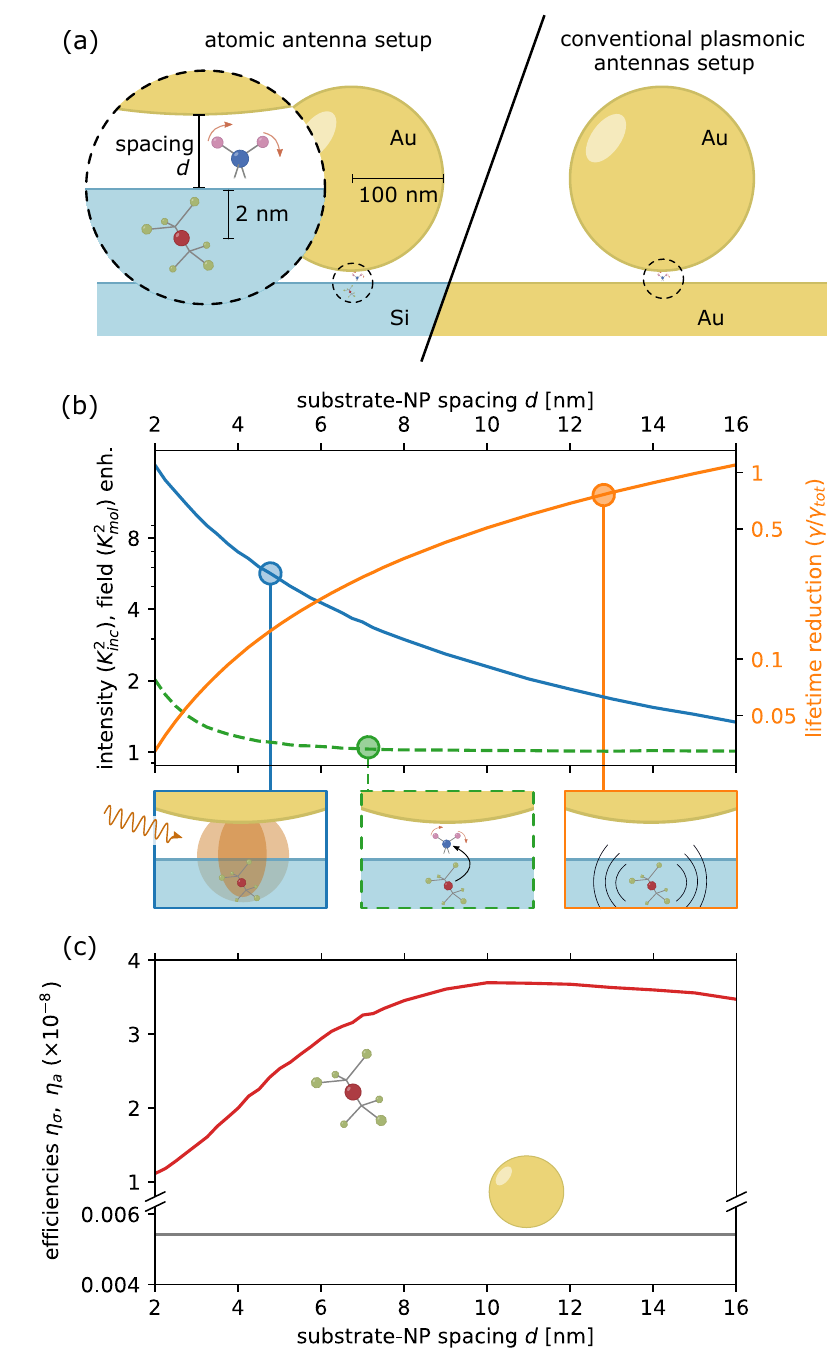}
	\caption{Effects of the nanoparticle (NP) on { parameters of the atomic antenna setup depicted in the left panel of (a)}. {(b)} Enhancements of the intensity of: incident field $K_\text{inc}^2$ at the antenna position (solid blue line), field from the antenna at the molecules position (dashed green line), and the reduction of the lifetime $\gamma/\gamma_\text{tot}$ (solid orange line) of the antenna dipole. {(c)} Corrected conversion efficiencies for the atomic antenna (solid red line; Eq.~\eqref{eq:efficiency.atomic.mod}) and conventional plasmonic system (solid gray line; Eq.~\eqref{eq:efficiency.plasmon}).}
	\label{fig:numerics}
\end{figure*}

For the entire range of the substrate-nanoparticle spacings, the adjusted efficiencies of these systems are about three orders of magnitude larger than for the conventional plasmonic systems. This implies that \textit{per input photon}, atomic antennas would yield about a thousand-fold increase in upconversion probability over the conventional plasmonic SERS. This dramatic contrast is the second message of our work.

\subsection{Nonlinearities in atomic antenna-enhanced Raman scattering}\label{sec:nonlinearities}

{ Previous sections explored the properties of Raman scattering in a linearized picture, which we derived in Subsection~\ref{subsec:coherent} by replacing the pseudospin operator with its expectation value. This approximation works well in the \textit{Heitler regime}, where the coherent driving amplitude $\Omega$ of the atomic antenna is far smaller than its total decay rate $\gamma$ \cite{carmichael2013statistical,walls2012quantum}.} 
Here we ask what happens beyond this regime: does the linearization scheme break down? Does the atomic antenna-enhanced Stokes emission exhibit a different dependence on laser intensity than a conventional plasmonic SERS setup?


To answer these questions, we use QuTip~\cite{johansson_qutip_2013,johansson_qutip_2012} to model the dynamics of the system described by the full (un-linearized) optomechanical Hamiltonian {with additional coupling between the atomic and plasmonic antennas and conventional SERS coupling}, and the dissipative terms described in Subsection~\ref{subsec:dissipation}. {Details of this model are given in Appendix~\ref{app:comsol}}.

\begin{figure*}[ht!]
	\centering
	\includegraphics[width=.7\linewidth]{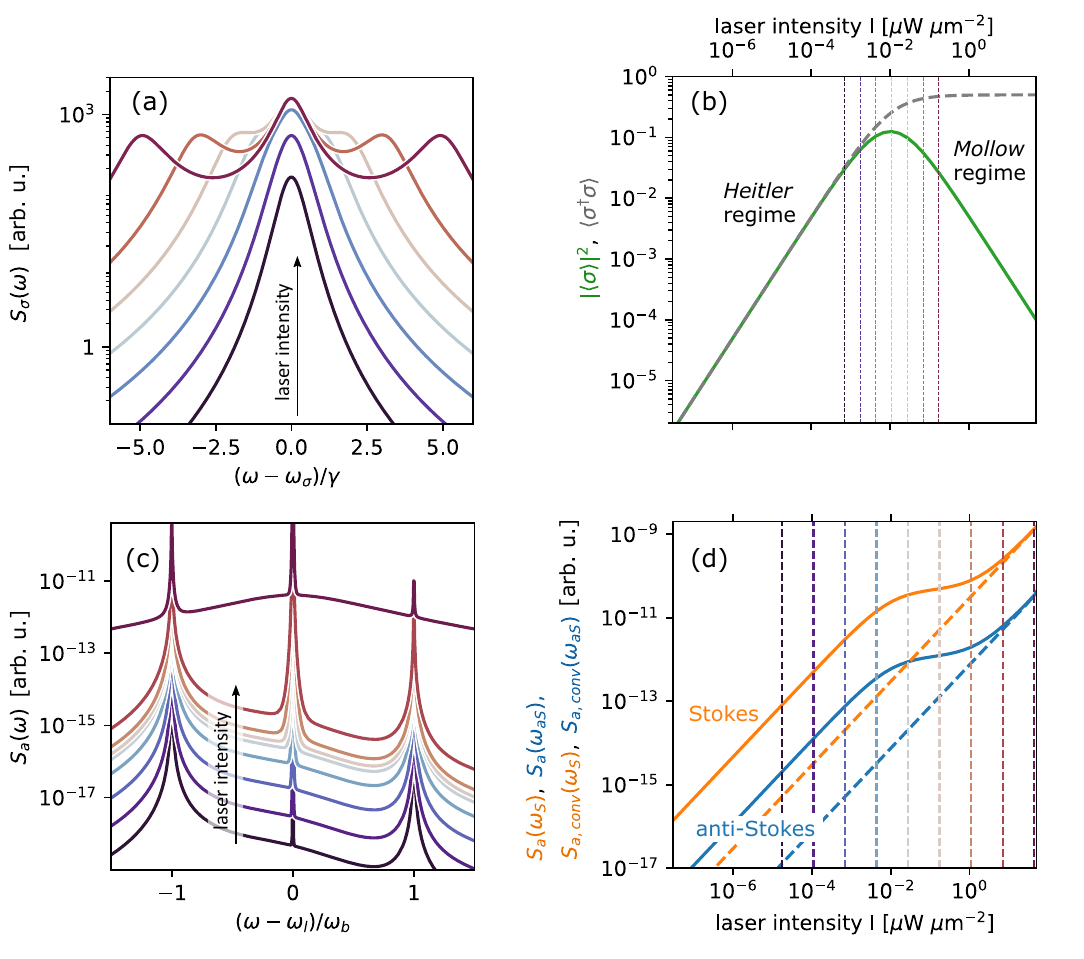}
	\caption{{Analysis of the (a) incoherent emission spectrum from the atomic antenna $S_\sigma$ (Eq.~\eqref{eq:spectrum.antenna}) and (b) the coherence $|\langle \hat{\sigma} \rangle|^2$ (solid green) and population $\langle \hat{\sigma}^\dag \hat{\sigma} \rangle$ (dashed orange) of the atomic antenna for increasing laser intensities. Colors of the spectra in (a) correspond to the vertical lines in (b) to denote intensities (explicitly listed in Appendix~\ref{app:comsol}). In (c) we plot the incoherent spectra of Raman emission $S_a$ (Eq.~\eqref{eq:emission.a}) for increasing laser intensities; in (d) we trace the Stokes (solid orange) and anti-Stokes (solid blue) peak intensities, and compare them with the same calculated in conventional SERS setup (dashed lines). Details of this model and parameters are given in Appendix~\ref{app:extra} and in Table~\ref{tab:estimates}.}}
	\label{fig:nonlinear}
\end{figure*}

{We start by analysing the response of the atomic antenna itself on the amplitude of the driving. First, 
in Fig.~\ref{fig:nonlinear}(a) we plot the incoherent emission spectrum of the antenna
\begin{equation}\label{eq:spectrum.antenna}
    S_\sigma(\omega)=\omega^4 \int_{-\infty}^\infty \mathrm{d} \tau e^{i\omega\tau} \langle \hat{\sigma}^\dag(\tau)\hat{\sigma}(0)\rangle,   
\end{equation}
for increasing laser power. Colors of the spectra in Fig.~\ref{fig:nonlinear}(a) correspond to the laser intensities marked with vertical dashed lines in Fig.~\ref{fig:nonlinear}(b), where we plot the steady-state coherence of the antenna $|\langle \hat{\sigma}\rangle|^2$ (solid blue line) and the population of its excited state $\langle \hat{\sigma}^\dag \hat{\sigma} \rangle$ (dashed orange line). For laser intensities beyond $I\sim 10^{-2}~\upmu\text{W}/\upmu\text{m}^2$ the antenna transitions from the Heitler to the \textit{Mollow regime}, is driven towards the maximally mixed state with $\mean{\hat{\sigma}^\dagger \hat{\sigma}}=0.5$, and its emission spectrum develops the characteristic three-peak structure of a Mollow triplet. 
}

In that regime the system cannot absorb more energy, and we expect that the intensity of the Raman processes should saturate {or --- due to the decrease of the coherence $\langle \hat{\sigma}\rangle$ and the effective optomechanical coupling --- weaken. To test this, in Fig.~\ref{fig:nonlinear}(c) we plot the spectra of light emitted from the system for increasing laser intensities, which we again color-code to the vertical lines in Fig.~\ref{fig:nonlinear}(d). That panel traces the peak intensities of the Stokes (solid orange line) and anti-Stokes (solid blue line) emission, calculated from the incoherent spectrum of the plasmonic cavity (Eq.~\eqref{eq:spectrum.final}).} 

{ This result clearly illustrates the breakdown of the linearization description, and suggests that the overall Raman response is governed by the population of the atomic antenna, whether it be coherent or incoherent. It is reminiscent of the features observed in other hybrid optomechanical systems, for example in resonant Raman scattering, where the vibronic coupling of the electronic transition (denoted by the pseudospin operator $\hat{\sigma}_m$, to differentiate it from the operator of the atomic antenna), and molecular vibrations of a molecule is governed by some version of the interaction Hamiltonian
\begin{equation}\label{eq:parametric}
    \hat{H}_\text{res} = g_\text{res}\hat{\sigma}_m^\dag \hat{\sigma}_m (\hat{b}+\hat{b}^\dag),
\end{equation}
accompanied by either a coherent driving of the electronic transition, either directly by the laser source, or mediated by a cavity mode. For example, in Ref.~[\citenum{neuman_quantum_2019}], Neuman and co-authors discuss the latter case, and find that Raman emission is proportional to the population of the excited electronic state $\mean{\hat{\sigma}_m^\dag \hat{\sigma}_m}$, which has the same functional dependence as that discussed in Fig.~\ref{fig:nonlinear}. In Appendix~\ref{app:spectrum} we analytically derive and analyze this correspondence in detail.

The interaction Hamiltonian given in Eq.~\eqref{eq:parametric} describes also a range of other hybrid mechanical systems, where a quantum dot \cite{PhysRevLett.92.075507, yeo2014strain} or a superconducting circuit \cite{lahaye2009nanomechanical} is coupled to mechanical vibrations. In the former scenario, when the quantum dot resonantly couples to the optical field of a cavity, this interaction Hamiltonian can be transformed into the three-way coupling between the pseudo-spin, optical, and mechanical modes, similar to that given in Eq.~\eqref{eq:interaction.hamiltonian}. 
}

{To complete the comparison with a conventional SERS setup, in panel Fig.~\ref{fig:nonlinear}(d) we also plot, with dashed orange and blue lines, the intensities of the Stokes and anti-Stokes emission from a plasmonic SERS system ($S_{a,\text{conv}}(\omega_S/\omega_{aS})$; calculated assuming vanishing $g_{0,\sigma}=0$). Below $I\sim 10^{-2}~\upmu\text{W}/\upmu\text{m}^2$ (the point of saturation of the atomic antenna-enhanced Stokes), its intensity is two to three orders of magnitude weaker than the antomic antenna-enhanced Raman scattering, consistent with the efficiency prediction shown in Fig.~\ref{fig:numerics}(c). Above $5~\upmu\text{W}/\upmu\text{m}^2$, the conventional SERS dominates the response.}


Even for the largest pump intensity for which the linearization approximation would be justified (see the value in Table~\ref{tab:estimates}), the phonon populations do not diverge from the thermal equilibrium values ($\nbth \approx 5\times 10^{-2}$), and the optomechanical pumping and decay rates $\Gamma_+$ and $\Gamma_-$ are several orders of magnitude smaller than $\Gamma$ ($\Gamma_+/\Gamma \sim 10^{-9}\gg \Gamma_-/\Gamma$), suggesting that we would observe neither the vibrational pumping, nor instability.

{Finally, we note that in the context of Surface Enhanced Resonant Raman Scattering (SERRS), Neuman and co-authors in Ref.~[\citenum{neuman_quantum_2019}] considered a system where the sidebands of Mollow triplet are frequency-matched to a Raman lines. Since the splitting between these sidebands is approximately given by coherent driving rate of the atomic antenna, such a setup would require an extremely strong coherent driving of the order of mechanical frequency $\sim$10-40 THz.}

{
\subsection{Feasibility}\label{subsec:feasibility}

The experimental realization of the system presented in this work would be arguably challenging, since on top of the usual challenges of positioning the Raman-active molecule in the gap of a substrate-nanoparticle system, it adds the requirement of spatially aligning them with a shallow deposited GeV. This problem could be addressed by using sufficiently high density of defects and molecules, or by swapping the nanoparticle for a metallic tip in a Tip-Enhanced Raman Scattering experiment.

Using the estimates for $10^{-8}$ efficiency of thes Stoke emission, and considering laser intensity of $I\sim 10^{-2}~\upmu\text{W}/\upmu\text{m}^2$ (of flux of $3\times 10^{10}$ pump photons/s) which saturates the GeV, we can expect the Stokes flux to reach about 300 photons/s, consistent with the estimates for observation of Surface- and Tip-Enhanced Raman Scattering from single molecules~\cite{nie_probing_1997,langer_present_2020,yi_surface-enhanced_2025,zhang_chemical_2013,berweger_signal_2010}.

}

\section{Summary and outlook}\label{sec:Summary}

Thanks to a large scattering cross section and very large quality factors, atomic antennas can be very efficiently populated at relatively low laser intensities. Achieving comparable populations of a plasmonic cavity mode would require at least five orders of magnitude more intense illumination. This large intensity requirement limits the application of SERS to spectroscopy of molecules which are more sensitive to photo-dissociation/photo-bleaching, or studies requiring low temperatures. Additionally, large laser power can induce melting, or reshaping of nanoparticles.

Here we showed that the ability to efficiently excited atomic antennas means they are a promising alternative platform for low-intensity Raman scattering, although they do offer slightly smaller values of the optomechanical coupling compared to the conventional plasmonic systems. The tradeoff between these effects is captured by the efficiency of Stokes emission, which estimates the ratio of input photons which are up- and down-converted through Raman processes, and is about three orders of magnitude larger for the atomic antenna-enhanced Raman systems.

The lower intensities required for the operation of that system will also likely reduce the noise associated with nonlinear effects in plasmonic substrates. 

{ 
We identify a clear signature of the atomic-antenna enhanced Raman scattering, that can be targeted for experimental verification --- the strongly nonlinear dependence of the intensity of Stokes emission on the laser intensity $I$, which illustrates a clear saturation, and a significant deviation from the conventionally observed in SERS spontaneous ($\propto I$), or vibrationally pumped ($\propto I^2$) emission.} 
Additionally, the sub-GHz narrowband response of the atomic antenna can be utilized to identify and filter atomic antenna-enhanced Raman scattering. By comparing photoluminescence with on- and off- resonant laser excitation of the atomic antenna, background signals can be accurately identified, similar to observations of F{\"o}rster resonant excitation transfer~\cite{li_atomic_2024}.  

\section*{Acknowledgements}
Authors acknowledge fruitful discussions with Javier Aizpurua and Ruben Esteban. 

\subsection*{Funding sources}
MKS acknowledges funding from the Australian Research Council Discovery Early Career Researcher Award (DE220101272). AAH acknowledges funding from Q-NEXT, supported by the US Department of Energy, Office of Science, National Quantum Information Science Research Centers.

\bibliography{references_main}

\begin{thebibliography}{59}%
\makeatletter
\providecommand \@ifxundefined [1]{%
 \@ifx{#1\undefined}
}%
\providecommand \@ifnum [1]{%
 \ifnum #1\expandafter \@firstoftwo
 \else \expandafter \@secondoftwo
 \fi
}%
\providecommand \@ifx [1]{%
 \ifx #1\expandafter \@firstoftwo
 \else \expandafter \@secondoftwo
 \fi
}%
\providecommand \natexlab [1]{#1}%
\providecommand \enquote  [1]{``#1''}%
\providecommand \bibnamefont  [1]{#1}%
\providecommand \bibfnamefont [1]{#1}%
\providecommand \citenamefont [1]{#1}%
\providecommand \href@noop [0]{\@secondoftwo}%
\providecommand \href [0]{\begingroup \@sanitize@url \@href}%
\providecommand \@href[1]{\@@startlink{#1}\@@href}%
\providecommand \@@href[1]{\endgroup#1\@@endlink}%
\providecommand \@sanitize@url [0]{\catcode `\\12\catcode `\$12\catcode `\&12\catcode `\#12\catcode `\^12\catcode `\_12\catcode `\%12\relax}%
\providecommand \@@startlink[1]{}%
\providecommand \@@endlink[0]{}%
\providecommand \url  [0]{\begingroup\@sanitize@url \@url }%
\providecommand \@url [1]{\endgroup\@href {#1}{\urlprefix }}%
\providecommand \urlprefix  [0]{URL }%
\providecommand \Eprint [0]{\href }%
\providecommand \doibase [0]{https://doi.org/}%
\providecommand \selectlanguage [0]{\@gobble}%
\providecommand \bibinfo  [0]{\@secondoftwo}%
\providecommand \bibfield  [0]{\@secondoftwo}%
\providecommand \translation [1]{[#1]}%
\providecommand \BibitemOpen [0]{}%
\providecommand \bibitemStop [0]{}%
\providecommand \bibitemNoStop [0]{.\EOS\space}%
\providecommand \EOS [0]{\spacefactor3000\relax}%
\providecommand \BibitemShut  [1]{\csname bibitem#1\endcsname}%
\let\auto@bib@innerbib\@empty
\bibitem [{\citenamefont {Jakob}\ \emph {et~al.}(2023)\citenamefont {Jakob}, \citenamefont {Deacon}, \citenamefont {Zhang}, \citenamefont {De~Nijs}, \citenamefont {Pavlenko}, \citenamefont {Hu}, \citenamefont {Carnegie}, \citenamefont {Neuman}, \citenamefont {Esteban}, \citenamefont {Aizpurua},\ and\ \citenamefont {Baumberg}}]{jakob_giant_2023}%
  \BibitemOpen
  \bibfield  {author} {\bibinfo {author} {\bibfnamefont {L.~A.}\ \bibnamefont {Jakob}}, \bibinfo {author} {\bibfnamefont {W.~M.}\ \bibnamefont {Deacon}}, \bibinfo {author} {\bibfnamefont {Y.}~\bibnamefont {Zhang}}, \bibinfo {author} {\bibfnamefont {B.}~\bibnamefont {De~Nijs}}, \bibinfo {author} {\bibfnamefont {E.}~\bibnamefont {Pavlenko}}, \bibinfo {author} {\bibfnamefont {S.}~\bibnamefont {Hu}}, \bibinfo {author} {\bibfnamefont {C.}~\bibnamefont {Carnegie}}, \bibinfo {author} {\bibfnamefont {T.}~\bibnamefont {Neuman}}, \bibinfo {author} {\bibfnamefont {R.}~\bibnamefont {Esteban}}, \bibinfo {author} {\bibfnamefont {J.}~\bibnamefont {Aizpurua}},\ and\ \bibinfo {author} {\bibfnamefont {J.~J.}\ \bibnamefont {Baumberg}},\ }\bibfield  {title} {\bibinfo {title} {Giant optomechanical spring effect in plasmonic nano- and picocavities probed by surface-enhanced {Raman} scattering},\ }\href {https://doi.org/10.1038/s41467-023-38124-1} {\bibfield  {journal} {\bibinfo  {journal} {Nature Communications}\ }\textbf
  {\bibinfo {volume} {14}},\ \bibinfo {pages} {3291} (\bibinfo {year} {2023})}\BibitemShut {NoStop}%
\bibitem [{\citenamefont {Benz}\ \emph {et~al.}(2016)\citenamefont {Benz}, \citenamefont {Schmidt}, \citenamefont {Dreismann}, \citenamefont {Chikkaraddy}, \citenamefont {Zhang}, \citenamefont {Demetriadou}, \citenamefont {Carnegie}, \citenamefont {Ohadi}, \citenamefont {de~Nijs}, \citenamefont {Esteban}, \citenamefont {Aizpurua},\ and\ \citenamefont {Baumberg}}]{benz_single-molecule_2016}%
  \BibitemOpen
  \bibfield  {author} {\bibinfo {author} {\bibfnamefont {F.}~\bibnamefont {Benz}}, \bibinfo {author} {\bibfnamefont {M.~K.}\ \bibnamefont {Schmidt}}, \bibinfo {author} {\bibfnamefont {A.}~\bibnamefont {Dreismann}}, \bibinfo {author} {\bibfnamefont {R.}~\bibnamefont {Chikkaraddy}}, \bibinfo {author} {\bibfnamefont {Y.}~\bibnamefont {Zhang}}, \bibinfo {author} {\bibfnamefont {A.}~\bibnamefont {Demetriadou}}, \bibinfo {author} {\bibfnamefont {C.}~\bibnamefont {Carnegie}}, \bibinfo {author} {\bibfnamefont {H.}~\bibnamefont {Ohadi}}, \bibinfo {author} {\bibfnamefont {B.}~\bibnamefont {de~Nijs}}, \bibinfo {author} {\bibfnamefont {R.}~\bibnamefont {Esteban}}, \bibinfo {author} {\bibfnamefont {J.}~\bibnamefont {Aizpurua}},\ and\ \bibinfo {author} {\bibfnamefont {J.~J.}\ \bibnamefont {Baumberg}},\ }\bibfield  {title} {\bibinfo {title} {Single-molecule optomechanics in “picocavities”},\ }\href {https://doi.org/10.1126/science.aah5243} {\bibfield  {journal} {\bibinfo  {journal} {Science}\ }\textbf {\bibinfo {volume}
  {354}},\ \bibinfo {pages} {726} (\bibinfo {year} {2016})}\BibitemShut {NoStop}%
\bibitem [{\citenamefont {Urbieta}\ \emph {et~al.}(2018)\citenamefont {Urbieta}, \citenamefont {Barbry}, \citenamefont {Zhang}, \citenamefont {Koval}, \citenamefont {Sánchez-Portal}, \citenamefont {Zabala},\ and\ \citenamefont {Aizpurua}}]{urbieta_atomic-scale_2018}%
  \BibitemOpen
  \bibfield  {author} {\bibinfo {author} {\bibfnamefont {M.}~\bibnamefont {Urbieta}}, \bibinfo {author} {\bibfnamefont {M.}~\bibnamefont {Barbry}}, \bibinfo {author} {\bibfnamefont {Y.}~\bibnamefont {Zhang}}, \bibinfo {author} {\bibfnamefont {P.}~\bibnamefont {Koval}}, \bibinfo {author} {\bibfnamefont {D.}~\bibnamefont {Sánchez-Portal}}, \bibinfo {author} {\bibfnamefont {N.}~\bibnamefont {Zabala}},\ and\ \bibinfo {author} {\bibfnamefont {J.}~\bibnamefont {Aizpurua}},\ }\bibfield  {title} {\bibinfo {title} {Atomic-{Scale} {Lightning} {Rod} {Effect} in {Plasmonic} {Picocavities}: {A} {Classical} {View} to a {Quantum} {Effect}},\ }\href {https://doi.org/10.1021/acsnano.7b07401} {\bibfield  {journal} {\bibinfo  {journal} {ACS Nano}\ }\textbf {\bibinfo {volume} {12}},\ \bibinfo {pages} {585} (\bibinfo {year} {2018})}\BibitemShut {NoStop}%
\bibitem [{\citenamefont {Wu}\ \emph {et~al.}(2021)\citenamefont {Wu}, \citenamefont {Yan},\ and\ \citenamefont {Lalanne}}]{wu_bright_2021}%
  \BibitemOpen
  \bibfield  {author} {\bibinfo {author} {\bibfnamefont {T.}~\bibnamefont {Wu}}, \bibinfo {author} {\bibfnamefont {W.}~\bibnamefont {Yan}},\ and\ \bibinfo {author} {\bibfnamefont {P.}~\bibnamefont {Lalanne}},\ }\bibfield  {title} {\bibinfo {title} {Bright {Plasmons} with {Cubic} {Nanometer} {Mode} {Volumes} through {Mode} {Hybridization}},\ }\href {https://doi.org/10.1021/acsphotonics.0c01569} {\bibfield  {journal} {\bibinfo  {journal} {ACS Photonics}\ }\textbf {\bibinfo {volume} {8}},\ \bibinfo {pages} {307} (\bibinfo {year} {2021})}\BibitemShut {NoStop}%
\bibitem [{\citenamefont {Zhang}\ \emph {et~al.}(2013)\citenamefont {Zhang}, \citenamefont {Zhang}, \citenamefont {Dong}, \citenamefont {Jiang}, \citenamefont {Zhang}, \citenamefont {Chen}, \citenamefont {Zhang}, \citenamefont {Liao}, \citenamefont {Aizpurua}, \citenamefont {Luo}, \citenamefont {Yang},\ and\ \citenamefont {Hou}}]{zhang_chemical_2013}%
  \BibitemOpen
  \bibfield  {author} {\bibinfo {author} {\bibfnamefont {R.}~\bibnamefont {Zhang}}, \bibinfo {author} {\bibfnamefont {Y.}~\bibnamefont {Zhang}}, \bibinfo {author} {\bibfnamefont {Z.~C.}\ \bibnamefont {Dong}}, \bibinfo {author} {\bibfnamefont {S.}~\bibnamefont {Jiang}}, \bibinfo {author} {\bibfnamefont {C.}~\bibnamefont {Zhang}}, \bibinfo {author} {\bibfnamefont {L.~G.}\ \bibnamefont {Chen}}, \bibinfo {author} {\bibfnamefont {L.}~\bibnamefont {Zhang}}, \bibinfo {author} {\bibfnamefont {Y.}~\bibnamefont {Liao}}, \bibinfo {author} {\bibfnamefont {J.}~\bibnamefont {Aizpurua}}, \bibinfo {author} {\bibfnamefont {Y.}~\bibnamefont {Luo}}, \bibinfo {author} {\bibfnamefont {J.~L.}\ \bibnamefont {Yang}},\ and\ \bibinfo {author} {\bibfnamefont {J.~G.}\ \bibnamefont {Hou}},\ }\bibfield  {title} {\bibinfo {title} {Chemical mapping of a single molecule by plasmon-enhanced {Raman} scattering},\ }\href {https://doi.org/10.1038/nature12151} {\bibfield  {journal} {\bibinfo  {journal} {Nature}\ }\textbf {\bibinfo {volume}
  {498}},\ \bibinfo {pages} {82} (\bibinfo {year} {2013})}\BibitemShut {NoStop}%
\bibitem [{\citenamefont {Lee}\ \emph {et~al.}(2019)\citenamefont {Lee}, \citenamefont {Crampton}, \citenamefont {Tallarida},\ and\ \citenamefont {Apkarian}}]{lee_visualizing_2019}%
  \BibitemOpen
  \bibfield  {author} {\bibinfo {author} {\bibfnamefont {J.}~\bibnamefont {Lee}}, \bibinfo {author} {\bibfnamefont {K.~T.}\ \bibnamefont {Crampton}}, \bibinfo {author} {\bibfnamefont {N.}~\bibnamefont {Tallarida}},\ and\ \bibinfo {author} {\bibfnamefont {V.~A.}\ \bibnamefont {Apkarian}},\ }\bibfield  {title} {{\selectlanguage {en}\bibinfo {title} {Visualizing vibrational normal modes of a single molecule with atomically confined light}},\ }\href {https://doi.org/10.1038/s41586-019-1059-9} {\bibfield  {journal} {\bibinfo  {journal} {Nature}\ }\textbf {\bibinfo {volume} {568}},\ \bibinfo {pages} {78} (\bibinfo {year} {2019})}\BibitemShut {NoStop}%
\bibitem [{\citenamefont {Rosławska}\ \emph {et~al.}(2024)\citenamefont {Rosławska}, \citenamefont {Kaiser}, \citenamefont {Romeo}, \citenamefont {Devaux}, \citenamefont {Scheurer}, \citenamefont {Berciaud}, \citenamefont {Neuman},\ and\ \citenamefont {Schull}}]{roslawska_submolecular-scale_2024}%
  \BibitemOpen
  \bibfield  {author} {\bibinfo {author} {\bibfnamefont {A.}~\bibnamefont {Rosławska}}, \bibinfo {author} {\bibfnamefont {K.}~\bibnamefont {Kaiser}}, \bibinfo {author} {\bibfnamefont {M.}~\bibnamefont {Romeo}}, \bibinfo {author} {\bibfnamefont {E.}~\bibnamefont {Devaux}}, \bibinfo {author} {\bibfnamefont {F.}~\bibnamefont {Scheurer}}, \bibinfo {author} {\bibfnamefont {S.}~\bibnamefont {Berciaud}}, \bibinfo {author} {\bibfnamefont {T.}~\bibnamefont {Neuman}},\ and\ \bibinfo {author} {\bibfnamefont {G.}~\bibnamefont {Schull}},\ }\bibfield  {title} {{\selectlanguage {en}\bibinfo {title} {Submolecular-scale control of phototautomerization}},\ }\href {https://doi.org/10.1038/s41565-024-01622-4} {\bibfield  {journal} {\bibinfo  {journal} {Nature Nanotechnology}\ }\textbf {\bibinfo {volume} {19}},\ \bibinfo {pages} {738} (\bibinfo {year} {2024})}\BibitemShut {NoStop}%
\bibitem [{\citenamefont {Shlesinger}\ \emph {et~al.}(2023)\citenamefont {Shlesinger}, \citenamefont {Palstra},\ and\ \citenamefont {Koenderink}}]{shlesinger_integrated_2023}%
  \BibitemOpen
  \bibfield  {author} {\bibinfo {author} {\bibfnamefont {I.}~\bibnamefont {Shlesinger}}, \bibinfo {author} {\bibfnamefont {I.~M.}\ \bibnamefont {Palstra}},\ and\ \bibinfo {author} {\bibfnamefont {A.~F.}\ \bibnamefont {Koenderink}},\ }\bibfield  {title} {\bibinfo {title} {Integrated {Sideband}-{Resolved} {SERS} with a {Dimer} on a {Nanobeam} {Hybrid}},\ }\href {https://doi.org/10.1103/PhysRevLett.130.016901} {\bibfield  {journal} {\bibinfo  {journal} {Physical Review Letters}\ }\textbf {\bibinfo {volume} {130}},\ \bibinfo {pages} {016901} (\bibinfo {year} {2023})}\BibitemShut {NoStop}%
\bibitem [{\citenamefont {Shlesinger}\ \emph {et~al.}(2021)\citenamefont {Shlesinger}, \citenamefont {Cognée}, \citenamefont {Verhagen},\ and\ \citenamefont {Koenderink}}]{shlesinger_integrated_2021}%
  \BibitemOpen
  \bibfield  {author} {\bibinfo {author} {\bibfnamefont {I.}~\bibnamefont {Shlesinger}}, \bibinfo {author} {\bibfnamefont {K.~G.}\ \bibnamefont {Cognée}}, \bibinfo {author} {\bibfnamefont {E.}~\bibnamefont {Verhagen}},\ and\ \bibinfo {author} {\bibfnamefont {A.~F.}\ \bibnamefont {Koenderink}},\ }\bibfield  {title} {\bibinfo {title} {Integrated {Molecular} {Optomechanics} with {Hybrid} {Dielectric}–{Metallic} {Resonators}},\ }\href {https://doi.org/10.1021/acsphotonics.1c00808} {\bibfield  {journal} {\bibinfo  {journal} {ACS Photonics}\ }\textbf {\bibinfo {volume} {8}},\ \bibinfo {pages} {3506} (\bibinfo {year} {2021})}\BibitemShut {NoStop}%
\bibitem [{\citenamefont {Dezfouli}\ \emph {et~al.}(2019)\citenamefont {Dezfouli}, \citenamefont {Gordon},\ and\ \citenamefont {Hughes}}]{dezfouli_molecular_2019}%
  \BibitemOpen
  \bibfield  {author} {\bibinfo {author} {\bibfnamefont {M.~K.}\ \bibnamefont {Dezfouli}}, \bibinfo {author} {\bibfnamefont {R.}~\bibnamefont {Gordon}},\ and\ \bibinfo {author} {\bibfnamefont {S.}~\bibnamefont {Hughes}},\ }\bibfield  {title} {\bibinfo {title} {Molecular {Optomechanics} in the {Anharmonic} {Cavity}-{QED} {Regime} {Using} {Hybrid} {Metal}–{Dielectric} {Cavity} {Modes}},\ }\href {https://doi.org/10.1021/acsphotonics.8b01091} {\bibfield  {journal} {\bibinfo  {journal} {ACS Photonics}\ }\textbf {\bibinfo {volume} {6}},\ \bibinfo {pages} {1400} (\bibinfo {year} {2019})}\BibitemShut {NoStop}%
\bibitem [{\citenamefont {Roelli}\ \emph {et~al.}(2016)\citenamefont {Roelli}, \citenamefont {Galland}, \citenamefont {Piro},\ and\ \citenamefont {Kippenberg}}]{roelli_molecular_2016}%
  \BibitemOpen
  \bibfield  {author} {\bibinfo {author} {\bibfnamefont {P.}~\bibnamefont {Roelli}}, \bibinfo {author} {\bibfnamefont {C.}~\bibnamefont {Galland}}, \bibinfo {author} {\bibfnamefont {N.}~\bibnamefont {Piro}},\ and\ \bibinfo {author} {\bibfnamefont {T.~J.}\ \bibnamefont {Kippenberg}},\ }\bibfield  {title} {\bibinfo {title} {Molecular cavity optomechanics as a theory of plasmon-enhanced {Raman} scattering},\ }\href {https://doi.org/10.1038/nnano.2015.264} {\bibfield  {journal} {\bibinfo  {journal} {Nature Nanotechnology}\ }\textbf {\bibinfo {volume} {11}},\ \bibinfo {pages} {164} (\bibinfo {year} {2016})}\BibitemShut {NoStop}%
\bibitem [{\citenamefont {Schmidt}\ \emph {et~al.}(2016)\citenamefont {Schmidt}, \citenamefont {Esteban}, \citenamefont {González-Tudela}, \citenamefont {Giedke},\ and\ \citenamefont {Aizpurua}}]{schmidt_quantum_2016}%
  \BibitemOpen
  \bibfield  {author} {\bibinfo {author} {\bibfnamefont {M.~K.}\ \bibnamefont {Schmidt}}, \bibinfo {author} {\bibfnamefont {R.}~\bibnamefont {Esteban}}, \bibinfo {author} {\bibfnamefont {A.}~\bibnamefont {González-Tudela}}, \bibinfo {author} {\bibfnamefont {G.}~\bibnamefont {Giedke}},\ and\ \bibinfo {author} {\bibfnamefont {J.}~\bibnamefont {Aizpurua}},\ }\bibfield  {title} {\bibinfo {title} {Quantum {Mechanical} {Description} of {Raman} {Scattering} from {Molecules} in {Plasmonic} {Cavities}},\ }\href {https://doi.org/10.1021/acsnano.6b02484} {\bibfield  {journal} {\bibinfo  {journal} {ACS Nano}\ }\textbf {\bibinfo {volume} {10}},\ \bibinfo {pages} {6291} (\bibinfo {year} {2016})}\BibitemShut {NoStop}%
\bibitem [{\citenamefont {Schmidt}\ \emph {et~al.}(2017)\citenamefont {Schmidt}, \citenamefont {Esteban}, \citenamefont {Benz}, \citenamefont {Baumberg},\ and\ \citenamefont {Aizpurua}}]{schmidt_linking_2017}%
  \BibitemOpen
  \bibfield  {author} {\bibinfo {author} {\bibfnamefont {M.~K.}\ \bibnamefont {Schmidt}}, \bibinfo {author} {\bibfnamefont {R.}~\bibnamefont {Esteban}}, \bibinfo {author} {\bibfnamefont {F.}~\bibnamefont {Benz}}, \bibinfo {author} {\bibfnamefont {J.~J.}\ \bibnamefont {Baumberg}},\ and\ \bibinfo {author} {\bibfnamefont {J.}~\bibnamefont {Aizpurua}},\ }\bibfield  {title} {{\selectlanguage {en}\bibinfo {title} {Linking classical and molecular optomechanics descriptions of {SERS}}},\ }\href {https://doi.org/10.1039/C7FD00145B} {\bibfield  {journal} {\bibinfo  {journal} {Faraday Discussions}\ }\textbf {\bibinfo {volume} {205}},\ \bibinfo {pages} {31} (\bibinfo {year} {2017})}\BibitemShut {NoStop}%
\bibitem [{\citenamefont {Lombardi}\ \emph {et~al.}(2018)\citenamefont {Lombardi}, \citenamefont {Schmidt}, \citenamefont {Weller}, \citenamefont {Deacon}, \citenamefont {Benz}, \citenamefont {De~Nijs}, \citenamefont {Aizpurua},\ and\ \citenamefont {Baumberg}}]{lombardi_pulsed_2018}%
  \BibitemOpen
  \bibfield  {author} {\bibinfo {author} {\bibfnamefont {A.}~\bibnamefont {Lombardi}}, \bibinfo {author} {\bibfnamefont {M.~K.}\ \bibnamefont {Schmidt}}, \bibinfo {author} {\bibfnamefont {L.}~\bibnamefont {Weller}}, \bibinfo {author} {\bibfnamefont {W.~M.}\ \bibnamefont {Deacon}}, \bibinfo {author} {\bibfnamefont {F.}~\bibnamefont {Benz}}, \bibinfo {author} {\bibfnamefont {B.}~\bibnamefont {De~Nijs}}, \bibinfo {author} {\bibfnamefont {J.}~\bibnamefont {Aizpurua}},\ and\ \bibinfo {author} {\bibfnamefont {J.~J.}\ \bibnamefont {Baumberg}},\ }\bibfield  {title} {{\selectlanguage {en}\bibinfo {title} {Pulsed {Molecular} {Optomechanics} in {Plasmonic} {Nanocavities}: {From} {Nonlinear} {Vibrational} {Instabilities} to {Bond}-{Breaking}}},\ }\href {https://doi.org/10.1103/PhysRevX.8.011016} {\bibfield  {journal} {\bibinfo  {journal} {Physical Review X}\ }\textbf {\bibinfo {volume} {8}},\ \bibinfo {pages} {011016} (\bibinfo {year} {2018})}\BibitemShut {NoStop}%
\bibitem [{\citenamefont {Schmidt}\ \emph {et~al.}(2021)\citenamefont {Schmidt}, \citenamefont {Esteban}, \citenamefont {Giedke}, \citenamefont {Aizpurua},\ and\ \citenamefont {González-Tudela}}]{schmidt_frequency-resolved_2021}%
  \BibitemOpen
  \bibfield  {author} {\bibinfo {author} {\bibfnamefont {M.~K.}\ \bibnamefont {Schmidt}}, \bibinfo {author} {\bibfnamefont {R.}~\bibnamefont {Esteban}}, \bibinfo {author} {\bibfnamefont {G.}~\bibnamefont {Giedke}}, \bibinfo {author} {\bibfnamefont {J.}~\bibnamefont {Aizpurua}},\ and\ \bibinfo {author} {\bibfnamefont {A.}~\bibnamefont {González-Tudela}},\ }\bibfield  {title} {\bibinfo {title} {Frequency-resolved photon correlations in cavity optomechanics},\ }\href {https://doi.org/10.1088/2058-9565/abe569} {\bibfield  {journal} {\bibinfo  {journal} {Quantum Science and Technology}\ }\textbf {\bibinfo {volume} {6}},\ \bibinfo {pages} {034005} (\bibinfo {year} {2021})}\BibitemShut {NoStop}%
\bibitem [{\citenamefont {Roelli}\ \emph {et~al.}(2020)\citenamefont {Roelli}, \citenamefont {Martin-Cano}, \citenamefont {Kippenberg},\ and\ \citenamefont {Galland}}]{roelli_molecular_2020}%
  \BibitemOpen
  \bibfield  {author} {\bibinfo {author} {\bibfnamefont {P.}~\bibnamefont {Roelli}}, \bibinfo {author} {\bibfnamefont {D.}~\bibnamefont {Martin-Cano}}, \bibinfo {author} {\bibfnamefont {T.~J.}\ \bibnamefont {Kippenberg}},\ and\ \bibinfo {author} {\bibfnamefont {C.}~\bibnamefont {Galland}},\ }\bibfield  {title} {{\selectlanguage {en}\bibinfo {title} {Molecular {Platform} for {Frequency} {Upconversion} at the {Single}-{Photon} {Level}}},\ }\href {https://doi.org/10.1103/PhysRevX.10.031057} {\bibfield  {journal} {\bibinfo  {journal} {Physical Review X}\ }\textbf {\bibinfo {volume} {10}},\ \bibinfo {pages} {031057} (\bibinfo {year} {2020})}\BibitemShut {NoStop}%
\bibitem [{\citenamefont {Chen}\ \emph {et~al.}(2021)\citenamefont {Chen}, \citenamefont {Roelli}, \citenamefont {Hu}, \citenamefont {Verlekar}, \citenamefont {Amirtharaj}, \citenamefont {Barreda}, \citenamefont {Kippenberg}, \citenamefont {Kovylina}, \citenamefont {Verhagen}, \citenamefont {Martínez},\ and\ \citenamefont {Galland}}]{chen_continuous-wave_2021}%
  \BibitemOpen
  \bibfield  {author} {\bibinfo {author} {\bibfnamefont {W.}~\bibnamefont {Chen}}, \bibinfo {author} {\bibfnamefont {P.}~\bibnamefont {Roelli}}, \bibinfo {author} {\bibfnamefont {H.}~\bibnamefont {Hu}}, \bibinfo {author} {\bibfnamefont {S.}~\bibnamefont {Verlekar}}, \bibinfo {author} {\bibfnamefont {S.~P.}\ \bibnamefont {Amirtharaj}}, \bibinfo {author} {\bibfnamefont {A.~I.}\ \bibnamefont {Barreda}}, \bibinfo {author} {\bibfnamefont {T.~J.}\ \bibnamefont {Kippenberg}}, \bibinfo {author} {\bibfnamefont {M.}~\bibnamefont {Kovylina}}, \bibinfo {author} {\bibfnamefont {E.}~\bibnamefont {Verhagen}}, \bibinfo {author} {\bibfnamefont {A.}~\bibnamefont {Martínez}},\ and\ \bibinfo {author} {\bibfnamefont {C.}~\bibnamefont {Galland}},\ }\bibfield  {title} {\bibinfo {title} {Continuous-wave frequency upconversion with a molecular optomechanical nanocavity},\ }\href {https://doi.org/10.1126/science.abk3106} {\bibfield  {journal} {\bibinfo  {journal} {Science}\ }\textbf {\bibinfo {volume} {374}},\ \bibinfo {pages} {1264}
  (\bibinfo {year} {2021})}\BibitemShut {NoStop}%
\bibitem [{\citenamefont {Xomalis}\ \emph {et~al.}(2021)\citenamefont {Xomalis}, \citenamefont {Zheng}, \citenamefont {Chikkaraddy}, \citenamefont {Koczor-Benda}, \citenamefont {Miele}, \citenamefont {Rosta}, \citenamefont {Vandenbosch}, \citenamefont {Martínez},\ and\ \citenamefont {Baumberg}}]{xomalis_detecting_2021}%
  \BibitemOpen
  \bibfield  {author} {\bibinfo {author} {\bibfnamefont {A.}~\bibnamefont {Xomalis}}, \bibinfo {author} {\bibfnamefont {X.}~\bibnamefont {Zheng}}, \bibinfo {author} {\bibfnamefont {R.}~\bibnamefont {Chikkaraddy}}, \bibinfo {author} {\bibfnamefont {Z.}~\bibnamefont {Koczor-Benda}}, \bibinfo {author} {\bibfnamefont {E.}~\bibnamefont {Miele}}, \bibinfo {author} {\bibfnamefont {E.}~\bibnamefont {Rosta}}, \bibinfo {author} {\bibfnamefont {G.~A.~E.}\ \bibnamefont {Vandenbosch}}, \bibinfo {author} {\bibfnamefont {A.}~\bibnamefont {Martínez}},\ and\ \bibinfo {author} {\bibfnamefont {J.~J.}\ \bibnamefont {Baumberg}},\ }\bibfield  {title} {\bibinfo {title} {Detecting mid-infrared light by molecular frequency upconversion in dual-wavelength nanoantennas},\ }\href {https://doi.org/10.1126/science.abk2593} {\bibfield  {journal} {\bibinfo  {journal} {Science}\ }\textbf {\bibinfo {volume} {374}},\ \bibinfo {pages} {1268} (\bibinfo {year} {2021})}\BibitemShut {NoStop}%
\bibitem [{\citenamefont {Schmidt}\ and\ \citenamefont {Steel}(2024{\natexlab{a}})}]{schmidt_molecular_2024}%
  \BibitemOpen
  \bibfield  {author} {\bibinfo {author} {\bibfnamefont {M.~K.}\ \bibnamefont {Schmidt}}\ and\ \bibinfo {author} {\bibfnamefont {M.~J.}\ \bibnamefont {Steel}},\ }\bibfield  {title} {\bibinfo {title} {Molecular optomechanics in the anharmonic regime: from nonclassical mechanical states to mechanical lasing},\ }\href {https://iopscience.iop.org/article/10.1088/1367-2630/ad32e4/meta} {\bibfield  {journal} {\bibinfo  {journal} {New Journal of Physics}\ }\textbf {\bibinfo {volume} {26}},\ \bibinfo {pages} {033041} (\bibinfo {year} {2024}{\natexlab{a}})}\BibitemShut {NoStop}%
\bibitem [{\citenamefont {Kalarde}\ \emph {et~al.}(2025)\citenamefont {Kalarde}, \citenamefont {Ciccarello}, \citenamefont {Muñoz}, \citenamefont {Feist},\ and\ \citenamefont {Galland}}]{kalarde_photon_2024}%
  \BibitemOpen
  \bibfield  {author} {\bibinfo {author} {\bibfnamefont {F.~M.}\ \bibnamefont {Kalarde}}, \bibinfo {author} {\bibfnamefont {F.}~\bibnamefont {Ciccarello}}, \bibinfo {author} {\bibfnamefont {C.~S.}\ \bibnamefont {Muñoz}}, \bibinfo {author} {\bibfnamefont {J.}~\bibnamefont {Feist}},\ and\ \bibinfo {author} {\bibfnamefont {C.}~\bibnamefont {Galland}},\ }\bibfield  {title} {\bibinfo {title} {Photon antibunching in single-molecule vibrational sum-frequency generation},\ }\href {https://doi.org/doi:10.1515/nanoph-2024-0469} {\bibfield  {journal} {\bibinfo  {journal} {Nanophotonics}\ }\textbf {\bibinfo {volume} {14}},\ \bibinfo {pages} {59} (\bibinfo {year} {2025})}\BibitemShut {NoStop}%
\bibitem [{\citenamefont {Zhang}\ \emph {et~al.}(2020)\citenamefont {Zhang}, \citenamefont {Aizpurua},\ and\ \citenamefont {Esteban}}]{zhang_optomechanical_2020}%
  \BibitemOpen
  \bibfield  {author} {\bibinfo {author} {\bibfnamefont {Y.}~\bibnamefont {Zhang}}, \bibinfo {author} {\bibfnamefont {J.}~\bibnamefont {Aizpurua}},\ and\ \bibinfo {author} {\bibfnamefont {R.}~\bibnamefont {Esteban}},\ }\bibfield  {title} {\bibinfo {title} {Optomechanical {Collective} {Effects} in {Surface}-{Enhanced} {Raman} {Scattering} from {Many} {Molecules}},\ }\href {https://doi.org/10.1021/acsphotonics.0c00032} {\bibfield  {journal} {\bibinfo  {journal} {ACS Photonics}\ }\textbf {\bibinfo {volume} {7}},\ \bibinfo {pages} {1676} (\bibinfo {year} {2020})}\BibitemShut {NoStop}%
\bibitem [{\citenamefont {Itoh}\ \emph {et~al.}(2023)\citenamefont {Itoh}, \citenamefont {Procházka}, \citenamefont {Dong}, \citenamefont {Ji}, \citenamefont {Yamamoto}, \citenamefont {Zhang},\ and\ \citenamefont {Ozaki}}]{itoh_toward_2023}%
  \BibitemOpen
  \bibfield  {author} {\bibinfo {author} {\bibfnamefont {T.}~\bibnamefont {Itoh}}, \bibinfo {author} {\bibfnamefont {M.}~\bibnamefont {Procházka}}, \bibinfo {author} {\bibfnamefont {Z.-C.}\ \bibnamefont {Dong}}, \bibinfo {author} {\bibfnamefont {W.}~\bibnamefont {Ji}}, \bibinfo {author} {\bibfnamefont {Y.~S.}\ \bibnamefont {Yamamoto}}, \bibinfo {author} {\bibfnamefont {Y.}~\bibnamefont {Zhang}},\ and\ \bibinfo {author} {\bibfnamefont {Y.}~\bibnamefont {Ozaki}},\ }\bibfield  {title} {\bibinfo {title} {Toward a {New} {Era} of {SERS} and {TERS} at the {Nanometer} {Scale}: {From} {Fundamentals} to {Innovative} {Applications}},\ }\href {https://doi.org/10.1021/acs.chemrev.2c00316} {\bibfield  {journal} {\bibinfo  {journal} {Chemical Reviews}\ }\textbf {\bibinfo {volume} {123}},\ \bibinfo {pages} {1552} (\bibinfo {year} {2023})}\BibitemShut {NoStop}%
\bibitem [{\citenamefont {Höppener}\ \emph {et~al.}(2024)\citenamefont {Höppener}, \citenamefont {Aizpurua}, \citenamefont {Chen}, \citenamefont {Gräfe}, \citenamefont {Jorio}, \citenamefont {Kupfer}, \citenamefont {Zhang},\ and\ \citenamefont {Deckert}}]{hoppener_tip-enhanced_2024}%
  \BibitemOpen
  \bibfield  {author} {\bibinfo {author} {\bibfnamefont {C.}~\bibnamefont {Höppener}}, \bibinfo {author} {\bibfnamefont {J.}~\bibnamefont {Aizpurua}}, \bibinfo {author} {\bibfnamefont {H.}~\bibnamefont {Chen}}, \bibinfo {author} {\bibfnamefont {S.}~\bibnamefont {Gräfe}}, \bibinfo {author} {\bibfnamefont {A.}~\bibnamefont {Jorio}}, \bibinfo {author} {\bibfnamefont {S.}~\bibnamefont {Kupfer}}, \bibinfo {author} {\bibfnamefont {Z.}~\bibnamefont {Zhang}},\ and\ \bibinfo {author} {\bibfnamefont {V.}~\bibnamefont {Deckert}},\ }\bibfield  {title} {{\selectlanguage {en}\bibinfo {title} {Tip-enhanced {Raman} scattering}},\ }\href {https://doi.org/10.1038/s43586-024-00323-5} {\bibfield  {journal} {\bibinfo  {journal} {Nature Reviews Methods Primers}\ }\textbf {\bibinfo {volume} {4}},\ \bibinfo {pages} {47} (\bibinfo {year} {2024})}\BibitemShut {NoStop}%
\bibitem [{\citenamefont {Zhang}\ \emph {et~al.}(2021)\citenamefont {Zhang}, \citenamefont {Esteban}, \citenamefont {Boto}, \citenamefont {Urbieta}, \citenamefont {Arrieta}, \citenamefont {Shan}, \citenamefont {Li}, \citenamefont {Baumberg},\ and\ \citenamefont {Aizpurua}}]{zhang_addressing_2021-1}%
  \BibitemOpen
  \bibfield  {author} {\bibinfo {author} {\bibfnamefont {Y.}~\bibnamefont {Zhang}}, \bibinfo {author} {\bibfnamefont {R.}~\bibnamefont {Esteban}}, \bibinfo {author} {\bibfnamefont {R.~A.}\ \bibnamefont {Boto}}, \bibinfo {author} {\bibfnamefont {M.}~\bibnamefont {Urbieta}}, \bibinfo {author} {\bibfnamefont {X.}~\bibnamefont {Arrieta}}, \bibinfo {author} {\bibfnamefont {C.}~\bibnamefont {Shan}}, \bibinfo {author} {\bibfnamefont {S.}~\bibnamefont {Li}}, \bibinfo {author} {\bibfnamefont {J.~J.}\ \bibnamefont {Baumberg}},\ and\ \bibinfo {author} {\bibfnamefont {J.}~\bibnamefont {Aizpurua}},\ }\bibfield  {title} {\bibinfo {title} {Addressing molecular optomechanical effects in nanocavity-enhanced {Raman} scattering beyond the single plasmonic mode},\ }\href {https://doi.org/10.1039/D0NR06649D} {\bibfield  {journal} {\bibinfo  {journal} {Nanoscale}\ }\textbf {\bibinfo {volume} {13}},\ \bibinfo {pages} {1938} (\bibinfo {year} {2021})}\BibitemShut {NoStop}%
\bibitem [{\citenamefont {Kamandar~Dezfouli}\ and\ \citenamefont {Hughes}(2017)}]{kamandar_dezfouli_quantum_2017}%
  \BibitemOpen
  \bibfield  {author} {\bibinfo {author} {\bibfnamefont {M.}~\bibnamefont {Kamandar~Dezfouli}}\ and\ \bibinfo {author} {\bibfnamefont {S.}~\bibnamefont {Hughes}},\ }\bibfield  {title} {\bibinfo {title} {Quantum {Optics} {Model} of {Surface}-{Enhanced} {Raman} {Spectroscopy} for {Arbitrarily} {Shaped} {Plasmonic} {Resonators}},\ }\href {https://doi.org/10.1021/acsphotonics.7b00157} {\bibfield  {journal} {\bibinfo  {journal} {ACS Photonics}\ }\textbf {\bibinfo {volume} {4}},\ \bibinfo {pages} {1245} (\bibinfo {year} {2017})}\BibitemShut {NoStop}%
\bibitem [{\citenamefont {Li}\ \emph {et~al.}(2024)\citenamefont {Li}, \citenamefont {Guo}, \citenamefont {Jin}, \citenamefont {Andreoli}, \citenamefont {Bilgin}, \citenamefont {Awschalom}, \citenamefont {Delegan}, \citenamefont {Heremans}, \citenamefont {Chang}, \citenamefont {Galli},\ and\ \citenamefont {High}}]{li_atomic_2024}%
  \BibitemOpen
  \bibfield  {author} {\bibinfo {author} {\bibfnamefont {Z.}~\bibnamefont {Li}}, \bibinfo {author} {\bibfnamefont {X.}~\bibnamefont {Guo}}, \bibinfo {author} {\bibfnamefont {Y.}~\bibnamefont {Jin}}, \bibinfo {author} {\bibfnamefont {F.}~\bibnamefont {Andreoli}}, \bibinfo {author} {\bibfnamefont {A.}~\bibnamefont {Bilgin}}, \bibinfo {author} {\bibfnamefont {D.~D.}\ \bibnamefont {Awschalom}}, \bibinfo {author} {\bibfnamefont {N.}~\bibnamefont {Delegan}}, \bibinfo {author} {\bibfnamefont {F.~J.}\ \bibnamefont {Heremans}}, \bibinfo {author} {\bibfnamefont {D.}~\bibnamefont {Chang}}, \bibinfo {author} {\bibfnamefont {G.}~\bibnamefont {Galli}},\ and\ \bibinfo {author} {\bibfnamefont {A.~A.}\ \bibnamefont {High}},\ }\bibfield  {title} {{\selectlanguage {en}\bibinfo {title} {Atomic optical antennas in solids}},\ }\href {https://doi.org/10.1038/s41566-024-01456-5} {\bibfield  {journal} {\bibinfo  {journal} {Nature Photonics}\ }\textbf {\bibinfo {volume} {18}},\ \bibinfo {pages} {1113} (\bibinfo {year}
  {2024})}\BibitemShut {NoStop}%
\bibitem [{\citenamefont {Schmidt}\ and\ \citenamefont {Steel}(2024{\natexlab{b}})}]{schmidt_molecular_2024-1}%
  \BibitemOpen
  \bibfield  {author} {\bibinfo {author} {\bibfnamefont {M.~K.}\ \bibnamefont {Schmidt}}\ and\ \bibinfo {author} {\bibfnamefont {M.~J.}\ \bibnamefont {Steel}},\ }\bibfield  {title} {\bibinfo {title} {Molecular optomechanics in the anharmonic regime: from nonclassical mechanical states to mechanical lasing},\ }\href {https://doi.org/10.1088/1367-2630/ad32e4} {\bibfield  {journal} {\bibinfo  {journal} {New Journal of Physics}\ }\textbf {\bibinfo {volume} {26}},\ \bibinfo {pages} {033041} (\bibinfo {year} {2024}{\natexlab{b}})}\BibitemShut {NoStop}%
\bibitem [{\citenamefont {Aspelmeyer}\ \emph {et~al.}(2014)\citenamefont {Aspelmeyer}, \citenamefont {Kippenberg},\ and\ \citenamefont {Marquardt}}]{aspelmeyer_cavity_2014}%
  \BibitemOpen
  \bibfield  {author} {\bibinfo {author} {\bibfnamefont {M.}~\bibnamefont {Aspelmeyer}}, \bibinfo {author} {\bibfnamefont {T.~J.}\ \bibnamefont {Kippenberg}},\ and\ \bibinfo {author} {\bibfnamefont {F.}~\bibnamefont {Marquardt}},\ }\bibfield  {title} {\bibinfo {title} {Cavity optomechanics},\ }\href {https://doi.org/10.1103/RevModPhys.86.1391} {\bibfield  {journal} {\bibinfo  {journal} {Reviews of Modern Physics}\ }\textbf {\bibinfo {volume} {86}},\ \bibinfo {pages} {1391} (\bibinfo {year} {2014})}\BibitemShut {NoStop}%
\bibitem [{\citenamefont {Roelli}\ \emph {et~al.}(2024)\citenamefont {Roelli}, \citenamefont {Hu}, \citenamefont {Verhagen}, \citenamefont {Reich},\ and\ \citenamefont {Galland}}]{roelli_nanocavities_2024}%
  \BibitemOpen
  \bibfield  {author} {\bibinfo {author} {\bibfnamefont {P.}~\bibnamefont {Roelli}}, \bibinfo {author} {\bibfnamefont {H.}~\bibnamefont {Hu}}, \bibinfo {author} {\bibfnamefont {E.}~\bibnamefont {Verhagen}}, \bibinfo {author} {\bibfnamefont {S.}~\bibnamefont {Reich}},\ and\ \bibinfo {author} {\bibfnamefont {C.}~\bibnamefont {Galland}},\ }\bibfield  {title} {\bibinfo {title} {Nanocavities for {Molecular} {Optomechanics}: {Their} {Fundamental} {Description} and {Applications}},\ }\href {https://doi.org/10.1021/acsphotonics.4c01548} {\bibfield  {journal} {\bibinfo  {journal} {ACS Photonics}\ }\textbf {\bibinfo {volume} {11}},\ \bibinfo {pages} {4486} (\bibinfo {year} {2024})}\BibitemShut {NoStop}%
\bibitem [{\citenamefont {Marquardt}\ \emph {et~al.}(2008)\citenamefont {Marquardt}, \citenamefont {Clerk},\ and\ \citenamefont {Girvin}}]{marquardt_quantum_2008}%
  \BibitemOpen
  \bibfield  {author} {\bibinfo {author} {\bibfnamefont {F.}~\bibnamefont {Marquardt}}, \bibinfo {author} {\bibfnamefont {A.}~\bibnamefont {Clerk}},\ and\ \bibinfo {author} {\bibfnamefont {S.}~\bibnamefont {Girvin}},\ }\bibfield  {title} {\bibinfo {title} {Quantum theory of optomechanical cooling},\ }\href {https://doi.org/10.1080/09500340802454971} {\bibfield  {journal} {\bibinfo  {journal} {Journal of Modern Optics}\ }\textbf {\bibinfo {volume} {55}},\ \bibinfo {pages} {3329} (\bibinfo {year} {2008})}\BibitemShut {NoStop}%
\bibitem [{\citenamefont {Clerk}\ \emph {et~al.}(2010)\citenamefont {Clerk}, \citenamefont {Devoret}, \citenamefont {Girvin}, \citenamefont {Marquardt},\ and\ \citenamefont {Schoelkopf}}]{clerk_introduction_2010}%
  \BibitemOpen
  \bibfield  {author} {\bibinfo {author} {\bibfnamefont {A.~A.}\ \bibnamefont {Clerk}}, \bibinfo {author} {\bibfnamefont {M.~H.}\ \bibnamefont {Devoret}}, \bibinfo {author} {\bibfnamefont {S.~M.}\ \bibnamefont {Girvin}}, \bibinfo {author} {\bibfnamefont {F.}~\bibnamefont {Marquardt}},\ and\ \bibinfo {author} {\bibfnamefont {R.~J.}\ \bibnamefont {Schoelkopf}},\ }\bibfield  {title} {\bibinfo {title} {Introduction to quantum noise, measurement, and amplification},\ }\href {https://doi.org/10.1103/RevModPhys.82.1155} {\bibfield  {journal} {\bibinfo  {journal} {Reviews of Modern Physics}\ }\textbf {\bibinfo {volume} {82}},\ \bibinfo {pages} {1155} (\bibinfo {year} {2010})}\BibitemShut {NoStop}%
\bibitem [{\citenamefont {Carmichael}(2013)}]{carmichael2013statistical}%
  \BibitemOpen
  \bibfield  {author} {\bibinfo {author} {\bibfnamefont {H.~J.}\ \bibnamefont {Carmichael}},\ }\href {https://link.springer.com/book/10.1007/978-3-540-28574-8} {\emph {\bibinfo {title} {Statistical methods in quantum optics 1: master equations and {Fokker-Planck} equations}}},\ Theoretical and Mathematical Physics\ (\bibinfo  {publisher} {Springer Science \& Business Media},\ \bibinfo {year} {2013})\BibitemShut {NoStop}%
\bibitem [{\citenamefont {Wilson-Rae}\ \emph {et~al.}(2007)\citenamefont {Wilson-Rae}, \citenamefont {Nooshi}, \citenamefont {Zwerger},\ and\ \citenamefont {Kippenberg}}]{wilson-rae_theory_2007}%
  \BibitemOpen
  \bibfield  {author} {\bibinfo {author} {\bibfnamefont {I.}~\bibnamefont {Wilson-Rae}}, \bibinfo {author} {\bibfnamefont {N.}~\bibnamefont {Nooshi}}, \bibinfo {author} {\bibfnamefont {W.}~\bibnamefont {Zwerger}},\ and\ \bibinfo {author} {\bibfnamefont {T.~J.}\ \bibnamefont {Kippenberg}},\ }\bibfield  {title} {\bibinfo {title} {Theory of {Ground} {State} {Cooling} of a {Mechanical} {Oscillator} {Using} {Dynamical} {Backaction}},\ }\href {https://doi.org/10.1103/PhysRevLett.99.093901} {\bibfield  {journal} {\bibinfo  {journal} {Physical Review Letters}\ }\textbf {\bibinfo {volume} {99}},\ \bibinfo {pages} {093901} (\bibinfo {year} {2007})}\BibitemShut {NoStop}%
\bibitem [{\citenamefont {Kneipp}\ \emph {et~al.}(1996)\citenamefont {Kneipp}, \citenamefont {Wang}, \citenamefont {Kneipp}, \citenamefont {Itzkan}, \citenamefont {Dasari},\ and\ \citenamefont {Feld}}]{PhysRevLett.76.2444}%
  \BibitemOpen
  \bibfield  {author} {\bibinfo {author} {\bibfnamefont {K.}~\bibnamefont {Kneipp}}, \bibinfo {author} {\bibfnamefont {Y.}~\bibnamefont {Wang}}, \bibinfo {author} {\bibfnamefont {H.}~\bibnamefont {Kneipp}}, \bibinfo {author} {\bibfnamefont {I.}~\bibnamefont {Itzkan}}, \bibinfo {author} {\bibfnamefont {R.~R.}\ \bibnamefont {Dasari}},\ and\ \bibinfo {author} {\bibfnamefont {M.~S.}\ \bibnamefont {Feld}},\ }\bibfield  {title} {\bibinfo {title} {Population pumping of excited vibrational states by spontaneous surface-enhanced raman scattering},\ }\href {https://doi.org/10.1103/PhysRevLett.76.2444} {\bibfield  {journal} {\bibinfo  {journal} {Phys. Rev. Lett.}\ }\textbf {\bibinfo {volume} {76}},\ \bibinfo {pages} {2444} (\bibinfo {year} {1996})}\BibitemShut {NoStop}%
\bibitem [{\citenamefont {Maher}\ \emph {et~al.}(2006)\citenamefont {Maher}, \citenamefont {Etchegoin}, \citenamefont {Le~Ru},\ and\ \citenamefont {Cohen}}]{maher2006conclusive}%
  \BibitemOpen
  \bibfield  {author} {\bibinfo {author} {\bibfnamefont {R.}~\bibnamefont {Maher}}, \bibinfo {author} {\bibfnamefont {P.}~\bibnamefont {Etchegoin}}, \bibinfo {author} {\bibfnamefont {E.}~\bibnamefont {Le~Ru}},\ and\ \bibinfo {author} {\bibfnamefont {L.}~\bibnamefont {Cohen}},\ }\bibfield  {title} {\bibinfo {title} {A conclusive demonstration of vibrational pumping under surface enhanced raman scattering conditions},\ }\href@noop {} {\bibfield  {journal} {\bibinfo  {journal} {The Journal of Physical Chemistry B}\ }\textbf {\bibinfo {volume} {110}},\ \bibinfo {pages} {11757} (\bibinfo {year} {2006})}\BibitemShut {NoStop}%
\bibitem [{\citenamefont {Vitkova}\ \emph {et~al.}(2022)\citenamefont {Vitkova}, \citenamefont {I. Walker},\ and\ \citenamefont {Sykulska-Lawrence}}]{vitkova_cryogenically_2022}%
  \BibitemOpen
  \bibfield  {author} {\bibinfo {author} {\bibfnamefont {A.}~\bibnamefont {Vitkova}}, \bibinfo {author} {\bibfnamefont {S.~J.}\ \bibnamefont {I. Walker}},\ and\ \bibinfo {author} {\bibfnamefont {H.}~\bibnamefont {Sykulska-Lawrence}},\ }\bibfield  {title} {{\selectlanguage {en}\bibinfo {title} {Cryogenically induced signal enhancement of {Raman} spectra of porphyrin molecules}},\ }\href {https://doi.org/10.1039/D2AY00538G} {\bibfield  {journal} {\bibinfo  {journal} {Analytical Methods}\ }\textbf {\bibinfo {volume} {14}},\ \bibinfo {pages} {3307} (\bibinfo {year} {2022})},\ \bibinfo {note} {publisher: Royal Society of Chemistry}\BibitemShut {NoStop}%
\bibitem [{\citenamefont {Yi}\ \emph {et~al.}(2025)\citenamefont {Yi}, \citenamefont {You}, \citenamefont {Hu}, \citenamefont {Wu}, \citenamefont {Liu}, \citenamefont {Yang}, \citenamefont {Zhang}, \citenamefont {Gu}, \citenamefont {Wang}, \citenamefont {Wang}, \citenamefont {Ma}, \citenamefont {Yang}, \citenamefont {Liu}, \citenamefont {Ru Fan}, \citenamefont {Zhan}, \citenamefont {Tian}, \citenamefont {Qiao}, \citenamefont {Wang}, \citenamefont {Luo}, \citenamefont {Meng}, \citenamefont {Mao}, \citenamefont {Li}, \citenamefont {Ren}, \citenamefont {Aizpurua}, \citenamefont {Ara Apkarian}, \citenamefont {N. Bartlett}, \citenamefont {Baumberg}, \citenamefont {J. Bell}, \citenamefont {G. Brolo}, \citenamefont {E. Brus}, \citenamefont {Choo}, \citenamefont {Cui}, \citenamefont {Deckert}, \citenamefont {F. Domke}, \citenamefont {Dong}, \citenamefont {Duan}, \citenamefont {Faulds}, \citenamefont {Frontiera}, \citenamefont {Halas}, \citenamefont {Haynes}, \citenamefont {Itoh}, \citenamefont {Kneipp},
  \citenamefont {Kneipp}, \citenamefont {Ru}, \citenamefont {Li}, \citenamefont {Yi Ling}, \citenamefont {Lipkowski}, \citenamefont {M. Liz-Marzán}, \citenamefont {Nam}, \citenamefont {Nie}, \citenamefont {Nordlander}, \citenamefont {Ozaki}, \citenamefont {Panneerselvam}, \citenamefont {Popp}, \citenamefont {E. Russell}, \citenamefont {Schlücker}, \citenamefont {Tian}, \citenamefont {Tong}, \citenamefont {Xu}, \citenamefont {Xu}, \citenamefont {Yang}, \citenamefont {Yao}, \citenamefont {Zhang}, \citenamefont {Zhang}, \citenamefont {Zhang}, \citenamefont {Zhao}, \citenamefont {Zenobi}, \citenamefont {C. Schatz}, \citenamefont {Graham},\ and\ \citenamefont {Tian}}]{yi_surface-enhanced_2025}%
  \BibitemOpen
  \bibfield  {author} {\bibinfo {author} {\bibfnamefont {J.}~\bibnamefont {Yi}}, \bibinfo {author} {\bibfnamefont {E.-M.}\ \bibnamefont {You}}, \bibinfo {author} {\bibfnamefont {R.}~\bibnamefont {Hu}}, \bibinfo {author} {\bibfnamefont {D.-Y.}\ \bibnamefont {Wu}}, \bibinfo {author} {\bibfnamefont {G.-K.}\ \bibnamefont {Liu}}, \bibinfo {author} {\bibfnamefont {Z.-L.}\ \bibnamefont {Yang}}, \bibinfo {author} {\bibfnamefont {H.}~\bibnamefont {Zhang}}, \bibinfo {author} {\bibfnamefont {Y.}~\bibnamefont {Gu}}, \bibinfo {author} {\bibfnamefont {Y.-H.}\ \bibnamefont {Wang}}, \bibinfo {author} {\bibfnamefont {X.}~\bibnamefont {Wang}}, \bibinfo {author} {\bibfnamefont {H.}~\bibnamefont {Ma}}, \bibinfo {author} {\bibfnamefont {Y.}~\bibnamefont {Yang}}, \bibinfo {author} {\bibfnamefont {J.-Y.}\ \bibnamefont {Liu}}, \bibinfo {author} {\bibfnamefont {F.}~\bibnamefont {Ru Fan}}, \bibinfo {author} {\bibfnamefont {C.}~\bibnamefont {Zhan}}, \bibinfo {author} {\bibfnamefont {J.-H.}\ \bibnamefont {Tian}}, \bibinfo {author}
  {\bibfnamefont {Y.}~\bibnamefont {Qiao}}, \bibinfo {author} {\bibfnamefont {H.}~\bibnamefont {Wang}}, \bibinfo {author} {\bibfnamefont {S.-H.}\ \bibnamefont {Luo}}, \bibinfo {author} {\bibfnamefont {Z.-D.}\ \bibnamefont {Meng}}, \bibinfo {author} {\bibfnamefont {B.-W.}\ \bibnamefont {Mao}}, \bibinfo {author} {\bibfnamefont {J.-F.}\ \bibnamefont {Li}}, \bibinfo {author} {\bibfnamefont {B.}~\bibnamefont {Ren}}, \bibinfo {author} {\bibfnamefont {J.}~\bibnamefont {Aizpurua}}, \bibinfo {author} {\bibfnamefont {V.}~\bibnamefont {Ara Apkarian}}, \bibinfo {author} {\bibfnamefont {P.}~\bibnamefont {N. Bartlett}}, \bibinfo {author} {\bibfnamefont {J.}~\bibnamefont {Baumberg}}, \bibinfo {author} {\bibfnamefont {S.~E.}\ \bibnamefont {J. Bell}}, \bibinfo {author} {\bibfnamefont {A.}~\bibnamefont {G. Brolo}}, \bibinfo {author} {\bibfnamefont {L.}~\bibnamefont {E. Brus}}, \bibinfo {author} {\bibfnamefont {J.}~\bibnamefont {Choo}}, \bibinfo {author} {\bibfnamefont {L.}~\bibnamefont {Cui}}, \bibinfo {author}
  {\bibfnamefont {V.}~\bibnamefont {Deckert}}, \bibinfo {author} {\bibfnamefont {K.}~\bibnamefont {F. Domke}}, \bibinfo {author} {\bibfnamefont {Z.-C.}\ \bibnamefont {Dong}}, \bibinfo {author} {\bibfnamefont {S.}~\bibnamefont {Duan}}, \bibinfo {author} {\bibfnamefont {K.}~\bibnamefont {Faulds}}, \bibinfo {author} {\bibfnamefont {R.}~\bibnamefont {Frontiera}}, \bibinfo {author} {\bibfnamefont {N.}~\bibnamefont {Halas}}, \bibinfo {author} {\bibfnamefont {C.}~\bibnamefont {Haynes}}, \bibinfo {author} {\bibfnamefont {T.}~\bibnamefont {Itoh}}, \bibinfo {author} {\bibfnamefont {J.}~\bibnamefont {Kneipp}}, \bibinfo {author} {\bibfnamefont {K.}~\bibnamefont {Kneipp}}, \bibinfo {author} {\bibfnamefont {E.~C.~L.}\ \bibnamefont {Ru}}, \bibinfo {author} {\bibfnamefont {Z.-P.}\ \bibnamefont {Li}}, \bibinfo {author} {\bibfnamefont {X.}~\bibnamefont {Yi Ling}}, \bibinfo {author} {\bibfnamefont {J.}~\bibnamefont {Lipkowski}}, \bibinfo {author} {\bibfnamefont {L.}~\bibnamefont {M. Liz-Marzán}}, \bibinfo {author}
  {\bibfnamefont {J.-M.}\ \bibnamefont {Nam}}, \bibinfo {author} {\bibfnamefont {S.}~\bibnamefont {Nie}}, \bibinfo {author} {\bibfnamefont {P.}~\bibnamefont {Nordlander}}, \bibinfo {author} {\bibfnamefont {Y.}~\bibnamefont {Ozaki}}, \bibinfo {author} {\bibfnamefont {R.}~\bibnamefont {Panneerselvam}}, \bibinfo {author} {\bibfnamefont {J.}~\bibnamefont {Popp}}, \bibinfo {author} {\bibfnamefont {A.}~\bibnamefont {E. Russell}}, \bibinfo {author} {\bibfnamefont {S.}~\bibnamefont {Schlücker}}, \bibinfo {author} {\bibfnamefont {Y.}~\bibnamefont {Tian}}, \bibinfo {author} {\bibfnamefont {L.}~\bibnamefont {Tong}}, \bibinfo {author} {\bibfnamefont {H.}~\bibnamefont {Xu}}, \bibinfo {author} {\bibfnamefont {Y.}~\bibnamefont {Xu}}, \bibinfo {author} {\bibfnamefont {L.}~\bibnamefont {Yang}}, \bibinfo {author} {\bibfnamefont {J.}~\bibnamefont {Yao}}, \bibinfo {author} {\bibfnamefont {J.}~\bibnamefont {Zhang}}, \bibinfo {author} {\bibfnamefont {Y.}~\bibnamefont {Zhang}}, \bibinfo {author} {\bibfnamefont {Y.}~\bibnamefont
  {Zhang}}, \bibinfo {author} {\bibfnamefont {B.}~\bibnamefont {Zhao}}, \bibinfo {author} {\bibfnamefont {R.}~\bibnamefont {Zenobi}}, \bibinfo {author} {\bibfnamefont {G.}~\bibnamefont {C. Schatz}}, \bibinfo {author} {\bibfnamefont {D.}~\bibnamefont {Graham}},\ and\ \bibinfo {author} {\bibfnamefont {Z.-Q.}\ \bibnamefont {Tian}},\ }\bibfield  {title} {{\selectlanguage {en}\bibinfo {title} {Surface-enhanced {Raman} spectroscopy: a half-century historical perspective}},\ }\bibfield  {journal} {\bibinfo  {journal} {Chemical Society Reviews}\ }\href {https://doi.org/10.1039/D4CS00883A} {10.1039/D4CS00883A} (\bibinfo {year} {2025}),\ \bibinfo {note} {publisher: Royal Society of Chemistry}\BibitemShut {NoStop}%
\bibitem [{\citenamefont {Shen}\ \emph {et~al.}(2023)\citenamefont {Shen}, \citenamefont {Zhang}, \citenamefont {Zhang}, \citenamefont {Zhang}, \citenamefont {Meng}, \citenamefont {Zheng}, \citenamefont {Lv}, \citenamefont {Wang}, \citenamefont {Boto}, \citenamefont {Shan},\ and\ \citenamefont {Aizpurua}}]{shen_optomechanical_2023}%
  \BibitemOpen
  \bibfield  {author} {\bibinfo {author} {\bibfnamefont {X.-M.}\ \bibnamefont {Shen}}, \bibinfo {author} {\bibfnamefont {Y.}~\bibnamefont {Zhang}}, \bibinfo {author} {\bibfnamefont {S.}~\bibnamefont {Zhang}}, \bibinfo {author} {\bibfnamefont {Y.}~\bibnamefont {Zhang}}, \bibinfo {author} {\bibfnamefont {Q.-S.}\ \bibnamefont {Meng}}, \bibinfo {author} {\bibfnamefont {G.}~\bibnamefont {Zheng}}, \bibinfo {author} {\bibfnamefont {S.}~\bibnamefont {Lv}}, \bibinfo {author} {\bibfnamefont {L.}~\bibnamefont {Wang}}, \bibinfo {author} {\bibfnamefont {R.~A.}\ \bibnamefont {Boto}}, \bibinfo {author} {\bibfnamefont {C.}~\bibnamefont {Shan}},\ and\ \bibinfo {author} {\bibfnamefont {J.}~\bibnamefont {Aizpurua}},\ }\bibfield  {title} {\bibinfo {title} {Optomechanical effects in nanocavity-enhanced resonant {Raman} scattering of a single molecule},\ }\href {https://doi.org/10.1103/PhysRevB.107.075435} {\bibfield  {journal} {\bibinfo  {journal} {Physical Review B}\ }\textbf {\bibinfo {volume} {107}},\ \bibinfo {pages} {075435}
  (\bibinfo {year} {2023})}\BibitemShut {NoStop}%
\bibitem [{\citenamefont {Gottscholl}\ \emph {et~al.}(2021)\citenamefont {Gottscholl}, \citenamefont {Diez}, \citenamefont {Soltamov}, \citenamefont {Kasper}, \citenamefont {Sperlich}, \citenamefont {Kianinia}, \citenamefont {Bradac}, \citenamefont {Aharonovich},\ and\ \citenamefont {Dyakonov}}]{gottscholl2021room}%
  \BibitemOpen
  \bibfield  {author} {\bibinfo {author} {\bibfnamefont {A.}~\bibnamefont {Gottscholl}}, \bibinfo {author} {\bibfnamefont {M.}~\bibnamefont {Diez}}, \bibinfo {author} {\bibfnamefont {V.}~\bibnamefont {Soltamov}}, \bibinfo {author} {\bibfnamefont {C.}~\bibnamefont {Kasper}}, \bibinfo {author} {\bibfnamefont {A.}~\bibnamefont {Sperlich}}, \bibinfo {author} {\bibfnamefont {M.}~\bibnamefont {Kianinia}}, \bibinfo {author} {\bibfnamefont {C.}~\bibnamefont {Bradac}}, \bibinfo {author} {\bibfnamefont {I.}~\bibnamefont {Aharonovich}},\ and\ \bibinfo {author} {\bibfnamefont {V.}~\bibnamefont {Dyakonov}},\ }\bibfield  {title} {\bibinfo {title} {Room temperature coherent control of spin defects in hexagonal boron nitride},\ }\href {https://doi.org/10.1126/sciadv.abf3630} {\bibfield  {journal} {\bibinfo  {journal} {Science Advances}\ }\textbf {\bibinfo {volume} {7}},\ \bibinfo {pages} {eabf3630} (\bibinfo {year} {2021})},\ \bibinfo {note} {publisher: American Association for the Advancement of Science}\BibitemShut
  {NoStop}%
\bibitem [{\citenamefont {Anderson}\ \emph {et~al.}(2022)\citenamefont {Anderson}, \citenamefont {Glen}, \citenamefont {Zeledon}, \citenamefont {Bourassa}, \citenamefont {Jin}, \citenamefont {Zhu}, \citenamefont {Vorwerk}, \citenamefont {Crook}, \citenamefont {Abe}, \citenamefont {Ul-Hassan}, \citenamefont {Ohshima}, \citenamefont {Son}, \citenamefont {Galli},\ and\ \citenamefont {Awschalom}}]{anderson2022five}%
  \BibitemOpen
  \bibfield  {author} {\bibinfo {author} {\bibfnamefont {C.~P.}\ \bibnamefont {Anderson}}, \bibinfo {author} {\bibfnamefont {E.~O.}\ \bibnamefont {Glen}}, \bibinfo {author} {\bibfnamefont {C.}~\bibnamefont {Zeledon}}, \bibinfo {author} {\bibfnamefont {A.}~\bibnamefont {Bourassa}}, \bibinfo {author} {\bibfnamefont {Y.}~\bibnamefont {Jin}}, \bibinfo {author} {\bibfnamefont {Y.}~\bibnamefont {Zhu}}, \bibinfo {author} {\bibfnamefont {C.}~\bibnamefont {Vorwerk}}, \bibinfo {author} {\bibfnamefont {A.~L.}\ \bibnamefont {Crook}}, \bibinfo {author} {\bibfnamefont {H.}~\bibnamefont {Abe}}, \bibinfo {author} {\bibfnamefont {J.}~\bibnamefont {Ul-Hassan}}, \bibinfo {author} {\bibfnamefont {T.}~\bibnamefont {Ohshima}}, \bibinfo {author} {\bibfnamefont {N.~T.}\ \bibnamefont {Son}}, \bibinfo {author} {\bibfnamefont {G.}~\bibnamefont {Galli}},\ and\ \bibinfo {author} {\bibfnamefont {D.~D.}\ \bibnamefont {Awschalom}},\ }\bibfield  {title} {\bibinfo {title} {Five-second coherence of a single spin with single-shot readout in
  silicon carbide},\ }\href {https://doi.org/10.1126/sciadv.abm5912} {\bibfield  {journal} {\bibinfo  {journal} {Science Advances}\ }\textbf {\bibinfo {volume} {8}},\ \bibinfo {pages} {eabm5912} (\bibinfo {year} {2022})},\ \bibinfo {note} {publisher: American Association for the Advancement of Science}\BibitemShut {NoStop}%
\bibitem [{\citenamefont {Bayliss}\ \emph {et~al.}(2022)\citenamefont {Bayliss}, \citenamefont {Deb}, \citenamefont {Laorenza}, \citenamefont {Onizhuk}, \citenamefont {Galli}, \citenamefont {Freedman},\ and\ \citenamefont {Awschalom}}]{PhysRevX.12.031028}%
  \BibitemOpen
  \bibfield  {author} {\bibinfo {author} {\bibfnamefont {S.~L.}\ \bibnamefont {Bayliss}}, \bibinfo {author} {\bibfnamefont {P.}~\bibnamefont {Deb}}, \bibinfo {author} {\bibfnamefont {D.~W.}\ \bibnamefont {Laorenza}}, \bibinfo {author} {\bibfnamefont {M.}~\bibnamefont {Onizhuk}}, \bibinfo {author} {\bibfnamefont {G.}~\bibnamefont {Galli}}, \bibinfo {author} {\bibfnamefont {D.~E.}\ \bibnamefont {Freedman}},\ and\ \bibinfo {author} {\bibfnamefont {D.~D.}\ \bibnamefont {Awschalom}},\ }\bibfield  {title} {\bibinfo {title} {Enhancing spin coherence in optically addressable molecular qubits through host-matrix control},\ }\href {https://doi.org/10.1103/PhysRevX.12.031028} {\bibfield  {journal} {\bibinfo  {journal} {Phys. Rev. X}\ }\textbf {\bibinfo {volume} {12}},\ \bibinfo {pages} {031028} (\bibinfo {year} {2022})}\BibitemShut {NoStop}%
\bibitem [{\citenamefont {Zhong}\ \emph {et~al.}(2015)\citenamefont {Zhong}, \citenamefont {Kindem}, \citenamefont {Miyazono},\ and\ \citenamefont {Faraon}}]{zhong2015nanophotonic}%
  \BibitemOpen
  \bibfield  {author} {\bibinfo {author} {\bibfnamefont {T.}~\bibnamefont {Zhong}}, \bibinfo {author} {\bibfnamefont {J.~M.}\ \bibnamefont {Kindem}}, \bibinfo {author} {\bibfnamefont {E.}~\bibnamefont {Miyazono}},\ and\ \bibinfo {author} {\bibfnamefont {A.}~\bibnamefont {Faraon}},\ }\bibfield  {title} {\bibinfo {title} {Nanophotonic coherent light--matter interfaces based on rare-earth-doped crystals},\ }\href {https://www.nature.com/articles/ncomms9206} {\bibfield  {journal} {\bibinfo  {journal} {Nature communications}\ }\textbf {\bibinfo {volume} {6}},\ \bibinfo {pages} {8206} (\bibinfo {year} {2015})}\BibitemShut {NoStop}%
\bibitem [{\citenamefont {Moerner}\ and\ \citenamefont {Kador}(1989)}]{PhysRevLett.62.2535}%
  \BibitemOpen
  \bibfield  {author} {\bibinfo {author} {\bibfnamefont {W.~E.}\ \bibnamefont {Moerner}}\ and\ \bibinfo {author} {\bibfnamefont {L.}~\bibnamefont {Kador}},\ }\bibfield  {title} {\bibinfo {title} {Optical detection and spectroscopy of single molecules in a solid},\ }\href {https://doi.org/10.1103/PhysRevLett.62.2535} {\bibfield  {journal} {\bibinfo  {journal} {Phys. Rev. Lett.}\ }\textbf {\bibinfo {volume} {62}},\ \bibinfo {pages} {2535} (\bibinfo {year} {1989})}\BibitemShut {NoStop}%
\bibitem [{\citenamefont {Arjona~Martínez}\ \emph {et~al.}(2022)\citenamefont {Arjona~Martínez}, \citenamefont {Parker}, \citenamefont {Chen}, \citenamefont {Purser}, \citenamefont {Li}, \citenamefont {Michaels}, \citenamefont {Stramma}, \citenamefont {Debroux}, \citenamefont {Harris}, \citenamefont {Hayhurst~Appel}, \citenamefont {Nichols}, \citenamefont {Trusheim}, \citenamefont {Gangloff}, \citenamefont {Englund},\ and\ \citenamefont {Atatüre}}]{arjona_martinez_photonic_2022}%
  \BibitemOpen
  \bibfield  {author} {\bibinfo {author} {\bibfnamefont {J.}~\bibnamefont {Arjona~Martínez}}, \bibinfo {author} {\bibfnamefont {R.~A.}\ \bibnamefont {Parker}}, \bibinfo {author} {\bibfnamefont {K.~C.}\ \bibnamefont {Chen}}, \bibinfo {author} {\bibfnamefont {C.~M.}\ \bibnamefont {Purser}}, \bibinfo {author} {\bibfnamefont {L.}~\bibnamefont {Li}}, \bibinfo {author} {\bibfnamefont {C.~P.}\ \bibnamefont {Michaels}}, \bibinfo {author} {\bibfnamefont {A.~M.}\ \bibnamefont {Stramma}}, \bibinfo {author} {\bibfnamefont {R.}~\bibnamefont {Debroux}}, \bibinfo {author} {\bibfnamefont {I.~B.}\ \bibnamefont {Harris}}, \bibinfo {author} {\bibfnamefont {M.}~\bibnamefont {Hayhurst~Appel}}, \bibinfo {author} {\bibfnamefont {E.~C.}\ \bibnamefont {Nichols}}, \bibinfo {author} {\bibfnamefont {M.~E.}\ \bibnamefont {Trusheim}}, \bibinfo {author} {\bibfnamefont {D.~A.}\ \bibnamefont {Gangloff}}, \bibinfo {author} {\bibfnamefont {D.}~\bibnamefont {Englund}},\ and\ \bibinfo {author} {\bibfnamefont {M.}~\bibnamefont {Atatüre}},\
  }\bibfield  {title} {\bibinfo {title} {Photonic {Indistinguishability} of the {Tin}-{Vacancy} {Center} in {Nanostructured} {Diamond}},\ }\href {https://doi.org/10.1103/PhysRevLett.129.173603} {\bibfield  {journal} {\bibinfo  {journal} {Physical Review Letters}\ }\textbf {\bibinfo {volume} {129}},\ \bibinfo {pages} {173603} (\bibinfo {year} {2022})}\BibitemShut {NoStop}%
\bibitem [{\citenamefont {Anger}\ \emph {et~al.}(2006)\citenamefont {Anger}, \citenamefont {Bharadwaj},\ and\ \citenamefont {Novotny}}]{anger_enhancement_2006}%
  \BibitemOpen
  \bibfield  {author} {\bibinfo {author} {\bibfnamefont {P.}~\bibnamefont {Anger}}, \bibinfo {author} {\bibfnamefont {P.}~\bibnamefont {Bharadwaj}},\ and\ \bibinfo {author} {\bibfnamefont {L.}~\bibnamefont {Novotny}},\ }\bibfield  {title} {\bibinfo {title} {Enhancement and {Quenching} of {Single}-{Molecule} {Fluorescence}},\ }\href {https://doi.org/10.1103/PhysRevLett.96.113002} {\bibfield  {journal} {\bibinfo  {journal} {Physical Review Letters}\ }\textbf {\bibinfo {volume} {96}},\ \bibinfo {pages} {113002} (\bibinfo {year} {2006})}\BibitemShut {NoStop}%
\bibitem [{\citenamefont {Kühn}\ \emph {et~al.}(2006)\citenamefont {Kühn}, \citenamefont {Håkanson}, \citenamefont {Rogobete},\ and\ \citenamefont {Sandoghdar}}]{kuhn_enhancement_2006}%
  \BibitemOpen
  \bibfield  {author} {\bibinfo {author} {\bibfnamefont {S.}~\bibnamefont {Kühn}}, \bibinfo {author} {\bibfnamefont {U.}~\bibnamefont {Håkanson}}, \bibinfo {author} {\bibfnamefont {L.}~\bibnamefont {Rogobete}},\ and\ \bibinfo {author} {\bibfnamefont {V.}~\bibnamefont {Sandoghdar}},\ }\bibfield  {title} {\bibinfo {title} {Enhancement of {Single}-{Molecule} {Fluorescence} {Using} a {Gold} {Nanoparticle} as an {Optical} {Nanoantenna}},\ }\href {https://doi.org/10.1103/PhysRevLett.97.017402} {\bibfield  {journal} {\bibinfo  {journal} {Physical Review Letters}\ }\textbf {\bibinfo {volume} {97}},\ \bibinfo {pages} {017402} (\bibinfo {year} {2006})}\BibitemShut {NoStop}%
\bibitem [{\citenamefont {Chen}\ \emph {et~al.}(2012)\citenamefont {Chen}, \citenamefont {Agio},\ and\ \citenamefont {Sandoghdar}}]{chen_metallodielectric_2012}%
  \BibitemOpen
  \bibfield  {author} {\bibinfo {author} {\bibfnamefont {X.-W.}\ \bibnamefont {Chen}}, \bibinfo {author} {\bibfnamefont {M.}~\bibnamefont {Agio}},\ and\ \bibinfo {author} {\bibfnamefont {V.}~\bibnamefont {Sandoghdar}},\ }\bibfield  {title} {{\selectlanguage {en}\bibinfo {title} {Metallodielectric {Hybrid} {Antennas} for {Ultrastrong} {Enhancement} of {Spontaneous} {Emission}}},\ }\href {https://doi.org/10.1103/PhysRevLett.108.233001} {\bibfield  {journal} {\bibinfo  {journal} {Physical Review Letters}\ }\textbf {\bibinfo {volume} {108}},\ \bibinfo {pages} {233001} (\bibinfo {year} {2012})}\BibitemShut {NoStop}%
\bibitem [{\citenamefont {Walls}\ and\ \citenamefont {Milburn}(2012)}]{walls2012quantum}%
  \BibitemOpen
  \bibfield  {author} {\bibinfo {author} {\bibfnamefont {D.}~\bibnamefont {Walls}}\ and\ \bibinfo {author} {\bibfnamefont {G.}~\bibnamefont {Milburn}},\ }\href {https://books.google.com.au/books?id=o6nrCAAAQBAJ} {\emph {\bibinfo {title} {Quantum Optics}}},\ Springer Study Edition\ (\bibinfo  {publisher} {Springer Berlin Heidelberg},\ \bibinfo {year} {2012})\BibitemShut {NoStop}%
\bibitem [{\citenamefont {Johansson}\ \emph {et~al.}(2013)\citenamefont {Johansson}, \citenamefont {Nation},\ and\ \citenamefont {Nori}}]{johansson_qutip_2013}%
  \BibitemOpen
  \bibfield  {author} {\bibinfo {author} {\bibfnamefont {J.~R.}\ \bibnamefont {Johansson}}, \bibinfo {author} {\bibfnamefont {P.~D.}\ \bibnamefont {Nation}},\ and\ \bibinfo {author} {\bibfnamefont {F.}~\bibnamefont {Nori}},\ }\bibfield  {title} {\bibinfo {title} {{QuTiP} 2: {A} {Python} framework for the dynamics of open quantum systems},\ }\href {https://doi.org/10.1016/j.cpc.2012.11.019} {\bibfield  {journal} {\bibinfo  {journal} {Computer Physics Communications}\ }\textbf {\bibinfo {volume} {184}},\ \bibinfo {pages} {1234} (\bibinfo {year} {2013})}\BibitemShut {NoStop}%
\bibitem [{\citenamefont {Johansson}\ \emph {et~al.}(2012)\citenamefont {Johansson}, \citenamefont {Nation},\ and\ \citenamefont {Nori}}]{johansson_qutip_2012}%
  \BibitemOpen
  \bibfield  {author} {\bibinfo {author} {\bibfnamefont {J.~R.}\ \bibnamefont {Johansson}}, \bibinfo {author} {\bibfnamefont {P.~D.}\ \bibnamefont {Nation}},\ and\ \bibinfo {author} {\bibfnamefont {F.}~\bibnamefont {Nori}},\ }\bibfield  {title} {\bibinfo {title} {{QuTiP}: {An} open-source {Python} framework for the dynamics of open quantum systems},\ }\href {https://doi.org/10.1016/j.cpc.2012.02.021} {\bibfield  {journal} {\bibinfo  {journal} {Computer Physics Communications}\ }\textbf {\bibinfo {volume} {183}},\ \bibinfo {pages} {1760} (\bibinfo {year} {2012})}\BibitemShut {NoStop}%
\bibitem [{\citenamefont {Neuman}\ \emph {et~al.}(2019)\citenamefont {Neuman}, \citenamefont {Esteban}, \citenamefont {Giedke}, \citenamefont {Schmidt},\ and\ \citenamefont {Aizpurua}}]{neuman_quantum_2019}%
  \BibitemOpen
  \bibfield  {author} {\bibinfo {author} {\bibfnamefont {T.}~\bibnamefont {Neuman}}, \bibinfo {author} {\bibfnamefont {R.}~\bibnamefont {Esteban}}, \bibinfo {author} {\bibfnamefont {G.}~\bibnamefont {Giedke}}, \bibinfo {author} {\bibfnamefont {M.~K.}\ \bibnamefont {Schmidt}},\ and\ \bibinfo {author} {\bibfnamefont {J.}~\bibnamefont {Aizpurua}},\ }\bibfield  {title} {{\selectlanguage {en}\bibinfo {title} {Quantum description of surface-enhanced resonant {Raman} scattering within a hybrid-optomechanical model}},\ }\href {https://doi.org/10.1103/PhysRevA.100.043422} {\bibfield  {journal} {\bibinfo  {journal} {Physical Review A}\ }\textbf {\bibinfo {volume} {100}},\ \bibinfo {pages} {043422} (\bibinfo {year} {2019})}\BibitemShut {NoStop}%
\bibitem [{\citenamefont {Wilson-Rae}\ \emph {et~al.}(2004)\citenamefont {Wilson-Rae}, \citenamefont {Zoller},\ and\ \citenamefont {Imamo\ifmmode~\bar{g}\else \={g}\fi{}lu}}]{PhysRevLett.92.075507}%
  \BibitemOpen
  \bibfield  {author} {\bibinfo {author} {\bibfnamefont {I.}~\bibnamefont {Wilson-Rae}}, \bibinfo {author} {\bibfnamefont {P.}~\bibnamefont {Zoller}},\ and\ \bibinfo {author} {\bibfnamefont {A.}~\bibnamefont {Imamo\ifmmode~\bar{g}\else \={g}\fi{}lu}},\ }\bibfield  {title} {\bibinfo {title} {Laser cooling of a nanomechanical resonator mode to its quantum ground state},\ }\href {https://doi.org/10.1103/PhysRevLett.92.075507} {\bibfield  {journal} {\bibinfo  {journal} {Phys. Rev. Lett.}\ }\textbf {\bibinfo {volume} {92}},\ \bibinfo {pages} {075507} (\bibinfo {year} {2004})}\BibitemShut {NoStop}%
\bibitem [{\citenamefont {Yeo}\ \emph {et~al.}(2014)\citenamefont {Yeo}, \citenamefont {De~Assis}, \citenamefont {Gloppe}, \citenamefont {Dupont-Ferrier}, \citenamefont {Verlot}, \citenamefont {Malik}, \citenamefont {Dupuy}, \citenamefont {Claudon}, \citenamefont {G{\'e}rard}, \citenamefont {Auff{\`e}ves} \emph {et~al.}}]{yeo2014strain}%
  \BibitemOpen
  \bibfield  {author} {\bibinfo {author} {\bibfnamefont {I.}~\bibnamefont {Yeo}}, \bibinfo {author} {\bibfnamefont {P.-L.}\ \bibnamefont {De~Assis}}, \bibinfo {author} {\bibfnamefont {A.}~\bibnamefont {Gloppe}}, \bibinfo {author} {\bibfnamefont {E.}~\bibnamefont {Dupont-Ferrier}}, \bibinfo {author} {\bibfnamefont {P.}~\bibnamefont {Verlot}}, \bibinfo {author} {\bibfnamefont {N.~S.}\ \bibnamefont {Malik}}, \bibinfo {author} {\bibfnamefont {E.}~\bibnamefont {Dupuy}}, \bibinfo {author} {\bibfnamefont {J.}~\bibnamefont {Claudon}}, \bibinfo {author} {\bibfnamefont {J.-M.}\ \bibnamefont {G{\'e}rard}}, \bibinfo {author} {\bibfnamefont {A.}~\bibnamefont {Auff{\`e}ves}}, \emph {et~al.},\ }\bibfield  {title} {\bibinfo {title} {Strain-mediated coupling in a quantum dot--mechanical oscillator hybrid system},\ }\href@noop {} {\bibfield  {journal} {\bibinfo  {journal} {Nature nanotechnology}\ }\textbf {\bibinfo {volume} {9}},\ \bibinfo {pages} {106} (\bibinfo {year} {2014})}\BibitemShut {NoStop}%
\bibitem [{\citenamefont {LaHaye}\ \emph {et~al.}(2009)\citenamefont {LaHaye}, \citenamefont {Suh}, \citenamefont {Echternach}, \citenamefont {Schwab},\ and\ \citenamefont {Roukes}}]{lahaye2009nanomechanical}%
  \BibitemOpen
  \bibfield  {author} {\bibinfo {author} {\bibfnamefont {M.}~\bibnamefont {LaHaye}}, \bibinfo {author} {\bibfnamefont {J.}~\bibnamefont {Suh}}, \bibinfo {author} {\bibfnamefont {P.}~\bibnamefont {Echternach}}, \bibinfo {author} {\bibfnamefont {K.~C.}\ \bibnamefont {Schwab}},\ and\ \bibinfo {author} {\bibfnamefont {M.~L.}\ \bibnamefont {Roukes}},\ }\bibfield  {title} {\bibinfo {title} {Nanomechanical measurements of a superconducting qubit},\ }\href@noop {} {\bibfield  {journal} {\bibinfo  {journal} {Nature}\ }\textbf {\bibinfo {volume} {459}},\ \bibinfo {pages} {960} (\bibinfo {year} {2009})}\BibitemShut {NoStop}%
\bibitem [{\citenamefont {Nie}\ and\ \citenamefont {Emory}(1997)}]{nie_probing_1997}%
  \BibitemOpen
  \bibfield  {author} {\bibinfo {author} {\bibfnamefont {S.}~\bibnamefont {Nie}}\ and\ \bibinfo {author} {\bibfnamefont {S.~R.}\ \bibnamefont {Emory}},\ }\bibfield  {title} {\bibinfo {title} {Probing {Single} {Molecules} and {Single} {Nanoparticles} by {Surface}-{Enhanced} {Raman} {Scattering}},\ }\href {https://doi.org/10.1126/science.275.5303.1102} {\bibfield  {journal} {\bibinfo  {journal} {Science}\ }\textbf {\bibinfo {volume} {275}},\ \bibinfo {pages} {1102} (\bibinfo {year} {1997})},\ \bibinfo {note} {publisher: American Association for the Advancement of Science}\BibitemShut {NoStop}%
\bibitem [{\citenamefont {Langer}\ \emph {et~al.}(2020)\citenamefont {Langer}, \citenamefont {Jimenez~de Aberasturi}, \citenamefont {Aizpurua}, \citenamefont {Alvarez-Puebla}, \citenamefont {Auguié}, \citenamefont {Baumberg}, \citenamefont {Bazan}, \citenamefont {Bell}, \citenamefont {Boisen}, \citenamefont {Brolo}, \citenamefont {Choo}, \citenamefont {Cialla-May}, \citenamefont {Deckert}, \citenamefont {Fabris}, \citenamefont {Faulds}, \citenamefont {García~de Abajo}, \citenamefont {Goodacre}, \citenamefont {Graham}, \citenamefont {Haes}, \citenamefont {Haynes}, \citenamefont {Huck}, \citenamefont {Itoh}, \citenamefont {Käll}, \citenamefont {Kneipp}, \citenamefont {Kotov}, \citenamefont {Kuang}, \citenamefont {Le~Ru}, \citenamefont {Lee}, \citenamefont {Li}, \citenamefont {Ling}, \citenamefont {Maier}, \citenamefont {Mayerhöfer}, \citenamefont {Moskovits}, \citenamefont {Murakoshi}, \citenamefont {Nam}, \citenamefont {Nie}, \citenamefont {Ozaki}, \citenamefont {Pastoriza-Santos}, \citenamefont
  {Perez-Juste}, \citenamefont {Popp}, \citenamefont {Pucci}, \citenamefont {Reich}, \citenamefont {Ren}, \citenamefont {Schatz}, \citenamefont {Shegai}, \citenamefont {Schlücker}, \citenamefont {Tay}, \citenamefont {Thomas}, \citenamefont {Tian}, \citenamefont {Van~Duyne}, \citenamefont {Vo-Dinh}, \citenamefont {Wang}, \citenamefont {Willets}, \citenamefont {Xu}, \citenamefont {Xu}, \citenamefont {Xu}, \citenamefont {Yamamoto}, \citenamefont {Zhao},\ and\ \citenamefont {Liz-Marzán}}]{langer_present_2020}%
  \BibitemOpen
  \bibfield  {author} {\bibinfo {author} {\bibfnamefont {J.}~\bibnamefont {Langer}}, \bibinfo {author} {\bibfnamefont {D.}~\bibnamefont {Jimenez~de Aberasturi}}, \bibinfo {author} {\bibfnamefont {J.}~\bibnamefont {Aizpurua}}, \bibinfo {author} {\bibfnamefont {R.~A.}\ \bibnamefont {Alvarez-Puebla}}, \bibinfo {author} {\bibfnamefont {B.}~\bibnamefont {Auguié}}, \bibinfo {author} {\bibfnamefont {J.~J.}\ \bibnamefont {Baumberg}}, \bibinfo {author} {\bibfnamefont {G.~C.}\ \bibnamefont {Bazan}}, \bibinfo {author} {\bibfnamefont {S.~E.~J.}\ \bibnamefont {Bell}}, \bibinfo {author} {\bibfnamefont {A.}~\bibnamefont {Boisen}}, \bibinfo {author} {\bibfnamefont {A.~G.}\ \bibnamefont {Brolo}}, \bibinfo {author} {\bibfnamefont {J.}~\bibnamefont {Choo}}, \bibinfo {author} {\bibfnamefont {D.}~\bibnamefont {Cialla-May}}, \bibinfo {author} {\bibfnamefont {V.}~\bibnamefont {Deckert}}, \bibinfo {author} {\bibfnamefont {L.}~\bibnamefont {Fabris}}, \bibinfo {author} {\bibfnamefont {K.}~\bibnamefont {Faulds}}, \bibinfo {author}
  {\bibfnamefont {F.~J.}\ \bibnamefont {García~de Abajo}}, \bibinfo {author} {\bibfnamefont {R.}~\bibnamefont {Goodacre}}, \bibinfo {author} {\bibfnamefont {D.}~\bibnamefont {Graham}}, \bibinfo {author} {\bibfnamefont {A.~J.}\ \bibnamefont {Haes}}, \bibinfo {author} {\bibfnamefont {C.~L.}\ \bibnamefont {Haynes}}, \bibinfo {author} {\bibfnamefont {C.}~\bibnamefont {Huck}}, \bibinfo {author} {\bibfnamefont {T.}~\bibnamefont {Itoh}}, \bibinfo {author} {\bibfnamefont {M.}~\bibnamefont {Käll}}, \bibinfo {author} {\bibfnamefont {J.}~\bibnamefont {Kneipp}}, \bibinfo {author} {\bibfnamefont {N.~A.}\ \bibnamefont {Kotov}}, \bibinfo {author} {\bibfnamefont {H.}~\bibnamefont {Kuang}}, \bibinfo {author} {\bibfnamefont {E.~C.}\ \bibnamefont {Le~Ru}}, \bibinfo {author} {\bibfnamefont {H.~K.}\ \bibnamefont {Lee}}, \bibinfo {author} {\bibfnamefont {J.-F.}\ \bibnamefont {Li}}, \bibinfo {author} {\bibfnamefont {X.~Y.}\ \bibnamefont {Ling}}, \bibinfo {author} {\bibfnamefont {S.~A.}\ \bibnamefont {Maier}}, \bibinfo {author}
  {\bibfnamefont {T.}~\bibnamefont {Mayerhöfer}}, \bibinfo {author} {\bibfnamefont {M.}~\bibnamefont {Moskovits}}, \bibinfo {author} {\bibfnamefont {K.}~\bibnamefont {Murakoshi}}, \bibinfo {author} {\bibfnamefont {J.-M.}\ \bibnamefont {Nam}}, \bibinfo {author} {\bibfnamefont {S.}~\bibnamefont {Nie}}, \bibinfo {author} {\bibfnamefont {Y.}~\bibnamefont {Ozaki}}, \bibinfo {author} {\bibfnamefont {I.}~\bibnamefont {Pastoriza-Santos}}, \bibinfo {author} {\bibfnamefont {J.}~\bibnamefont {Perez-Juste}}, \bibinfo {author} {\bibfnamefont {J.}~\bibnamefont {Popp}}, \bibinfo {author} {\bibfnamefont {A.}~\bibnamefont {Pucci}}, \bibinfo {author} {\bibfnamefont {S.}~\bibnamefont {Reich}}, \bibinfo {author} {\bibfnamefont {B.}~\bibnamefont {Ren}}, \bibinfo {author} {\bibfnamefont {G.~C.}\ \bibnamefont {Schatz}}, \bibinfo {author} {\bibfnamefont {T.}~\bibnamefont {Shegai}}, \bibinfo {author} {\bibfnamefont {S.}~\bibnamefont {Schlücker}}, \bibinfo {author} {\bibfnamefont {L.-L.}\ \bibnamefont {Tay}}, \bibinfo {author}
  {\bibfnamefont {K.~G.}\ \bibnamefont {Thomas}}, \bibinfo {author} {\bibfnamefont {Z.-Q.}\ \bibnamefont {Tian}}, \bibinfo {author} {\bibfnamefont {R.~P.}\ \bibnamefont {Van~Duyne}}, \bibinfo {author} {\bibfnamefont {T.}~\bibnamefont {Vo-Dinh}}, \bibinfo {author} {\bibfnamefont {Y.}~\bibnamefont {Wang}}, \bibinfo {author} {\bibfnamefont {K.~A.}\ \bibnamefont {Willets}}, \bibinfo {author} {\bibfnamefont {C.}~\bibnamefont {Xu}}, \bibinfo {author} {\bibfnamefont {H.}~\bibnamefont {Xu}}, \bibinfo {author} {\bibfnamefont {Y.}~\bibnamefont {Xu}}, \bibinfo {author} {\bibfnamefont {Y.~S.}\ \bibnamefont {Yamamoto}}, \bibinfo {author} {\bibfnamefont {B.}~\bibnamefont {Zhao}},\ and\ \bibinfo {author} {\bibfnamefont {L.~M.}\ \bibnamefont {Liz-Marzán}},\ }\bibfield  {title} {\bibinfo {title} {Present and {Future} of {Surface}-{Enhanced} {Raman} {Scattering}},\ }\href {https://doi.org/10.1021/acsnano.9b04224} {\bibfield  {journal} {\bibinfo  {journal} {ACS Nano}\ }\textbf {\bibinfo {volume} {14}},\ \bibinfo {pages} {28}
  (\bibinfo {year} {2020})}\BibitemShut {NoStop}%
\bibitem [{\citenamefont {Berweger}\ and\ \citenamefont {Raschke}(2010)}]{berweger_signal_2010}%
  \BibitemOpen
  \bibfield  {author} {\bibinfo {author} {\bibfnamefont {S.}~\bibnamefont {Berweger}}\ and\ \bibinfo {author} {\bibfnamefont {M.~B.}\ \bibnamefont {Raschke}},\ }\bibfield  {title} {{\selectlanguage {en}\bibinfo {title} {Signal limitations in tip-enhanced {Raman} scattering: the challenge to become a routine analytical technique}},\ }\href {https://doi.org/10.1007/s00216-009-3085-1} {\bibfield  {journal} {\bibinfo  {journal} {Analytical and Bioanalytical Chemistry}\ }\textbf {\bibinfo {volume} {396}},\ \bibinfo {pages} {115} (\bibinfo {year} {2010})}\BibitemShut {NoStop}%
\bibitem [{Note1()}]{Note1}%
  \BibitemOpen
  \bibinfo {note} {Once could imagine a different scenario, where the pump was tuned to match a particular Stokes or anti-Stokes line to the transition in GeV. However, since the linewidth of the GeV transition is far smaller than the typical decay rate of a molecular vibration, the former could not be treated as a bath, complicating the analytical treatment.}\BibitemShut {Stop}%
\bibitem [{\citenamefont {Johnson}\ and\ \citenamefont {Christy}(1972)}]{christy}%
  \BibitemOpen
  \bibfield  {author} {\bibinfo {author} {\bibfnamefont {P.~B.}\ \bibnamefont {Johnson}}\ and\ \bibinfo {author} {\bibfnamefont {R.~W.}\ \bibnamefont {Christy}},\ }\bibfield  {title} {\bibinfo {title} {Optical constants of the noble metals},\ }\href {https://doi.org/10.1103/PhysRevB.6.4370} {\bibfield  {journal} {\bibinfo  {journal} {Phys. Rev. B}\ }\textbf {\bibinfo {volume} {6}},\ \bibinfo {pages} {4370} (\bibinfo {year} {1972})}\BibitemShut {NoStop}%
\end{thebibliography}%
\clearpage
\newpage

\appendix

\begin{widetext}

\section{Derivation of the interaction Hamiltonian }\label{app:derivation}

{

\subsection{Complete Hamiltonian of the system}\label{subsec:complete.Hamiltonian}

\begin{figure}[ht!]
	\centering
	\includegraphics[width=.4\linewidth]{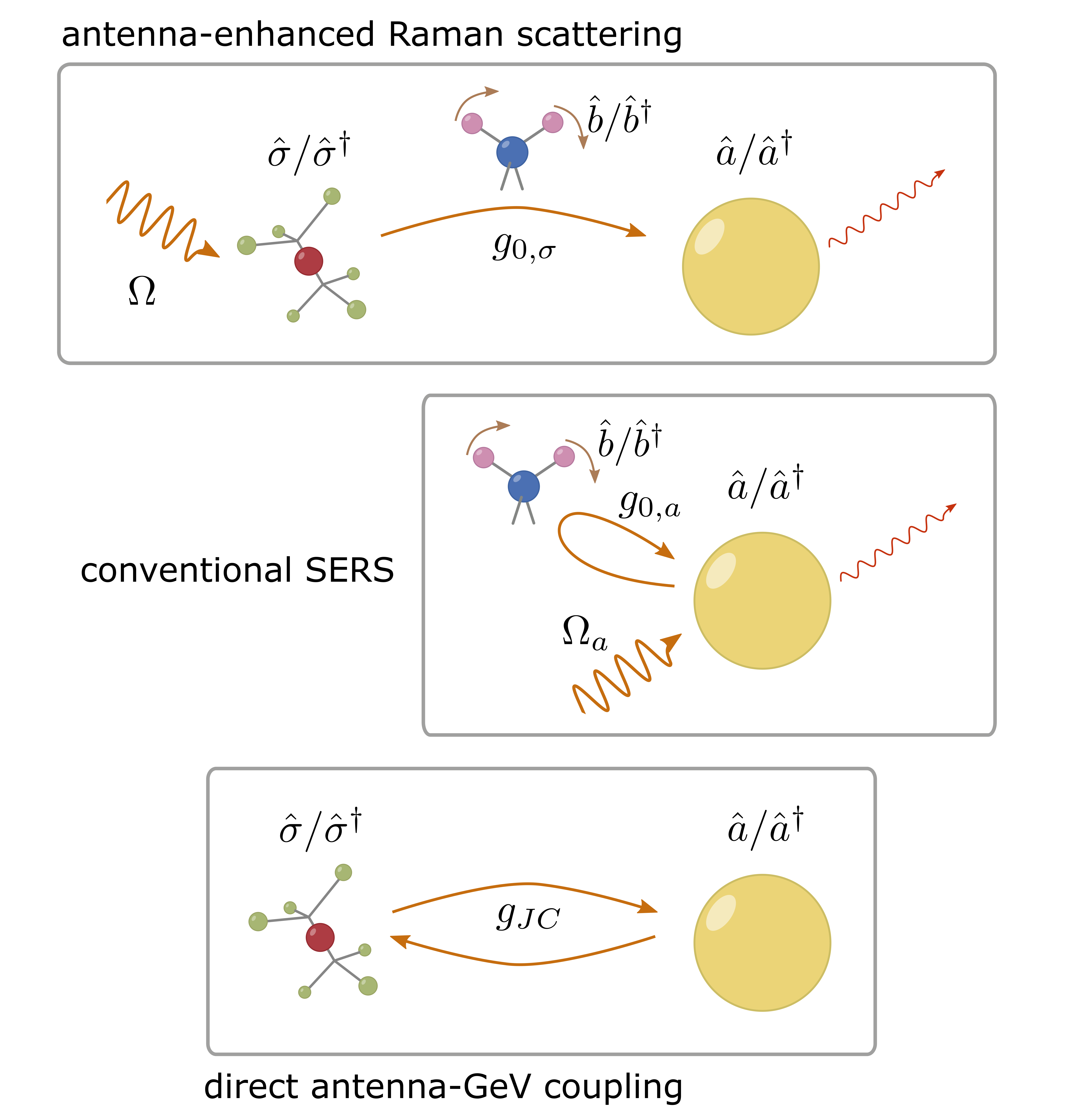}
	\caption{{Schematics for the effects discussed in Section~\ref{subsec:complete.Hamiltonian}.}}
	\label{fig:schematic}
\end{figure}

In Fig.~\ref{fig:schematic} we represent all the coherent processes that govern the dynamics of the system comprised of a single-mode plasmonic cavity, an atomic antenna modeled as a two-level system --- both exposed to a coherent illumination by a laser --- and a Raman-active molecule.
\begin{enumerate}
\item coherent driving of the atomic antenna
\begin{equation}
    \hat{H}_\text{coh}= \hbar\Omega\left(e^{-i\omega_lt} \hat{\sigma} + e^{i\omega_lt} \hat{\sigma}^\dag\right),
\end{equation}
and of the optical cavity mode 
\begin{equation}
    \hat{H}_{\text{coh},a}= \hbar\Omega_a\left(e^{-i\omega_lt} \hat{a} + e^{i\omega_lt} \hat{a}^\dag\right);
\end{equation}
The two amplitudes $\Omega$ and $\Omega_a$ can be derived from the laser intensity $I$ and the system parameters, as discussed in Appendix~\ref{app:conventional} and Section \ref{subsec:coherent}; throughout the manuscript, we assume laser to be tuned to the dominant transition in GeV,
\item the direct coupling between the optical cavity and the atomic antenna $\hat{H}_{a-\text{GeV}}$
\begin{align}\label{eq:HJC}
    \hat{H}_{a-\text{GeV}} = - \hat{\mathbf{E}}_{a}(\mathbf{r}_\sigma) \cdot \mathbf{d}_\sigma \approx \hbar g_{JC}(\hat{a} \hat{\sigma}^\dagger + \hat{a}^\dagger \hat{\sigma}),
\end{align}
where we performed the rotating wave approximation. 

In Section~\ref{sec:formulation}, we assume that this resonant coupling will result solely in the enhancement of the emission rate from the molecule via the Purcell effect. The magnitude of this effect for a system with parameters described Table~\ref{tab:estimates} can be calculated numerically. From the numerical simulations, the maximum Purcell effect from the plasmonic antenna (see the discussion in Section~\ref{subsec:modeling} and Section~\ref{subsec:modelling.comsol}) can be estimated as $F_P\approx 30$. If we assume that this effect is due to the coupling with a detuned single optical mode, $F_P$ should be given as 
\begin{equation}
    F_P = \frac{4g_{JC}^2\kappa}{[\kappa^2 + (\omega_\sigma - \omega_a)^2]\gamma},
\end{equation}
giving $g_{JC}/2\pi \approx 0.38$~THz. For larger spacing, this would drop to $0.1$~THz. 

These estimates place the effect deep in the weak coupling regime. If we nevertheless wanted to diagonalise the system made up of the optical cavity and atomic antenna, the mixing angle would be $g_{JC}/(\omega_\sigma - \omega_a)$ would be of the order of 0.01 (assuming detunings as discussed in the main text), justifying the perturbative treatment of the optomechanical coupling.


\item interaction between the Raman dipole and the electric fields of the optical cavity, and the atomic antenna $\hat{H}_{I}$  
\begin{align}
    \hat{H}_{I} = -\frac{1}{2}\hat{\mathbf{p}}_R \cdot \left[\hat{\mathbf{E}}_a(\mathbf{r}_m)+ \hat{\mathbf{E}}_\sigma(\mathbf{r}_m)\right].
\end{align}
The Raman dipole $\hat{\mathbf{p}}_R$ is induced by both the electric fields of the optical cavity and the atomic antenna oscillating at the pump frequency $\omega_l$, as $\hat{\mathbf{p}}_R =  R:\left[\hat{\mathbf{E}}_a(\mathbf{r}_m)+\hat{\mathbf{E}}_\sigma(\mathbf{r}_m)\right]$, and the above interaction Hamiltonian can be expanded as
\begin{align}
    \hat{H}_{I} = -\frac{1}{2}\hat{x}\bigg[&(R:\hat{\mathbf{E}}_a) \cdot \hat{\mathbf{E}}_a \nonumber \\
    & + (R:\hat{\mathbf{E}}_a) \cdot \hat{\mathbf{E}}_\sigma\nonumber \\
    & + (R:\hat{\mathbf{E}}_\sigma) \cdot \hat{\mathbf{E}}_a\nonumber \\
    &\left. + (R:\hat{\mathbf{E}}_\sigma) \cdot \hat{\mathbf{E}}_\sigma\right].
\end{align}
The first term describes the conventional optomechanical coupling, and processes where the cavity photons drive the Raman processes in the molecule, and the scattered Stokes and anti-Stokes photons couple into the cavity. This mechanism, together with the coherent pumping term $H_{\text{coh},a}$, will be referred to as the \textit{conventional plasmonic SERS}. Parameters of this setup are discussed in Section~\ref{app:conventional}.

The second line describes a similar processes, but with the Raman emission coupled to the atomic antenna, instead of the optical cavity mode. Since the pump laser is tuned to the dominant transition in GeV at 602 nm, and linewidth of this transition is several orders of magnitude smaller than the molecular vibration, this process cannot be frequency-matched. \footnote{Once could imagine a different scenario, where the pump was tuned to match a particular Stokes or anti-Stokes line to the transition in GeV. However, since the linewidth of the GeV transition is far smaller than the typical decay rate of a molecular vibration, the former could not be treated as a bath, complicating the analytical treatment.}  

The third line describes the effect denoted schematically in Fig.~\ref{fig:1}, where the incident illumination is resonantly scattered on the atomic antenna, drives the Raman scattering on the molecule, and the Raman photons couple to the spectrally broad optical cavity mode. We focus on this mechanism throughout this work.

Finally, in the last line we consider an extremely weak effect, where the Raman dipole is induced by light resonantly scattered on the atomic antenna, and the Raman emission coupling back to the antenna. Since the linewidth of the transition in GeV is about 5 orders of magnitude smaller than the vibrational frequency, necessarily either one of these processes will be off-resonant and thus its contribution will be negligible.
\end{enumerate}

\subsection{Derivation of the linearized interaction Hamiltonian}

We now focus on the atomic antenna-enhanced Raman scattering.} 
The linearized optomechanical Hamiltonian reads
\begin{align}
    \hat{H}_{I,\text{lin}}= -&\frac{1}{2} R_{ij} Q^0 \left(\hat{b}+\hat{b}^\dag\right) \left[\omega_\sigma^2 \mu_0 (\mathbf{G}(\mathbf{r}_\sigma,\mathbf{r}_m,\omega_\sigma) \cdot \mathbf{d}_\sigma)_i \alpha_\sigma + \text{h.c.}\right] \left[(\mathbf{e}_a)_j \hat{a} + (\mathbf{e}_a)_j^* \hat{a}^\dag\right].
\end{align}
In the rotating wave approximation (RWA), we drop the co-rotating terms like $\hat{a} \alpha_\sigma$ and $\hat{a}^\dag \alpha_\sigma^*$, and write 
\begin{align}
    \hat{H}_{I,\text{lin}} &= -\frac{1}{2} R_{ij} Q^0 \omega_\sigma^2 \mu_0  \left[\hat{a}^\dag\alpha_\sigma  (\mathbf{e}_a)_j (\mathbf{G}(\mathbf{r}_\sigma,\mathbf{r}_m,\omega_\sigma) \cdot \mathbf{d}_\sigma)^*_i + \text{h.c.}\right] \left(\hat{b}+\hat{b}^\dag\right).
\end{align}
Taking the simplest model for an isotropic Raman tensor $R_{ij}=R \delta_{ij}$ we find
\begin{align}
    \hat{H}_{I,\text{lin}} &= -\frac{1}{2} R Q^0 \omega_\sigma^2 \mu_0  \left[\hat{a}^\dag\alpha_\sigma  \mathbf{e}_a \cdot (\mathbf{G}(\mathbf{r}_\sigma,\mathbf{r}_m,\omega_\sigma) \cdot \mathbf{d}_\sigma)^* + \text{h.c.}\right] \left(\hat{b}+\hat{b}^\dag\right).
\end{align}
We can choose the phase of the incident laser, or $\alpha_\sigma$, in such a way that $\alpha_\sigma \mathbf{e}_a\cdot (\mathbf{G}\mathbf{d}_\sigma)^*$ becomes a real number, arriving at the interaction Hamiltonian given in Eq.~\eqref{eq:hamiltonian.1}.

\section{Power, flux and efficiency of Raman scattering}\label{app:powers}

In the main text, we introduced the expressions for the emission spectra as given by the Fourier transform of a two-time correlator, multiplied by $\omega^4$. This allowed us to trace the general trends of the emission intensities. However, to learn about the efficiencies of the Raman processes, here we carry out a more careful analysis.

We start by calculating the power carried via the Stokes emission. To this end, we adapt the expresssion for the total power emitted from a two-level system via a dipolar electric transition $\mathbf{d}_\sigma$~\cite{carmichael2013statistical}:
\begin{equation}
P = 2\varepsilon_0 c \int \text{d}\phi \int \text{d}\theta \sin(\theta) r^2 \int \text{d}\omega \frac{f(r,\theta,\omega)}{2\pi}  \int \text{d}\tau e^{i\omega \tau} \langle \hat{\sigma}^\dag(0) \hat{\sigma}(\tau)\rangle
\end{equation}
where
\begin{equation}
f(r,\theta, \omega) = \left(\frac{\omega^2 |\mathbf{d}_\sigma|}{4\pi \varepsilon_0 c^2} \frac{\sin(\theta)}{r}\right)^2,
\end{equation}
carries information about the scattering patern expressed in spherical coordinates $(r,\theta,\phi)$, assuming that the dipole is oriented along the $\hat{z}$ axis. With the plasmonic cavity mode emitting as an electric dipole $d_a = |\mathbf{d}_a|$, and described using operators $\hat{a}$, we rewrite these expressions as
\begin{equation}
P = 2\varepsilon_0 c \int \text{d}\phi \int \text{d}\theta \sin(\theta) r^2 \int \text{d}\omega \frac{f(r,\theta,\omega)}{2\pi} \int \text{d}\tau e^{i\omega \tau} \langle \hat{a}^\dag(0) \hat{a}(\tau)\rangle
\end{equation}
and
\begin{equation}
f(r,\theta, \omega) = \left(\frac{\omega^2 |{d}_a|}{4\pi \varepsilon_0 c^2} \frac{\sin(\theta)}{r}\right)^2.
\end{equation}

We now integrate the frequencies around the Stokes emission frequency $\omega_S$, using Eq.~\eqref{eq:spectrum.final}, to turn the last part of the expression for the power to
\begin{align}
&\int \text{d}\omega \frac{f(r,\theta,\omega)}{2\pi}  \int \text{d}\tau e^{i\omega \tau} \langle \hat{a}^\dag(0) \hat{a}(\tau)\rangle \nonumber \\
&\approx \frac{f(r,\theta,\omega_S)}{2\pi} \int \text{d}\omega \int \text{d}\tau e^{i\omega \tau} \langle \hat{a}^\dag(0) \hat{a}(\tau)\rangle \nonumber\\
&\approx f(r,\theta,\omega_S) 2 |g_\sigma|^2 \frac{n_b+1}{(\Delta_a+\omega_b)^2 + (\kappa/2)^2},
\end{align}
and the power of the Stokes emission, due to the atomic antenna-enhanced Raman scattering, to
\begin{align}
P_\sigma &= 4\varepsilon_0 c |g_\sigma|^2 \frac{n_b+1}{(\Delta_a+\omega_b)^2 + (\kappa/2)^2} \int \text{d}\theta \sin(\theta) r^2 f(r,\theta,\omega_S) \int \text{d}\phi  \nonumber \\
&= \frac{\omega_S^4 d_a^2}{4\pi^2 \varepsilon_0 c^3} |g_\sigma|^2 \frac{n_b+1}{(\Delta_a+\omega_b)^2 + (\kappa/2)^2} \int \text{d}\theta \sin^3(\theta) \int \text{d}\phi.
\end{align}
The flux of Stokes photons is calculated as $P_\sigma/(\hbar \omega_S)$. 

We can compare it to the flux of pump photons $I \sigma_\text{spot}/(\hbar \omega_l)$, where $\sigma_\text{spot}$ denotes the cross-section of illuminating beam, and identify the efficiency of down-conversion as
\begin{align}
\eta_\sigma &= \frac{P_\sigma/(\hbar \omega_S)}{I \sigma_\text{spot}/(\hbar \omega_l)} = \frac{\omega_l\omega_S^3 d_a^2}{4\pi^2 \varepsilon_0 c^3} \frac{|g_\sigma|^2}{I \sigma_\text{spot}} \frac{n_b+1}{(\Delta_a+\omega_b)^2 + (\kappa/2)^2}  \int \text{d}\theta \sin^3(\theta) \int \text{d}\phi.
\end{align}

If the vibrational populations remain unchanged (as is the case of an atomic antenna-enhanced Raman scattering, or SERS below the vibrational pumping regime), this efficiency can be rewritten as
\begin{equation}
\eta_\sigma = \eta' \frac{g_{0,\sigma}^2 |\langle \hat{\sigma} \rangle|^2}{I},
\end{equation}
where 
\begin{align}\label{eq:etaprime}
\eta' = &\frac{\omega_l \omega_S^3 d_a^2}{4\pi^2 \varepsilon_0 c^3} \frac{1}{\sigma_\text{spot}} \frac{n_b+1}{(\Delta_a+\omega_b)^2 + (\kappa/2)^2} \int \text{d}\theta \sin^3(\theta) \int \text{d}\phi.
\end{align}
To find the value of this parameter, we use the parameters from Table~\ref{tab:estimates}, and set $\sigma_\text{spot} = 10^{-12}~\text{m}^2$. The dipolar moment $d_a$ is estimated by equating the coherent driving amplitude $\hbar \Omega$ given in Eq.~\eqref{eq:plasmon.driving} with the expression for the coupling between incident laser field $\mathbf{E}_\text{inc}$ with a dipolar scatterer $\mathbf{d}_a\cdot \mathbf{E}_\text{inc}$. We get can estimate $\eta' \approx 1.3\times 10^{-22}$~kg/s.

We can also derive an almost identical expression for the conventional plasmonic system, which will take the form
\begin{equation}
\eta_a = \eta' \frac{g_{0,a}^2 |\langle \hat{a} \rangle|^2}{I}.
\end{equation}

\section{Derivation of the emission spectrum}\label{app:spectrum}

{ 
In this section we derive the emission spectra from the systems under weak, and strong illumination.

\subsection{Raman emission in the weakly pumped, linearized regime}}

To calculate the two-time correlator in Eq.~\eqref{eq:emission.a}, we consider the full Hamiltonian of the system, including the linearized interaction Hamiltonian, and free-energy terms: 
\begin{equation}
    \hat{H} = \hbar\omega_b \hat{b}^\dag \hat{b} + \hbar\underbrace{(\omega_a-\omega_l)}_{\Delta_a}\hat{a}^\dag \hat{a} - \hbar g_\sigma \left(\hat{a} + \hat{a}^\dag\right) (\hat{b}+\hat{b}^\dag).
\end{equation}
The Langevin equation for $\hat{a}$ reads
\begin{equation}
    \frac{\textrm{d}}{\textrm{d}t} \hat{a}(t) = \left(-i\Delta_a -\frac{\kappa}{2}\right) \hat{a}(t) + ig_\sigma [\hat{b}(t)+\hat{b}^\dag(t)] + \sqrt{\kappa}\hat{a}_\text{in}(t),
\end{equation}
with the formal solution
\begin{align}
    \hat{a}(t) &= ig_\sigma\int_{-\infty}^t \textrm{d}s~e^{\left(-i\Delta_a -\frac{\kappa}{2}\right)(t-s)} [\hat{b}(s)+\hat{b}^\dag(s)] \nonumber \\
    & ~~~+ \underbrace{\int_{-\infty}^t \textrm{d}s~e^{\left(-i\Delta_a -\frac{\kappa}{2}\right)(t-s)} \sqrt{\kappa}\hat{a}_\text{in}(s)}_{\hat{A}_\text{in}(t)}.
\end{align}
Let us now remove the fast oscillation at $\omega_b$ from the mechanical operators by introducing $\hat{b}(s) = \hat{\tilde{b}}(s) e^{-i\omega_b s}$:
\begin{align}
    \hat{a}(t) &= ig_\sigma\int_{-\infty}^t \textrm{d}s~\left[e^{\left(-i\Delta_a -\frac{\kappa}{2}\right)(t-s)}e^{-i\omega_b s} \hat{\tilde{b}}(s) + e^{\left(-i\Delta_a -\frac{\kappa}{2}\right)(t-s)}e^{i\omega_b s} \hat{\tilde{b}}^\dag(s)\right] +\hat{A}_\text{in}(t)\nonumber \\
    &= ig_\sigma \left[e^{-i\omega_b t} \int_{-\infty}^t \textrm{d}s~e^{\left(-i\Delta_a + i\omega_b -\frac{\kappa}{2} \right)(t-s)} \tilde{b}(s) + e^{i\omega_b t} \int_{-\infty}^t \textrm{d}s~e^{\left(-i\Delta_a -i\omega_b -\frac{\kappa}{2}\right)(t-s)}\tilde{b}^\dag(s) \right]+\hat{A}_\text{in}(t)\nonumber \\
    &\approx ig_\sigma \left[\frac{\hat{b}(t)}{-i\Delta_a + i\omega_b -\frac{\kappa}{2}}  + \frac{\hat{b}^\dag(t)}{-i\Delta_a -i\omega_b -\frac{\kappa}{2}}\right]+\hat{A}_\text{in}(t).
\end{align}
Notably, the last step requires that the dynamics described by the integral kernel be much faster than the dynamics of $\hat{\tilde{b}}(s)$ (determined by the effective dissipation rate of the mechanical mode $\Gamma$), allowing us to approximate it as $\hat{\tilde{b}}(t)$. This \textit{Markov approximation} therefore requires that $\Gamma\ll \kappa$. Another way to think about this approximation is by comparing the width of the emission Stokes/anti-Stokes lines to the spectrum of the cavity into which the molecule emits. For $\Gamma/2\pi=1$~THz, and $\kappa/2\pi = 50$~THz, this condition holds trivially.

Within the framework of the Markov approximation, the two-time correlator takes the form
\begin{align}
    \braket{\hat{a}^\dag(0) \hat{a}(s)} \approx & |g_\sigma|^2 \left\langle\left[\frac{\hat{b}^\dag(0)}{i\Delta_a - i\omega_b -\frac{\kappa}{2}}  + \frac{\hat{b}(0)}{i\Delta_a +i\omega_b -\frac{\kappa}{2}}\right] \left[\frac{\hat{b}(s)}{-i\Delta_a + i\omega_b -\frac{\kappa}{2}}  + \frac{\hat{b}^\dag(s)}{-i\Delta_a -i\omega_b -\frac{\kappa}{2}}\right]\right \rangle\nonumber \\
    & + \langle \hat{A}_\text{in}^\dagger(0) \hat{A}_\text{in}(s)\rangle\nonumber \\
    \approx & |g_\sigma|^2 \left[\frac{\braket{\hat{b}^\dag(0)\hat{b}(s)}}{(\Delta_a - \omega_b)^2 +(\kappa/2)^2} + \frac{\braket{\hat{b}(0)\hat{b}^\dag(s)}}{(\Delta_a + \omega_b)^2 +(\kappa/2)^2}\right].
\end{align}
In the above transformations, we have assumed no correlation between the input optical noise and the mechanical mode, and vanishing two-time correlators of the optical noise. 

Let us now consider the simplified version of the description for the dynamics of the mechanical mode, where we assume that the vibrations are at thermal equilibrium with the environment. We therefore neglect the effects of Stokes and anti-Stokes emission which create and annihilate phonons. This effect can be easily incorporated into this picture by changing the steady-state phonon population, and effective decay rate $\Gamma$. 
Using the properties of the noise terms which describe the thermal environment with population $n_b^\text{th}$, that the two-time correlators are
\begin{align}\label{eq:nb}
    \braket{\hat{b}^\dag(0)\hat{b}(s)} = e^{\left(-i\omega_b - \Gamma/2\right)s} n_b^\text{th},
\end{align}
\begin{align}\label{eq:nb.2}
    \braket{\hat{b}(0)\hat{b}^\dag(s)} = e^{\left(i\omega_b - \Gamma/2\right)s} (n_b^\text{th}+1).
\end{align}
Since we have moved to the frame rotating with $\omega_l$, we need to account for it in the expression for the spectrum, which now is calculated in the lab frame. Additionally, swapping the integration $\int_{-\infty}^\infty \textrm{d}s\rightarrow 2\Re \int_0^\infty \textrm{d}s$, we find
\begin{align}\label{eq:spectrum}
    S_a(\omega) &=2\omega^4 \Re \int_0^\infty \textrm{d}s~|g_\sigma|^2\left[\frac{ e^{\left(-i\omega_b - \frac{\Gamma}{2}\right)s} n_b^\text{th}}{(\Delta_a - \omega_b)^2 +(\kappa/2)^2} + \frac{e^{\left(-i\omega_b - \frac{\Gamma}{2}\right)s} (n_b^\text{th}+1)}{(\Delta_a + \omega_b)^2 +(\kappa/2)^2}\right] e^{i(\omega-\omega_l) s} \nonumber \\ 
    &=2\omega^4 \left[\frac{\eta \omega^4 |g_\sigma|^2 n_b}{(\Delta_a - \omega_b)^2 +(\kappa/2)^2}\Re \left( \frac{1}{-i\omega_b - i\omega_l+ i\omega- \frac{\Gamma}{2}}\right) \right. \nonumber  \\
    &\quad\quad\quad\quad\left.+ 
    \frac{|g_\sigma|^2 (n_b+1)}{(\Delta_a + \omega_b)^2 +(\kappa/2)^2} \Re \left(\frac{1}{i\omega_b - i\omega_l +i\omega- \frac{\Gamma}{2}}\right)\right]\nonumber \\ 
    &=2\omega^4 |g_\sigma|^2 \left[\frac{n_b}{(\Delta_a - \omega_b)^2 +(\kappa/2)^2} \frac{\Gamma}{(\omega_b+\omega_l - \omega)^2+(\Gamma/2)^2}\right. \nonumber \\
    &\quad\quad\quad\quad\quad\quad\left.+
    \frac{n_b+1}{(\Delta_a + \omega_b)^2 +(\kappa/2)^2} \frac{\Gamma}{(\omega_b-\omega_l +\omega)^2+(\Gamma/2)^2}\right].
\end{align}

{
\subsection{Raman emission beyond the linearization regime}

In Section~\ref{sec:nonlinearities}, we discuss the emission spectra of the system under strong illumination --- in the Mollow regime --- where the coherence of the atomic antenna $\mean{\hat{\sigma}}$ rapidly decays with the increasing illumination intensity. Simultaneously, we observe in Fig.~\ref{fig:nonlinear}(c,d), that the intensity of Raman emission appears to saturate. This suggests that the Raman emission becomes decoupled from the atomic coherence, in a clear departure from the linearization picture discussed above, and in Section~\ref{subsec:linearisation}. 

To understand this behavior, we note the similarity of our system to that discussed by Neuman and co-authors\cite{neuman_quantum_2019}, to treat Surface-Enhanced Resonant Raman Scattering under strong illumination. Adopting the approach used in that work, we start with the interaction Hamiltonian (Eq.~\eqref{eq:interaction.hamiltonian}), split the operator $\hat{\sigma} \rightarrow \alpha_\sigma + \delta \hat{\sigma}$, and perform the rotating wave approximation to remove terms oscillating at twice the optical frequencies:
\begin{equation}
    \hat{H}_I \approx -\hbar g_{0,\sigma} (\alpha_\sigma + \delta \hat{\sigma})\hat{a}^\dagger (\hat{b} + \hat{b}^\dagger) + \text{h.c.}.
\end{equation}
In the absence of any correlations in the optical input noise $\hat{a}_\text{in}$, we can neglect that term in writing down the Heisenberg equation for the optical cavity operator $\hat{a}$:
\begin{equation}
    \frac{\text{d}}{\text{d}t}\hat{a}(t) = \left( - i\Delta_a-\frac{\kappa}{2}\right) \hat{a}(t) + ig_{0,\sigma} (\alpha_\sigma + \delta \hat{\sigma})(\hat{b} + \hat{b}^\dagger),
\end{equation}
with formal solution
\begin{align}
    \hat{a}(t) = &~ig_{0,\sigma} \int_{-\infty}^t e^{(-i\Delta_a-\kappa/2)(t-s)}\alpha_\sigma(s)[\hat{b}(s) + \hat{b}^\dagger(s)] \nonumber \\
    &+ ig_{0,\sigma} \int_{-\infty}^t e^{(-i\Delta_a-\kappa/2)(t-s)}\delta \hat{\sigma}(s)[\hat{b}(s) + \hat{b}^\dagger(s)].
\end{align}
In the Mollow regime, the first term can be made arbitrarily small by choosing sufficiently large laser intensity, and so we focus on the second line of the above expression. Using again operators $\hat{\tilde{b}}(s)$, which evolve at the rate far smaller than $\kappa$, we can perform the Markov approximation and find
\begin{align}
    \hat{a}(t) \approx &~ig_{0,\sigma} \left[ \frac{\delta \hat{\sigma}(t)\hat{b}(t)}{-i\Delta_a + i\omega_b-\kappa/2}  + \frac{\delta \hat{\sigma}(t)\hat{b}^\dagger(t)}{-i\Delta_a - i\omega_b-\kappa/2} \right].
\end{align}
As before, the emission spectrum from these fluctuations can be calculated from the Fourier transform of the two-time correlator
\begin{align}
    \mean{\hat{a}^\dagger(0) \hat{a}(t)} \approx &\left|\frac{g_{0,\sigma}}{\kappa/2 + i(-\Delta_a + \omega_b)}\right|^2 \mean{\delta \hat{\sigma}^\dagger (0)\hat{b}^\dagger(0)\delta \hat{\sigma} (t)\hat{b}(t)} \nonumber \\
    &+\left|\frac{g_{0,\sigma}}{\kappa/2 - i(-\Delta_a + \omega_b)}\right|^2 \mean{\delta \hat{\sigma}^\dagger (0)\hat{b}(0)\delta \hat{\sigma} (t)\hat{b}^\dagger(t)} \nonumber \\
    = &\frac{g_{0,\sigma}^2}{\kappa^2/4 + (-\Delta_a + \omega_b)^2} \mean{\delta \hat{\sigma}^\dagger (0)\hat{b}^\dagger(0)\delta \hat{\sigma} (t)\hat{b}(t)} \nonumber \\
    &+\frac{g_{0,\sigma}^2}{\kappa^2/4 + (-\Delta_a - \omega_b)^2} \mean{\delta \hat{\sigma}^\dagger (0)\hat{b}(0)\delta \hat{\sigma} (t)\hat{b}^\dagger(t)},
\end{align}
where we dropped terms proportional to the squeezing of the mechanical mode. If we additionally factorize the correlators, evaluate the correlators assuming decoupled dynamics, and use correlators given in Eqs.~\eqref{eq:nb} and \eqref{eq:nb.2}, we find
\begin{align}
    &\mean{\delta \hat{\sigma}^\dagger (0)\hat{b}^\dagger(0)\delta \hat{\sigma} (t)\hat{b}(t)} \approx \mean{\delta \hat{\sigma}^\dagger (0)\delta \hat{\sigma} (t)}\mean{\hat{b}^\dagger(0)\hat{b}(t)} \approx \mean{\delta \hat{\sigma}^\dagger (0)\delta \hat{\sigma} (0)}\nbth e^{(-i\omega_b-\Gamma/2) t} \nonumber \\
    &\mean{\delta \hat{\sigma}^\dagger (0)\hat{b}(0)\delta \hat{\sigma} (t)\hat{b}^\dagger(t)} \approx \mean{\delta \hat{\sigma}^\dagger (0)\delta \hat{\sigma} (t)}\mean{\hat{b}(0)\hat{b}^\dagger(t)} \approx \mean{\delta \hat{\sigma}^\dagger (0)\delta \hat{\sigma} (0)}(\nbth+1) e^{(i\omega_b-\Gamma/2) t}.
\end{align}
Finally, the emission spectrum takes on a form very similar to that obtained through the linearization procedure (Eq.~\eqref{eq:spectrum}):
\begin{align}\label{eq:spectrum.incoh}
    S_a(\omega) &=2\omega^4 g_{0,\sigma}^2 \mean{\delta \hat{\sigma}^\dagger\delta \hat{\sigma}} \left[\frac{n_b}{(\Delta_a - \omega_b)^2 +(\kappa/2)^2} \frac{\Gamma}{(\omega_b+\omega_l - \omega)^2+(\Gamma/2)^2}\right. \nonumber \\
    &\quad\quad\quad\quad\quad\quad\quad\quad\quad\left.+
    \frac{n_b+1}{(\Delta_a + \omega_b)^2 +(\kappa/2)^2} \frac{\Gamma}{(\omega_b-\omega_l +\omega)^2+(\Gamma/2)^2}\right],
\end{align}
except for the change of $|g_\sigma|^2= g_{0,\sigma}^2 |\alpha_\sigma|^2 \rightarrow g_{0,\sigma}^2 \mean{\delta \hat{\sigma}^\dagger\delta \hat{\sigma}}$. This observation would suggest that the overall intensity of the Raman scattering should depend explicitly \textit{only} on the population of the atomic antenna, irrespective of whether it is in a coherent, or a maximally mixed state. We should note that the derivation carried out in this, and the previous subsection, rely on the same approximation, of decoupling the state of the atomic antenna from other components of the system. This approximation would likely break down only for the stronger single-photon coupling rates $g_{0,\sigma}$.

}

\section{Notes on the estimates for Table~1}\label{app:parameters}

To estimate the coherent driving amplitude and population of the atomic antenna, we need to estimate the dipole moment $\mathbf{d}_\sigma$ associated with the driven transition. With the transition frequency in GeV of $\omega_\sigma/2\pi=498$~THz (602 nm), we can estimate $|\mathbf{d}_\sigma|^2$ from the radiative contribution to the decay rate $\gamma_0/2\pi \approx 25$~MHz~\cite{li_atomic_2024}, which for a dipolar electric transition dipole should be given by
\begin{equation}
    \gamma_0 = n\frac{\omega_\sigma^3 |\mathbf{d}_\sigma|^2}{3\pi \varepsilon_0 \hbar c^3},
\end{equation}
where $n$ is the refractive index of the medium ($n=2.42$ for diamond). From here we get 
\begin{equation}
    |\mathbf{d}_\sigma|^2 = \gamma_0\frac{3\pi\varepsilon_0 \hbar c^3}{n\omega_\sigma^3} \approx 1.2\times 10^{-57}~(\text{Cm})^2,
\end{equation}
or $|\mathbf{d}_\sigma| \approx 2.2\times 10^{-29}~\text{Cm}\approx 6.7$~D. 

To calculate the optomechanical coupling in atomic antenna-enhanced systems, we take the free-space Green's function, neglecting any inhomogeneities (such as interface between the dielectric and air, or off-resonant effects of metallic nanoantenna) and polarisation, and taking only its near-field component 
 \begin{equation}
     G_\text{nf}(\mathbf{r}_\sigma,\mathbf{r}_m) \approx \frac{1}{4\pi k^2 |\mathbf{r}_\sigma-\mathbf{r}_m|^3},
 \end{equation}     
{and use it to estimate the electric field from the antenna at the position of the molecule
\begin{equation}\label{eq:e.field.GeV}
    |\mathbf{e}_\sigma(\mathbf{r}_m)| \approx \omega_\sigma^2 \mu_0 G(\mathbf{r}_\sigma,\mathbf{r}_m)|\mathbf{d}_\sigma|.
\end{equation}
}
 
Finally, as an example of a Raman-active molecule, we consider a $\omega_b/2\pi = 40.5$~THz ($1351~\text{cm}^{-1}$) vibrational mode of the R6G~\cite{roelli_molecular_2016} with Raman activity in the Gaussian units given as $3.5\times 10^2 \varepsilon_0^2 \text{\AA}^4 \text{amu}^{-1}$, and here assumed to be equal to $R^2$; accounting for the transformation between Gaussian and SI units (see the SM of Ref.~[\citenum{roelli_molecular_2016}]) we find
\begin{equation}
    R Q^0 \approx 4\pi \varepsilon_0 2\times 10^{-30}~\text{m}^3.
\end{equation}

\section{Parameters of conventional plasmonic SERS}\label{app:conventional}

In this Appendix we repeat the definitions of the parameters of a conventional molecular optomechanical system, in which a single plasmonic cavity mode (with parameters denoted in the main text) mediates both the enhancement of the optical driving, and the emission from the Raman-active molecule. 
\begin{itemize}
    \item coherent driving amplitude of the cavity mode can be estimated as
        \begin{equation}\label{eq:plasmon.driving}
            \Omega_a = \sqrt{\frac{I \sigma_\text{ext}}{Q 4\hbar}},
        \end{equation}
        where $Q = \omega_a/\kappa$ denotes the quality factor of the mode; this expression differs by a factor of $\sqrt{2}$ from the formulation given in~Ref.~[\citenum{roelli_nanocavities_2024}]. When decoupled from (or weakly coupled to) the Raman-active molecule, and driven on resonance, the cavity settles into a coherent state with amplitude
        \begin{equation}\label{eq:plasmon.amplitude}
            \alpha_a = \frac{2}{\kappa} \sqrt{\frac{I \sigma_\text{ext}}{Q 4\hbar}}.
        \end{equation}
        Thus, reaching $|\alpha_a| = 1$ would require laser intensity 
        \begin{equation}\label{eq:plasmon.threshold}
            I = \frac{\kappa^2}{4} \frac{Q 4\hbar}{\sigma_\text{ext}}.
        \end{equation}
    \item in the approximation where the electric fields of the mode are given as 
        \begin{equation}\label{eq:plasmon.field}
             |\mathbf{e}_a| = \sqrt{\frac{\hbar \omega_a}{2\varepsilon_0 V_\text{eff}}},
        \end{equation}
        the optomechanical coupling takes on a simple form
        \begin{equation}\label{eq:plasmon.coupling}
             g_{0,a} = \frac{Q^0 R \omega_a}{2\varepsilon_0 V_\text{eff}},
        \end{equation}
        and the effective optomechanical coupling is defined as $g_a = g_{0,a}\alpha_a$.
    \item the emission from the Raman dipole is, similarly as in the case of an atomic antenna, mediated by the plasmonic cavity mode. Therefore, the Stokes and anti-Stokes emission intensities are governed by expressions similar to those listed in Eq.~\eqref{eq:Stokes.b}, except for replacing the effective coupling $g_\sigma \rightarrow g_a$
        \begin{align}\label{eq:Stokes.b.conv}
            &S_{a,\text{conv}}(\omega_S) =\frac{2\omega_S^4 |g_a|^2 (n_b+1)}{[(\Delta_a + \omega_b)^2 +(\kappa/2)^2]\Gamma}, \\ 
            &S_{a,\text{conv}}(\omega_{aS}) = \frac{2\omega_{aS}^4 |g_a|^2 n_b}{[(\Delta_a - \omega_b)^2 +(\kappa/2)^2]\Gamma}.
        \end{align}      
\end{itemize}
The derivations of the above expressions can be found in Refs.~[\citenum{roelli_molecular_2016,schmidt_quantum_2016,roelli_nanocavities_2024,schmidt_linking_2017}].

\section{Numerical modelling}\label{app:comsol}

{\subsection{Modelling the classical response of the setup}\label{subsec:modelling.comsol}}

To calculate the characteristics of an atomic antenna-enhanced Raman setup, we have implemented two COMSOL models which share the same geometry: a gold nanoparticle with radius of 100 nm, and permittivity $\varepsilon_\text{Au} = -9.53+1.51i$ (value taken from Ref.~\citenum{christy}, at 602 nm), separated by distance $d$ from the flat diamond surface with permittivity $\varepsilon_\text{diam.} = n_\text{diam.}^2 = 5.86$. The computational domain was a sphere with radius of 3~$\mu$m, with a \textit{perfectly matching layer} (PML) of 1~$\mu$m thickness. Meshing was chosen to ensure convergence, and increasingly fine towards the gap, where the element size was maxed as 0.2~nm.

We used that geometry in two settings:
\begin{itemize}
    \item for the calculation of $K_\text{inc}$ we assumed a background field associated with incident $p$-polarized planewave at 60 degree incidence, and observed the electric field at where we expect the atomic antenna to be positioned  --- on the axis of symmetry of the system, $2$~nm below the diamond surface,
    \item for the calculation of $K_\text{mol}$ and $\gamma_\text{tot}$, an electric dipole positioned at the site of the atomic antenna, oriented vertically.
\end{itemize}

{
In Fig.~\ref{fig:Purcell} we plot the Purcell enhancement (solid orange line), split between the absorption (dashed green line) and radiation (dashed blue line), as a function of spacing between the substrate and the nanoparticle.

\begin{figure}[ht!]
	\centering
	\includegraphics[width=.5\linewidth]{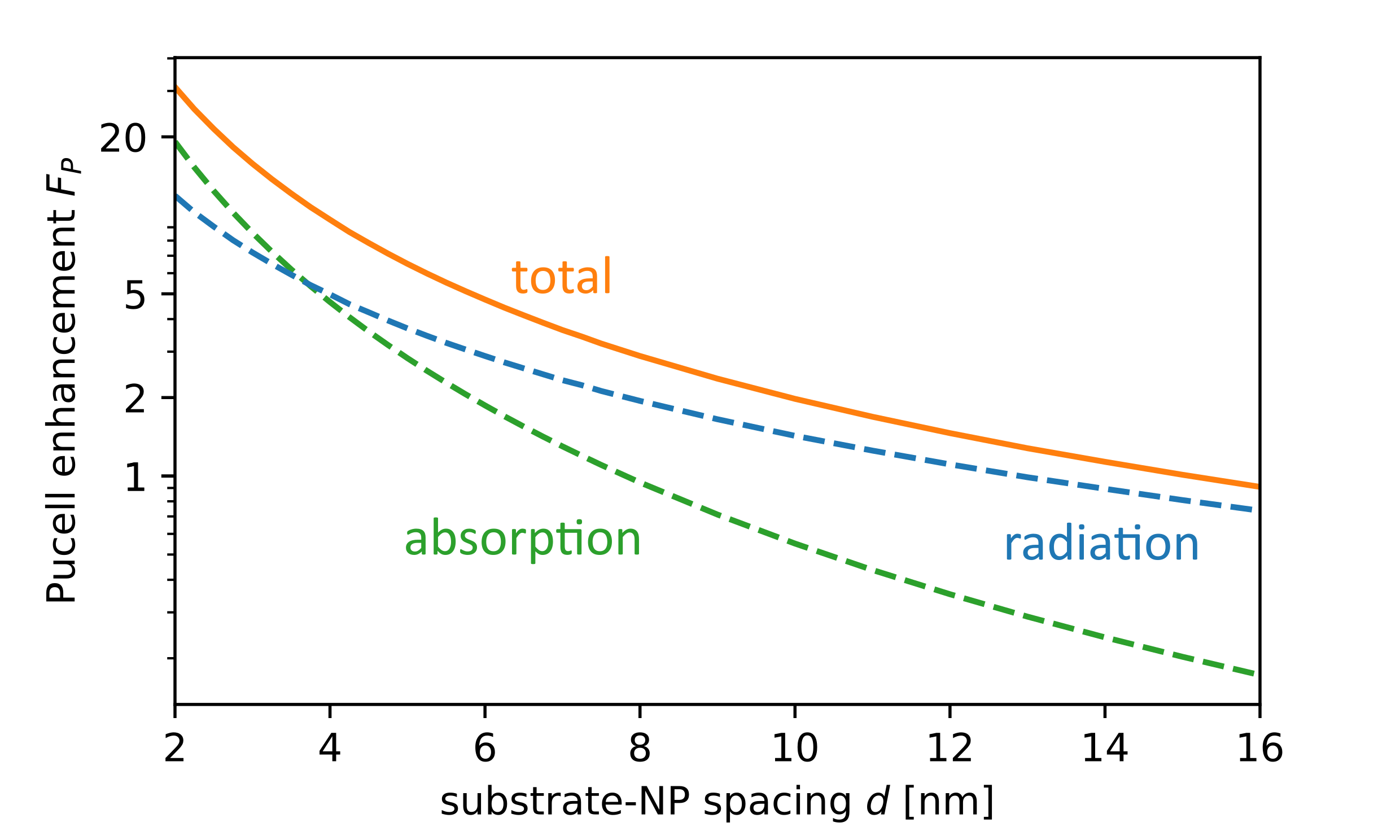}
	\caption{{Numerically calculated Purcell factor (solid orange line), as a function of spacing between the substrate and the nanoparticle (see the schematic in Fig.~2(a) and Section~\ref{subsec:modelling.comsol} for details of this model). Absorptive and radiative channels are depicted with the dashed green blue lines, respectively}}
	\label{fig:Purcell}
\end{figure}

}

{
\subsubsection{Derivation of Eq.~(19)}

In bulk diamond, the total decay rate of the GeV $\gamma = \gamma_\text{nr}+\gamma_\text{r}$ is determined by its radiation (with rate $\gamma_\text{r}=\gamma_0$), and nonradiative internal processes (with rate estimated from experiments as $\gamma_\text{nr}=3\gamma_0$ at 50 K)~\cite{li_atomic_2024}. When the defect is coupled to a structured environment, the internal nonradiative processes still occur at the same rate $\gamma_\text{nr}$. Conversely, the rate of emission from the emitter $\gamma_\text{r}$ is enhanced via the Purcell effect to $F_P\gamma_\text{r}$ (whether this enhancement is related to the increase of the radiative density of states, or absorption in the environment), and the total decay rate becomes 
\begin{align}
        \gamma \rightarrow \gamma_\text{tot} &= F_P \gamma_0 \, + 3\gamma_0,
\end{align}
which is the first line in Eq.~\eqref{eq:total.rate}.

In our work, the Purcell factor is numerically calculated as a ratio of the power absorbed in the environment and radiated from an electric dipole (positioned at the atomic antenna, and characterised with the quantum efficiency of 1) with, and without the nanoparticle:
\begin{align}
        F_P=\frac{(P_\text{rad}+P_\text{abs})~\text{(with nanoparticle)}}{P_\text{rad}^0~\text{(without nanoparticle)}},
\end{align}
where in the absence of nanoparticle we explicitly denote radiated power as $P_\text{rad}^0 = P_0$, and note that there are no losses in the medium that could yield absorption.

While the Purcell factor could alternatively be calculated from the Green’s function of the system, we found that the former approach is more numerically stable.

{\subsection{Modelling the quantum dynamics of the system}\label{app:extra}}

To understand the response of the system, comprising atomic antenna ($\hat{\sigma}$), plasmonic cavity mode ($\hat{a}$) and Raman-active molecule with a single vibrational mode ($\hat{b}$), in more detail (especially beyond the linearization regime), we numerically model its dynamics using QuTiP solvers~\cite{johansson_qutip_2013,johansson_qutip_2012}. The coherent dynamics is governed by the full Hamiltonian, accounting for all the terms discussed in Section~\ref{subsec:complete.Hamiltonian} and in Fig.~\ref{fig:schematic}:
\begin{align}
    \hat{H} =~&\hbar\omega_b \hat{b}^\dag\hat{b} + \hbar(\omega_\sigma-\omega_l) \hat{\sigma}^\dag \hat{\sigma} + \hbar(\omega_a-\omega_l) \hat{a}^\dag\hat{a} \nonumber \\
    &+ \hbar\Omega\left(\hat{\sigma} + \hat{\sigma}^\dag\right)+ \hbar\Omega_a\left(\hat{a} + \hat{a}^\dag\right)\nonumber \\
    &- \hbar g_{0,\sigma}\left(\hat{\sigma}^\dagger \hat{a}+\hat{a}^\dagger \hat{\sigma}\right)\left(\hat{b} + \hat{b}^\dag\right)- \hbar g_{0,a}\hat{a}^\dagger \hat{a}\left(\hat{b} + \hat{b}^\dag\right)\nonumber \\
    &+\hbar g_{JC}\left(\hat{\sigma}^\dagger \hat{a}+\hat{a}^\dagger \hat{\sigma}\right).
\end{align}
Besides the coherent driving ($\propto \Omega$), and optomechanical coupling mediated by the atomic antenna ($\propto g_{0,\sigma}$), we also include terms describing pumping of the cavity mode ($\propto \Omega_a$), conventional Raman interaction ($\propto g_{0,a}$), and the Jaynes-Cummings coupling between the GeV and cavity mode (see discussion in Section~\ref{subsec:complete.Hamiltonian}).

The incoherent dynamics is described as in the main text (Section~\ref{subsec:dissipation}).

For the calculations discussed in the main text, we use coupling, frequency and dissipation parameters as defined in Table~1 and take
$g_{JC}/2\pi = 10^{11}$~Hz (which would yield the Purcell factor around 2, identified for the spacings between the substrate and nanoparticle larger than 12 nm; see Section~\ref{subsec:complete.Hamiltonian}). To calculate the spectra in Fig.~\ref{fig:nonlinear}(a,c), we use the following laser intensities:
\begin{align}
    \text{for Fig.~\ref{fig:nonlinear}(a):}~~I =~&6.8\times 10^{-4},~1.7\times 10^{-3},~4.3\times 10^{-3},~1.1\times 10^{-2},~2.7\times 10^{-2},\nonumber \\
    &6.9\times 10^{-2},~0.17~\upmu\text{W}/\upmu\text{m}^2;\nonumber \\
    \text{for Fig.~\ref{fig:nonlinear}(c):}~~I =~&1.7\times 10^{-5},~1.1\times 10^{-4},~6.9\times 10^{-4},~4.4\times 10^{-3},~2.7\times 10^{-2},\nonumber \\
    &0.17,~1.2,~6.9,~43.4~\upmu\text{W}/\upmu\text{m}^2\nonumber.
\end{align}
These values are marked with vertical dashed lines in Fig.~\ref{fig:nonlinear}(b,d).

To identify the Raman scattering from conventional SERS and calculate the emission spectra $S_{a,\text{conv}}$, we run the same simulations setting $g_{0,\sigma}=0$.
}

\end{widetext}


\end{document}